\documentclass[12pt,epsfig]{report}
\usepackage{graphicx,color}
\usepackage{pdfpages}
\usepackage{float}
\usepackage[dvipsnames]{xcolor}
\usepackage{amsmath}
\usepackage{mathtools}
\usepackage{amsfonts,amsmath,amssymb}
\usepackage{amssymb}
\usepackage{epsfig}
\usepackage{appendix}
\usepackage{pdfpages}
\usepackage{latexsym}
\usepackage{fontenc}
\usepackage{hyperref}
\hypersetup{
	colorlinks=true,
	linkcolor=blue,
	filecolor=RawSienna,      
	urlcolor=blue,
	citecolor=blue
}
\graphicspath{ {figures/} }
\usepackage{appendix}
\usepackage[english]{babel}
\usepackage{physics}
\usepackage{bigints} 
\usepackage{relsize}
\usepackage{braket}
\usepackage{blindtext}
\usepackage[titles]{tocloft}  
\NewDocumentCommand{\tens}{t_}
{%
	\IfBooleanTF{#1}
	{\tensop}
	{\otimes}%
}
\NewDocumentCommand{\tensop}{m}
{%
	\mathbin{\mathop{\otimes}\displaylimits_{#1}}%
}

\usepackage{changepage} 

\usepackage{afterpage}

\usepackage{placeins}

\usepackage{wrapfig}
\usepackage[top=1in, bottom=1in, left=0.75in, right=0.75in]{geometry}
\newcommand{\dsl}{\pa \kern-0.5em /}

\newcommand{\pa}{\partial}

\RequirePackage[most]{tcolorbox}

\begin{document}
\begin{center}
\pagenumbering{roman}
\thispagestyle{empty}
{\Huge \bf Some Studies On Exact Solutions}
\vskip 0.5cm
{\Huge \bf Of Models In Noncommutative Spaces}
\vskip 5.0 true cm
\large Thesis submitted for the degree of\\
Doctor of Philosophy (Science)\\
in\\
Physics (Theoretical)
\vskip 5.0 true cm
 by \\
{\large \bf Manjari Dutta}
\vskip 1.5 true cm
Department of Physics\\
University of Calcutta\\
2024
\end{center}

\newpage
\begin{center}
	~~
	\vskip 3cm
	{\Huge   \textsl{Dedicated to}}
	
\end{center}
\vskip 2cm
{\textit{From the Golden Past\,:}}
\begin{center}
	~~
	{\Large   \textbf{{Late Narottam Dutta \vskip .30cm Late Manik Mazumder}}}\vskip .20cm (My grand fathers)
	
\end{center}
{\textit{From the Vibrant Present\,:}}
\begin{center}
	~~
	{\Large   \textbf{{Mr. Madhusudan Dutta, Mrs. Shampa Dutta }}}\vskip .20cm (My parents)
	
\end{center}
\begin{center}
	~~
	{\Large   \textbf{{Dr. Sunandan Gangopadhyay, \vskip .30cm Dr. Shreemoyee Ganguly }}}\vskip .20cm (My academic parents)
	
\end{center}

\thispagestyle{empty}
\newpage

\begin{center}
	{\Large  \textsf{Acknowledgements}}\vskip .30cm
\noindent \textbf{\textit{``Empathy" is a true essence of ``Love", ``Love" is for the sake of ``Life"\\ and ``Life" is lifeless without growth.\\ I would like to take this opportunity to express my heartfelt gratitude to those people, irrespective of whether they already belong to my life or have not yet come into it, who are always earnestly eager to see me move forward in academic life and empathetically consider my academic progress as their own progress, as well as my academic success as their own success.}}
\end{center}
\textsl{First, I would like to express my heartfelt gratitude to are my supervisor Prof.\,Sunandan Gangopadhyay (Sir) and my collaborator Prof.\,Shreemoyee\,Ganguly (Madam) whom I unofficially consider as another supervisor. They are not only the people who professionally take care my academic works but also the only source of unconditional love and support in my present professional life. As the first student of their academic collaboration, they always treat me like their academic child. To help me matured academically, they always put in as much effort as my parents would and share in my academic successes as my parents would. They have not only supervised my work but also taught me various things such as paper writing, preparing presentations and delivering talks. It is not possible to describe their contribution to my academic life and the impression I have received from their guidance in just a few words. Undoubtedly, they are one of the precious academic gifts from my life to me.\\
I acknowledge S. N. Bose National Centre for Basic Sciences for providing research funding to me as an internal research fellow. I would like to convey my respectful acknowledgement to Prof. Samit Kumar Roy, former Director of S. N. Bose National Centre for Basic Sciences, for granting me the precious opportunity to conduct research here since I started my research as an institute fellow during his tenure. In this regard, I would also like to respectfully mention Prof. Tanusri Saha Dasgupta, the current director of the institute, under whose tenure I have spent my academic life over the past years.\\
Now, I would like to take the opportunity to acknowledge my course work instructors like Prof. Subhrangshu Sekhar Manna (computational methods in physics), Prof. Amitabha Lahiri (classical dynamics), Prof. Sakuntala Chatterjee (quantum mechanics I), Prof. Samir Kumar Paul (mathematical methods), Prof. Manu Mathur (electromagnetic theory), Prof. Rabin Banerjee (quantum mechanics II), Prof. Jaydeb Chakrabarti (statistical mechanics), Prof. Rajib Kumar Mitra and Prof. Anjan Barman (atomic and molecular physics), Prof. Manoranjan Kumar (condensed matter physics), Prof. Ramkrishna Das and Prof. Soumen Mandal (astrophysics and astronomy), Prof. Partha Guha (advanced mathematical methods), Dr. Manik Banik (quantum information theory). Here, I also respectfully mention my supervisor (Sir) as a very good course instructor (advanced quantum mechanics, quantum field theory, general theory of relativity) who is always eager to see his students deliver presentations as lively as he conducts his classes. In this regard, I would like to express my gratitude to Prof. Archan S. Majumdar for granting me the opportunity to pursue my summer research project under his supervision, and to Prof. Biswajit Chakraborty for supporting me during my academic crisis at this center.\\
It is my great pleasure to thank my academic (\textit{S-Gravity}) group members who have made this five year academic journey pleasant and memorable for me. First, I would like to deeply acknowledge my collaborator, Mr.\,Soham Sen (beloved junior and genius brother), with whom any academic discussion enriches my understanding greatly due to his deep knowledge and insights in various aspects. His expertise and understanding always elevate our discussions, making him a mentor like figure to me. Over the past years, his guidance has significantly enriched my academic journey. Within my campus, whenever I face any difficulties, he is always there for me. In addition, I would like to express my sincere thanks to another collaborator, Mr.\,Arnab Mukherjee (college friend and faithful colleague) for engaging in useful discussions with him many times. Apart from my collaborators, I would like to acknowledge my adorable seniors, such as Debabrata (Ghorai) Da, Rituparna (Mandal) Di, and Ashis (Saha) Da, who have assisted me in operating the software Mathematica and Texmaker. Ritu Di always affectionately treats me like her younger sister and patiently accepts my temper at times, and Ashis Da provided me with his own thesis template to write this thesis. I would also like to mention other supportive seniors, such as Abhijit (Dutta) Da, Ashmita (Das) Di, Sukanta (Bhattacharyya) Da, Sourav (Karar) Da, Diganta (Parai) Da, and Suchetana (Pal) Di, for being my well wishers and for sharing many memorable moments with me. In this group, I also have various colourful friends like Anish Das, Ankur Srivastav, and Neeraj Kumar, who not only guide me in handling official formalities properly but also gift me beautiful moments with them. Especially, during my M. Sc. days, I had many useful academic discussions with Ankur and Neeraj. Next, I would like to mention my other loving junior brother, Anirban Roy Chowdhury (Motu), who always tries to support and help me whenever I need it, even during my COVID period on campus. I often use to enjoy the senior like attitude of this sweet brother. Undoubtedly, he is one of my most faithful juniors here. I am also grateful to my other juniors, like Jayarshi Bhattacharya and Souvik Paul, for their useful suggestions and assistances on many academic and technical occasions. Finally, I extend my warm welcome to Ms. Arpita Jana and Mr. Gopinath Guin, the newest members of our group and wish them a successful research career ahead. It would be very unfair if I do not acknowledge the soulful and skilful singing abilities of my supervisor (Sir) and Soham, which are the sources of my mental refreshment I receive from my group.\\
It is my good fortune to share a soulful hug with Ms. Nivedita Pan (my spiritual sister) and Mrs. Ria Saha Das (my ``Kirti"), who are an inseparable part of my campus life. Despite being under different supervisors, we share a very strong spiritual connection. On campus, they always play the roles of my right and left hands through their unconditional love and endless mental support. They are two of the very few people who witness all my true emotions, handle them sensitively and become genuinely happy whenever I make any remarkable progress in my academic life. Indeed, they are my lifetime achievement from this institute. Although I do not meet with Amrita Mondal and Didhiti Bhattacharya frequently, I have gained many useful academic support and informations from them. Additionally, my spiritual discussions with them are truly memorable to me. It is my pleasure to mention another close friend, Anwesha Chakraborty, who is also my college friend. I highly appreciate her research loving attitude, and the suggestions I have received from her have often been useful to me. I also feel really fortunate to have faithful classmates like Parthapratim Mahapatra, Harmit Joshi, Samir Rom, Shubham Purwar (Dash) and Riju Pal, with whom I have shared so many colourful and funny memories, both inside and outside the classroom. During our MSc days, we overcame many academic difficulties together, and their presence in my life is truly a source of mental support for me. Mr. Rajdeep Biswas (king of my adorable and faithful brothers group) deserves a special place in my campus life. I really enjoy his funny madness with his classic friends' group (Shinjini, Ishita, Soumen, Abhik, Animesh, Koushik, Soham, Ankita) and appreciate his serious actions when needed. On this campus, he is always there for me whenever I need him. Among the seniors from the other academic group, I would like to mention some names specially, such as Riddhi (Chatterjee) Di, Partha (Nandy) Da, and Ruchi (Pandey) Di. Riddhi di and Partha da have really helped me during my summer research project by providing useful notes and discussing crucial academic topics with me. Ruchi di, being my closest neighbour in our working bay, was always there to assist me with various official formalities. I am really grateful to them. Samiran (Chowdhury) Da also deserves special mention; from him, I have received the spirit of acknowledging people soulfully, as he did not overlook anyone inside or outside the campus in the acknowledgement section of his own thesis. I should also mention other snb guys who will be forever in my heart, such as, Arghya da, Shubhadip (Lalu) da, Biplab da, Arnab (Sarkar) da, Anuvab da, Debasish da, Priyanka di, Anulekha di, Anindita di, Kajal di, Monalisa di, Shouri di, Suraka di, Sanchi di, Indranil da, Prantik da, Arundhuti di, Shreya di, Anita di, Jayita di, Saili di, Susmita di, Ria (Ghosh) di, Lopamudra di, Indrani di, Sumana di (Khusi), Sumanti di, Anirban (Damra) da, Imadul da, Rafiqul da, Siddhartha (Sidhu),Rahul (Karmakar), Shashank, Shantanu, Anupam, Swarnali, Sudip, Megha, Sayan, Achintya, Anuj, Raghavendra, Ardhendu, Krishnendu (kaka), Shubhajit, Kanchan, Sreya, Soma, Ashesh, Aritra, Parama, Madhurita, Madhumita, Monalisa, Riya (Barik), Ananya, Saheli, Anushree, Prapti, Sayari, Avanti, Supriti, Rupayan, Shweta etc. I also deeply acknowledge all the volunteers (including Abhik, Indrani etc.) who used to served me during my COVID period. Here, I should mention Santwana and her vibrant little child, Aitihya, who are family members of Narayan Da, one of the research scholars at our center, for sharing very special and precious memories with them. Next, I want to thank the SNB staffs, such as Nibedita (Konar) madam, Chandrakana di, Rupam da, Jaydeep da, Sanchari di, Debalina di as they always tried to cooperate with me whenever I went there for any official purpose. The security group of the centre, the caretaker (Kamakhya da) and cleaner aunties (Babli masi, Piyali di, Sandhya masi, Pinki di, Shila masi) at my hostel, the cooking team (Kajal da, Ramu da, Sanjay da, Ganesh da, Mano da, Asim da etc.) of our SNB mess, and the (Arup da, Mrinmoy da, Hori da etc.) Bhagirathi canteen are indeed an inseparable part of our daily life because, thanks to their service, we can perform our jobs smoothly without any other non academic headaches. Outside the campus, the evening atmosphere at Dulal da's tea shop in IC block always evokes refreshing memories for us SNB fellows, as we used to go there for evening snacks and some relaxation together.   \\
Next, I want to address those people who were directly related to my academic life in the past, before I joined here. In this regard, I want to acknowledge my college teachers, Dr. Tapas Dutta, Dr. Gauranga Sinha Mahapatra, Dr. Shankhya Das, Dr. Prasenjit Sarkar, as well as tuition teachers from my college and school life: Dr. Arun Kumar Mukherjee (popularly known as AKM), Dr. Timir Baran Chakraborty, Dr. Partha Ghosh, Bapi sir (Pranab Ganguly), Pradipta sir, Manta kaka (Asit Baran Bhowmik), and school teachers such as Koushik (Mal) Babu, Sudeshna (Mukherjee) Madam, Somdutta (Koley) Madam and Debojyoti (Ghosh) sir. They always gave me their best to build a solid foundation for my academic life in science stream.\\
\noindent Now, I want to mention those friends who are not directly involved in my academic life but always provide support, motivation, and inspiration to proceed further in my career. Mrs. Kuheli Mukherjee Chakraborty, despite being a so called ordinary home maker of my age, carries a guardian like concern about my academic life and is always there, with her endless love, to cheer me up spiritually when I lack confidence or become frustrated and hopeless in my academic life. She never wants me to stop in my career. Pallabi Shit, my college friend, definitely deserves my honest acknowledgement for being mentally and spiritually connected with my academic successes. In this regard, I would also like to mention some other relevant names like Ankita Bhattacharya, Nandita Pan, Prajnya Singha and Piyali Adhikary etc.\\
Here, I want to address the most sensitive individuals who are empathetically connected with my academic life: My Parents and family members (including Mrs. Rina Mazumder, my Grandmother and Mr. Anirban Mazumder, my Maternal uncle). As I mentioned at the beginning of this acknowledgement section, they are not only the principal source of unconditional love and support in my life but also consistently eager to see me progress in my academic journey, considering my achievements as their own. Over the last five years, they have been wholeheartedly involved in this thesis work. My adorable cousin (Archisman) must deserve a special mention for being well aware of my academic journey and being my well wisher. I also convey my deep respect to the other family members (my uncles, aunts, cousins, grandaunts etc.) whom I cannot specifically mention here, but who are my well wishers forever.
\\
At the last and most important part of this acknowledgement section, I take the opportunity to express my respectful gratitude to Sri Abinendranath Chakravarty, an Indian personality (promotes ``\textsl{Unity in Diversity}" ), for suggesting an interesting and important direction in my academic life, which now serves as the principal motivation in my current academic endeavours.}
\newpage


\newpage
\begin{abstract}
This thesis explores exactly solvable quantum mechanical models in noncommutative spaces under time dependent backgrounds. It starts by examining a two dimensional damped harmonic oscillator in a time varying noncommutative space. Using the Lewis invariant method and the Ermakov-Pinney equation associated with this method, the study derives analytical eigenfunctions and explicit solutions for the noncommutative parameters under specific damping factors and oscillator frequencies. These solutions allow for exact expressions for the Lewis phase, linking the Hamiltonian's eigenstates to those of the Lewis invariant operator. The Hamiltonian's expectation value is computed, revealing time varying energy profiles based on different damping factors and oscillator frequencies. Next, the study extends the model by introducing an externally applied time varying magnetic field within a dynamic noncommutative framework. The system is solved using the same Lewis treatment, yielding explicit solutions for the noncommutative parameters and the Lewis phase. The study examines how the combined effects of damping and the magnetic field impact the system's energy profile, comparing the results with those obtained without the magnetic field. The third study investigates Berry's geometric phase in periodic models of the harmonic oscillators solvable in time varying noncommutative frameworks. Two systems are considered in this regard. One with a time reversal symmetry breaking scale invariant term in the Hamiltonian, and another where the scale invariant term arises from transforming noncommutative to commutative variables. The geometric phase is derived from the Lewis phase using the Lewis Riesenfeld treatment, and the Berry phase is explicitly calculated. The study finds out that a time reversal symmetry breaking scale invariant term cannot ensure a nontrivial Berry phase always. The final part explores a simple harmonic oscillator in a two dimensional dynamic noncommutative framework, a generalization of standard noncommutativity. The study solves the time dependent system, deriving exact analytical eigenfunctions and energy expressions, and investigates their dynamics graphically. It also establishes generalized uncertainty relations for both noncommutative and standard canonical operators. The results are consistent with previous findings for similar models under standard noncommutativity. 
\end{abstract}

\newpage

\includepdf[pages=-]{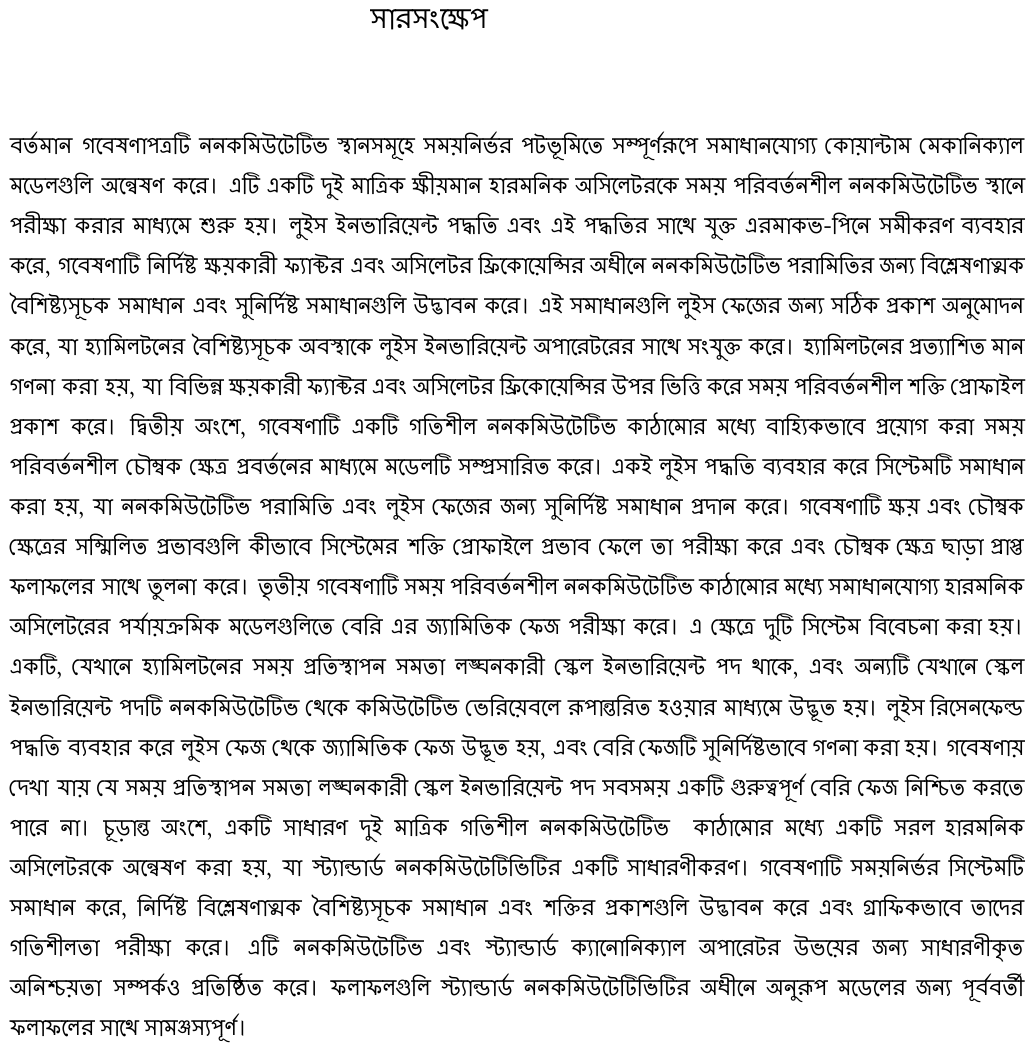}

\newpage

\begin{center}
 {\large \bf List of publications on which the thesis is based}
\end{center}
\begin{enumerate}
\item \textbf{Manjari Dutta}, Shreemoyee Ganguly and Sunandan Gangopadhyay {\it ``Exact solutions of a damped harmonic oscillator in a time dependent noncommutative space"},\\ \href{https://doi.org/10.1007/s10773-020-04637-4}{Int. Jnl. Theor. Phys. , 59, (2020)~3852.}

\item \textbf{Manjari Dutta}, Shreemoyee Ganguly and Sunandan Gangopadhyay {\it ``Investigation of a harmonic oscillator in a magnetic field with damping and time dependent noncommutativity"}, \href{https://doi.org/10.1088/1402-4896/ac2b4c}{Phys. Scr. 96 (2021) 125224.} 

\item \textbf{Manjari Dutta}, Shreemoyee Ganguly and Sunandan Gangopadhyay {\it ``Explicit form of Berry phase for time dependent harmonic oscillators in noncommutative space"},\\
\href{https://doi.org/10.1088/1402-4896/ac8dca}{Phys. Scr. 97 (2022) 105204.} 
\item \textbf{Manjari Dutta}, Shreemoyee Ganguly and Sunandan Gangopadhyay {\it ``Quantum Harmonic Oscillator in a Time Dependent Noncommutative Background"},\\
\href{https://doi.org/10.1007/s10773-024-05707-7}{Int J Theor Phys 63, 169 (2024)}.
\end{enumerate}

\vskip 1.0cm
\begin{center}
 {\large \bf List of publications not included in this thesis}
\end{center}

\begin{enumerate}
\item  Arnab Mukherjee, Sunandan Gangopadhyay and \textbf{Manjari Dutta}{\it ``Photon velocity, power spectrum in Unruh effect with modified dispersion relation"},\\
\href{https://doi.org/10.1209/0295-5075/129/30002}{2020 EPL 129 30002}.
\item \textbf{Manjari Dutta}, Shreemoyee Ganguly and Sunandan Gangopadhyay {\it ``A study of an Exotic Quantum Damped
Harmonic Oscillator"},\\
\href{http://doi.org/10.15864/ajps.1402}{AJPSA, Vol 1, Issue 4, 2021.}
\item \textbf{Manjari Dutta}, Shreemoyee Ganguly and Sunandan Gangopadhyay {\it ``Periodicity of damped harmonic oscillator affected by magnetic field in time
dependent noncommutative space"},\\
\href{http://doi.org/10.15864/ajps.2204}{AJPSA, Vol 2, Issue 2, 2022.} 
\item Soham Sen, \textbf{Manjari Dutta} and Sunandan Gangopadhyay{\it ``Lewis and Berry phases for a gravitational wave interacting with a quantum harmonic oscillator"}, \\
\href{https://doi.org/10.1088/1402-4896/ad1234}{Phys. Scr. 99 (2023) 015007}. 
\end{enumerate}



\newpage
\tableofcontents
\newpage
\thispagestyle{empty}
\listoffigures

\chapter{Introduction}
\pagenumbering{arabic}
The exactly solvable quantum mechanical models in noncommutative (NC)\footnote{We will consider NC phase space in our discussion but will generally refer to it as NC space.} spaces are currently an interesting area of study. The concept of NC space time or discrete space time, is crucial for unifying the quantum mechanics and theory of gravitation. Difficulties arise when a particle attempts to probe a very small distances at the Planck scale. According to the Heisenberg's uncertainty principle \cite{hei1}, the particle probes this extremely small length scale with a large amount of energy, resulting in the breakdown of space time manifold. Consequently, the laws of gravitation becomes irrelevant at the absence of the conceptual space time. However, the concept of minimum measurable length can prevent the probing particle from measuring beyond a certain length in space time manifold. Therefore, the consideration of such length is necessary for unifying the quantum field theory and general relativity. In 1930, Heisenberg first proposed the formalism of NC quantum field theory in a letter he wrote to Peierls \cite{hei}. This idea was later formalized in the pioneering work of Snyder \cite{Synder}, which sparked significant interest among theoretical physicists in studying quantum mechanical systems in NC spaces. In \cite{Synder}, an example of a discrete space time which remains invariant under Lorentz transformation, was provided and it was demonstrated that Lorentz invariance does not necessitate the usual assumption of continuous space time. Following this, Snyder's treatment was extended to curved space time, specifically de Sitter space, as shown by \cite{Yang}. Although initially ignored, the NC nature of space time later gained strong support from string theory \cite{amati}-\cite{pol2}, a leading theory of quantum gravity. String theory introduces the concept of strings having finite length $l_s$, which also defines the minimum distance in the space time manifold. Consequently, it is impossible to measure distances smaller than $l_s$. As demonstrated in a seminal work \cite{sw}, in a specific low energy regime, string theory can be interpreted as an effective quantum field theory in NC framework \cite{Doug, con}. Furthermore, nearly all quantum gravity theories  suggests the necessity of considering the concept of a minimum observable length, which is predicted to be near the Planck length,
\begin{equation}
l_p\,=\,\sqrt{\dfrac{G\,\hbar}{c^3}}\,=\,1.616 \times 10^{-35}~m~;
\end{equation}
and the concept of discrete space time, which can prevent gravitational collapse at this extremely small length scale. Thus, the concept of NC space time can also ensure gravitational stability~\cite{Dop1, Dop2}.\\
\noindent The impact of discrete space time have been explored in the context of single field inflation \cite{bal, nau}. Specifically, for curvature perturbations in Friedmann-Lemaître-Robertson-Walker space time, the two and three point correlation functions were computed, revealing that the power spectrum exhibited anisotropy due to NC geometry. Notably, the presence of a fundamental length scale in NC space time results in the violation of Lorentz symmetry. This violation extends to Poincaré symmetry and consequently affects the CPT theorem, microcausality, and spin statistics. However, the introduction of a deformed coproduct, as discussed in the literature \cite{chai, bloh, dimi, oeck}, can restore Poincaré symmetry. In relativistic quantum field theory \cite{bal1}, physicists have extensively explored such deformed coproducts, leading to investigations of its applications in various studies \cite{bal2, sgfgs2}.\\
\noindent Additionally, loop quantum gravity \cite{ast, rov} offers another approach to understanding quantum space time. Consequently, in the literature~\cite{suss}-\cite{GW3}, there have been numerous studies on quantum mechanical systems in the framework of NC space. \\
\noindent In the context of a simple model of noncommutative space, we consider a quantum mechanical space in two dimensions, where the commutator brackets among the noncommutative canonical coordinate operators are given by $[X, Y] = i\theta$, with $\theta$ being a real, positive constant. Apart from the configuration space, studies regarding the noncommutativity may also include NC commutation relations in momentum space such as $[P_x, P_y]=i\,\Omega$, where $\Omega$ is a real constant. The NC momentum operators in the system imply the existence of a magnetic field acting perpendicular to the configuration plane. However, in literature \cite{kem1, kem2, fr1, fr2, fr3, fr4}, there exist more interesting forms of NC algebra that can be expressed as functions of the canonical coordinate and momentum operators. In the literature on noncommutativity, the standard Bopp-shift relations~\cite{mez} have been found convenient for transforming a system from NC space to conventional quantum mechanical phase space. Nevertheless, a more generalized version of the Bopp-shift relations has been introduced in recent literature \cite{spb}.\\
\noindent Unlike the previous study where the NC parameters were held constant, our earlier research adopted both the coordinate mapping relations by introducing a time dependence in the NC parameters. The concept of dynamic noncommutativity can be considered a consequence of the renormalization group flow of $G$, Newton's gravitational constant, when the inverse of the energy scale is proportional to cosmic time. Such a form of Newton's constant arises from the solution of an exact functional renormalization group flow equation that is associated with an effective action that is dependent on the energy scale \cite{reu1, reu2, reu3}. Given that the minimum observable length is directly related to Newton's gravitational constant, it is reasonable to consider it also varies with time, similar to Newton's constant. String theory in a time dependent background \cite{sei1, sei2, Sen} also provides a motivation to look for dynamic NC parameters. In this context, time dependent NC parameters are emerged from a model system placed in a dynamic background, specifically a space time metric of the plane wave type that is supported by a Neveu–Schwarz two form potential \cite{dol}.
However, \cite{cal} also addresses both space and time dependent noncommutativity, considering Yang-Mills theories formulated on a spacetime dependent NC background.
The literature also includes other studies, especially regarding the exact quantum theory in dynamic NC framework. In \cite{strek2}, a follow up to \cite{strek1}, a harmonic oscillator in two dimensional dynamic NC background was studied, and its exact propagator was derived. The investigation of a simple prototype time independent model in such a varying NC space was initially pursued by \cite{Dey}. Employing Lewis's method of invariants \cite{Lewis3, Lewis2, Lewis}, the model of a time independent harmonic oscillator was solved exactly in two dimensional dynamic NC space. Since this method involves a nonlinear differential equation, namely Ermakov-Pinney (EP) equation \cite{Erm}-\cite{Pin}, it acts as a constraint relation governing the parameters existing in the Hamiltonian eigenfunctions. Interestingly, a specific integrability condition can aid in deriving explicit analytical forms of the parameters those adhere to the EP equation. On the other hand, the explicit solution of the EP equation presents the Hamiltonian eigenfunctions explicitly. Hence, a class of exact solutions for the model system can be generated by following this procedure. In \cite{Dey}, the generated explicit solution of the model system analysed the uncertainty relations, revealing that time dependence arises in their lower bounds. Moreover, in the Glauber coherent states, the uncertainties were also determined. This study indicates the potential for proceeding with the quantum theory in dynamic NC background. Hence, it motivates exploration of various models on different types of NC backgrounds, an area that has been relatively less explored until now. The crucial aspect to investigate in these studies is whether the models with various dynamic noncommutativities still permit achieving the exact solution of the Hamiltonian. Following the example set by \cite{Dey}, these studies should aim to discover exact solutions for prototype quantum mechanical systems within time dependent NC backgrounds, which promise to significantly enrich our understanding.\\
\noindent Motivated by this goal, we undertook our own investigations into exactly solvable models within dynamic NC spaces, employing the Lewis method of invariants for various analyses. In our various works, we address three key aspects related to the study of exactly solvable models. Firstly, we focus on developing effective procedures to obtain a class of explicit solutions in closed form for model systems and analyse these solutions using physical quantities such as energy expectation values, which can also be represented graphically. Secondly, the most intriguing aspect of this thesis involves separating the geometric component, under adiabatic approximation, from the Lewis phase factor which enables the construction of the eigenstate in an exact form for the periodic Hamiltonian. Lastly, we conduct a comparative study of an exactly solvable prototype model under two different types of noncommutativity. While our initial works \cite{SG1} and \cite{SG2} are related to the first aspect, the third study \cite{SG3} is under the second aspect and the final work \cite{SG4} corresponds to the last aspect. The thesis is organized in the following manner.\\
\noindent In chapter $\ref{chap-LR}$, we provide an overview of the Lewis-Riesenfeld theory, a useful method for solving time dependent Schrödinger equations exactly. We begin by introducing the concept of a time dependent hermitian invariant operator associated with a time dependent Hamiltonian. Following this, we thoroughly explain how the hermitian invariant constructs the eigenstate of a time dependent Hamiltonian, including a discussion on the Lewis phase.\\
\noindent To illustrate the application of this theory, we examine the Hamiltonian of a one dimensional harmonic oscillator having a time varying angular frequency. We derive the structural form of the Lewis invariant operator and the Lewis phase factor for this system, and establish the exact eigenstate of the Hamiltonian using the Lewis approach.\\
\noindent We then explore another aspect of this theory, showing the methodology of deducing Berry's geometric phase from the Lewis phase. For this purpose, we consider a time dependent periodic Hamiltonian of a generalized harmonic oscillator and use the Lewis technique to obtain its geometric phase.\\
\noindent In chapter \ref{paper1}, we explore an exactly solvable two dimensional model in the NC framework by choosing a model of a damped quantum harmonic oscillator in NC space \cite{SG1}, where the NC parameters vary with time. Initially, we establish the Hamiltonian that includes a damping factor and a time dependent angular frequency in NC space. We then transform it into standard commutative variables using the Bopp-shift relations. Following this, Lewis's method of invariants is employed to determine the exact eigenstate of the Hamiltonian. We begin by constructing the Ermakov-Lewis invariant system, which involves the Lewis invariant operator coupled with the nonlinear EP equation~\cite{Erm,Pin}. For simplicity in calculations, we subsequently express the entire system in terms of polar coordinates. Finally, we solve the invariant system using the ladder operator approach and determine the structural form of the Lewis phase factor, closely following the procedure outlined in \cite{Dey}.
It is important to note that the Hamiltonian eigenfunction is a combination of the invariant's eigenfunction and the Lewis phase factor. Hence, it is also coupled with the nonlinear EP equation.~\cite{Erm,Pin}. In this context, we briefly review the procedure for deriving exact analytical solutions of the EP equation, as initially demonstrated in \cite{Dey} using the Chiellini integrability condition~\cite{man1, man2, chill}. Subsequently, to generate a class of explicit eigenfunctions corresponding to each solution set of the EP equation, we carefully select the parameters (damping factor and angular frequency) which constructed the original Hamiltonian of the damped system in NC space. These parameters are chosen in such a way that they are consistent with all the constraint relations present in the system and enable us to achieve the system's solution in closed form. The analytical structures derived for the noncommutative parameters ensure that the Lewis phase, which is to be integrated out, is exactly integrable in various dissipative scenarios.\\
\noindent Next, we demonstrate how to calculate the matrix element of the position operator with finite, arbitrary power, with respect to the exact eigenfunctions of the Hamiltonian. Additionally, we derive the expectation value of the momentum operator and its square form with respect to the Hamiltonian eigenstate. To analyse the analytically generated solutions through a physical quantity, we use the obtained expectation values to compute the energy expectation values and explore their dynamical behaviour, both analytically and graphically, for various types of damping scenarios. Finally, we conclude the chapter with a summary of the entire discussion for clarity and convenience.\\
\noindent In chapter \ref{paper2}, we discussed the continuation of our previous work. Here \cite{SG2}, we focus on an exactly solvable model of a previously studied quantum harmonic oscillator with damping, now including an externally applied magnetic field that varies over time, while maintaining time dependent noncommutativity. We begin by formulating the two dimensional Hamiltonian for the previously considered model 
with a time varying magnetic field. This Hamiltonian is then transformed into commutative variables, as was done in our earlier research. We proceed by solving the Hamiltonian using the Lewis-Riesenfeld theory. Consequently, the time dependent parameters associated with the eigenfunction of this Hamiltonian follow the nonlinear EP equation~\cite{Erm,Pin}.\\
\noindent Subsequently, similar to our previous work, to generate a class of explicit eigenfunctions corresponding to each solution set of the EP equation, we carefully select the dynamic variables (applied magnetic field, angular frequency and damping factor) that define the original Hamiltonian in NC space. Again, our choices make us able to have a class of explicit analytical solutions in a closed form.\\
\noindent Next, corresponding to those explicit solutions, the expressions for the energy expectation value are established analytically and presented graphically. In these graphs, the influence of various forms of externally applied magnetic fields (increasing, decaying, and constant) on the dynamical behaviour of the previously discussed dissipative model is examined. Finally, we conclude the chapter with a summary.\\
\noindent In chapter \ref{paper3}, we explore another interesting aspect of this study concerning exactly solvable models in time dependent NC space. In the work \cite{SG3} discussed in this chapter, we address periodic model systems in NC space to derive Berry's geometric phase from the Lewis phase factor, which allows for the exact solution of these models. We formulate two periodic Hamiltonians for harmonic oscillators in two distinct types of NC space with time dependent backgrounds. The first Hamiltonian includes an explicit scale invariant term necessary for generating a geometric phase shift, while the second type lacks this term in its original form. To transform these Hamiltonians from NC space to conventional quantum mechanical phase space, the Bopp-shift relations \cite{mez} are employed for the first type, while a modified form of the Bopp-shift relation \cite{spb} is used for the second. We then apply Lewis prescription and continue our study concerning the exact results, up to finding the system's analytical eigenfunctions, which also include the exact integral form of the Lewis phase factor.\\
\noindent Subsequently, we incorporate the adiabatic approximation into the exact form of the Lewis phase factor, deriving a generic form of the geometric phase acquired by the periodic system, as demonstrated by \cite{dit} in their work. We further explore the explicit presence of the Berry phase in both systems, varying time dependent parameters, including NC parameters periodically over time. Additionally, we conduct a detailed analysis supported by graphical illustrations and a significant finding from this study is concluded at the end of this chapter.\\
\noindent In Chapter \ref{paper4}, we explored the latest work \cite{SG4} of our research. We investigate a time independent model of a simple harmonic oscillator in a two dimensional dynamic NC space, governed by the modified version of standard Bopp-shift relations \cite{spb}. Starting with the construction of the Hamiltonian in NC framework, we subsequently transform it into a conventional quantum mechanical phase space, resulting in a Hamiltonian comprising a scale invariant term and Zeeman term. The Hamiltonian is then solved using the Lewis-Riesenfeld treatment. As a result, the dynamic coefficients existing in the Hamiltonian eigenfunctions satisfy the nonlinear EP equation \cite{Erm,Pin}.\\
\noindent We extend the EP solution set outlined in \cite{Dey}, under the Chiellini integrability condition \cite{chill}, to determine the explicit analytical form of the EP variable that arises due to the modification introduced in noncommutativity. Subsequently, we compute the expectation value of the Hamiltonian using its eigenstate. This expression allows us to examine the time evolution of the system's energy expectation value across various types of EP solution sets, both through analytical methods and graphical representations. We also compare these energy dynamics with those observed in standard NC scenarios. Additionally, the generalized form of the uncertainty relations is established for both the noncommutative and standard canonical operators. Finally, we  summarize our discussion at the end of the chapter.


\chapter{Lewis-Riesenfeld treatment for time dependent Hamiltonian}\label{chap-LR}
Lewis-Reisenfeld theory, also known as Lewis method of invariant is an exact quantum theory developed by Lewis \textit{et} al. \cite{Lewis3, Lewis2, Lewis} for the quantum system whose Hamiltonians are explicitly time dependent. A Hamiltonian of this nature can arise from a system that is embedded in an environment that is fundamentally unknown. To effectively describe such systems, we can use $H(t)$, the explicitly time dependent Hamiltonians that, in classical physics, produce the proper equations of motion. In quantum physics, these systems are very elegantly handled by Lewis \textit{et} al. \cite{Lewis3, Lewis2, Lewis} since they outlined a very useful and elegant method to solve a time dependent Schr\"{o}dinger equation (considering $\hbar\,=\,1$) exactly: 
\begin{equation}
i\,\partial_t\,\ket{\psi}\,=\,H(t)\,\ket{\psi}~.\label{3sch}
\end{equation}



\noindent The core aspect of this theory of general systems is to establish the connection between the eigenstates of an explicitly time dependent invariant and the solutions of the Schr\"{o}dinger equation. For each eigenstate of an invariant, there exists a time dependent phase transformation that converts the eigenstate into a solution of the Schr\"{o}dinger equation. This phase factor can be obtained appropriately by solving a simple first order differential equation. The time dependent invariant and the phase factor are respectively known as the Lewis invariant and the Lewis phase in literature.\\
\noindent We will first explain this theory for a general time dependent Hamiltonian $H(t)$ and then demonstrate its application to a harmonic oscillator system, where the Hamiltonian varies with time due to a time dependent angular frequency. Additionally, by considering an example of a periodic Hamiltonian, we will discuss another interesting aspect of this theory, deriving Berry's geometric phase from the Lewis phase under the adiabatic approximation.   
\section{Concept of Lewis invariant}
Lewis invariant operator is basically an explicitly time dependent non trivial hermitian operator $I(t)$ which is itself invariant with respect to time. Hence, the operator obeys the following relations
\begin{equation}
\dot{I}(t)=\,i\partial_t\,I\,+\,[I, H] = 0\label{invprop}
\end{equation}
and
\begin{equation}
I^{\,\dagger}\,=\,I~.
\end{equation}
Though this hermitian invariant is generally utilized to find out the exact eigenstate of a time dependent Hamiltonian, the most significant feature of $I(t)$ is revealed out when it acts on an eigenstate of the time dependent Schr\"{o}dinger state $\ket{\psi(t)}$. We obtain the following,
\begin{align*}
i\,\partial_t\,(I\,\ket{\psi})\,&=\,i\,\left(\partial_t\,I\,\right)\,\ket{\psi}\,+\,i\,I\,\left(\partial_t\,\ket{\psi}\right)\\
&=\,-[I, H \,]\,\ket{\psi}\,+\,i\,I\,\left(\partial_t\,\ket{\psi}\right)\\
&=\,\left(HI-IH\right)\,\ket{\psi}\,+\,\,I\,\left(i\,\partial_t\,\ket{\psi}\right)\\
&=\,\left(HI-IH\right)\,\ket{\psi}\,+\,\,IH\,\,\ket{\psi}\\
&=H\,(\,I\,\ket{\psi})~.
\end{align*}
It clearly concludes that a time dependent Schr\"{o}dinger state produces another Schr\"{o}dinger state after being operated by the hermitian invariant operator $I(t)$. In this way, we can generate a class of Schr\"{o}dinger state vectors. Nevertheless, this characteristic of the hermitian invariant $I(t)$ is more fundamental than its another feature of finding the exact eigenstate of a time dependent Hamiltonian because, the second feature is valid when $I(t)$ does not include the time differentiation operator in its structure. Next, we describe the nice and explicit method of deriving the exact eigenstate of $H(t)$ by using the Lewis invariant $I(t)$.

\section{Lewis phase and exact eigenstate of Hamiltonian}
Lewis Reisenfeld theory depicts that if the hermitian invariant does not include any time differentiation form in its structure, it is possible to find out the appropriate phases of the eigenstates of $I(t)$ so that those eigenstates, after being multiplied to the found phase factors, can be the eigenstates of the Hamiltonian $H(t)$ also. \\
With the aim of illustrating the method, we begin by assuming that $I(t)$ has complete set of orthonormal eigenstates $\ket{\lambda, \kappa}$ associated with the real, time independent eigenvalues $\lambda\,:$
\begin{eqnarray}
I(t)\ket{\lambda, \kappa}\,=\,\lambda\,\ket{\lambda, \kappa}~,~\langle \lambda^{'}, \kappa{'}\,|\,\lambda, \kappa \rangle\,=\,\delta_{\lambda^{'},\lambda}\,\delta_{\kappa^{'},\kappa}~;\label{inves}
\end{eqnarray}
where $\kappa$ is a quantum number other than $\lambda$.\\
Next, in an effort to verify the time independent nature of the eigenvalues $\lambda$ of the hermitian invariant $I(t)$, we first differentiate Eqn.(\ref{inves}) and take a scalar product of it with $\ket{\lambda, \kappa}$. Since, $\bra{\lambda, \kappa}\,I\partial_t\,\ket{\lambda, \kappa}=\lambda\,\bra{\lambda, \kappa}\,\partial_t\,\ket{\lambda, \kappa}$, we can directly write
\begin{align}
\partial_t\,\lambda\,=\,\bra{\lambda, \kappa}\,\partial_t\,I\,\ket{\lambda, \kappa}~.\label{lambda}
\end{align} 
Now, using Eqn.(\ref{invprop}), the matrix element $\bra{\lambda^{'}, \kappa{'}}\,\partial_t\,I\,\ket{\lambda, \kappa}$ as well as $\partial_t\,\lambda$ can be derived in the following way,
\begin{align}
\partial_t\,\lambda &=\bra{\lambda, \kappa}\,\partial_t\,I\,\ket{\lambda, \kappa}=i\bra{\lambda, \kappa}(IH-HI)\ket{\lambda, \kappa}\nonumber\\
&=i\lambda\left[\bra{\lambda, \kappa}H\ket{\lambda, \kappa}-\bra{\lambda, \kappa}H\ket{\lambda, \kappa}\right]=0\label{lambdacons}
\end{align}
Since, the eigenvalues of the hermitian invariant $I$ is time independent, it is prominent from Eqn.(\ref{inves}) that the eigenstates of $I$ must be time dependent. \\   
With the above result in our hand, we proceed to establish the relation among the eigenstates of the hermitian invariant, $I(t)$ and the eigenstates of the Hamiltonian, $H(t)$. We again differentiate Eqn.(\ref{inves}) and take a scalar product of it with $\ket{\lambda^{'}, \kappa{'}}$ and have the following expression
\begin{align}
\bra{\lambda^{'}, \kappa^{'}}\,\partial_t\,I\,\ket{\lambda, \kappa}+\bra{\lambda^{'}, \kappa^{'}}\,I\,\partial_t\,\ket{\lambda, \kappa}\,=\,\lambda\,\bra{\lambda^{'}, \kappa^{'}}\,\partial_t\,\ket{\lambda, \kappa}~. \label{delti1}
\end{align}
The value of the matrix element $\bra{\lambda^{'}, \kappa^{'}}\,\partial_t\,I\,\ket{\lambda, \kappa}$ can also be obtained from Eqn.(\ref{invprop}) as  
\begin{align}
\bra{\lambda^{'}, \kappa^{'}}\,\partial_t\,I\,\ket{\lambda, \kappa}\,=\,i\,(\lambda^{'}-\lambda)\,\bra{\lambda^{'}, \kappa^{'}}\,H\,\ket{\lambda, \kappa}~;\label{delti2}
\end{align} 
which is substituted in Eqn.(\ref{delti1}) and it takes the following form as
\begin{align}
i\,(\lambda^{'}-\lambda)\,\bra{\lambda^{'}, \kappa^{'}}\,H\,\ket{\lambda, \kappa}+(\lambda^{'}-\lambda)\bra{\lambda^{'}, \kappa^{'}}\,\partial_t\,\ket{\lambda, \kappa}=0\label{nodelti}
\end{align}
The above relation reduces to the Schr\"{o}dinger equation only for the condition $\lambda^{'}\,\neq\,\lambda\,:$
\begin{align}
i\bra{\lambda^{'}, \kappa^{'}}\,\partial_t\,\ket{\lambda, \kappa}=\bra{\lambda^{'}, \kappa^{'}}\,H\,\ket{\lambda, \kappa}
~.\label{nolambda}
\end{align}
Hence, $\ket{\lambda, \kappa}$ cannot belong to the eigenstates of the Hamiltonian, $H(t)$. Since, there is no restriction to multiply an arbitrary time dependent phase factor with $\ket{\lambda, \kappa}$, a time dependent gauge transformation is performed on $\ket{\lambda, \kappa}$ and a new set of eigenstates of $I(t)$ are defined in a following way    
\begin{align}
\ket{\lambda, \kappa}_{\alpha}\,=\,e^{i\,\alpha_{\lambda, \kappa}(t)}\,\ket{\lambda, \kappa}~; \label{hames}
\end{align}
where $\alpha_{\lambda\kappa}(t)$ is real time dependent phase factor. As the invariant $I(t)$ is not supposed to include any time derivative operators, it would not interact with the multiplied phase factor $\alpha_{\lambda, \kappa}(t)$. Hence, in the manner of the eigenstates $\ket{\lambda, \kappa}$, the newly defined set of  eigenstates $\ket{\lambda, \kappa}_{\alpha}$ also holds the relation derived in Eqn.(\ref{nolambda}) for the condition $\lambda^{'}\neq\lambda\,:$
\begin{align}
i\,{}_\alpha\bra{\lambda^{'}, \kappa^{'}}\,\partial_t\,\ket{\lambda, \kappa}_{\alpha}={}_{\alpha}\bra{\lambda^{'}, \kappa^{'}}\,H\,\ket{\lambda, \kappa}_{\alpha}~.
\label{nolambdaalpha}
\end{align}      
At the same time, the above equality among the matrix elements should also be valid for the condition $\lambda^{'}=\lambda\,$ as the vector states $\ket{\lambda, \kappa}_{\alpha}$ are formed to be the Schr\"{o}dinger solution set. In this regard, Eqn.(\ref{hames}) is directly substituted to the structure of the above equation where $\lambda^{'}$ is made equal to $\lambda\,$. It gives rise to  the following differential equation
\begin{align}
\dot{\alpha}_{\lambda\kappa}\,\delta_{\kappa\kappa^{'}}=\bra{\lambda, \kappa^{'}}\,i\,\partial_t-H\,\ket{\lambda, \kappa}~.\label{l.phasek}
\end{align}
The above equation implies that the invariant's eigenstates $\ket{\lambda,\kappa}$ are to be selected such that the matrix elements except the diagonal ones vanish for $\kappa\neq\kappa^{'}$. Thus, to prepare the Schr\"{o}dinger solution as defined in Eqn.(\ref{hames}), the phase factor $\dot{\alpha}_{\lambda\kappa}(t)$ which is also known as Lewis phase in literature, simply obeys 
\begin{align}
\dot{\alpha}_{\lambda\kappa}\,=\bra{\lambda, \kappa}\,i\,\partial_t-H\,\ket{\lambda, \kappa}~.\label{l.phase}
\end{align}
Therefore, the general form of the eigenstate of the Hamiltonian $H(t)$ reads
\begin{align}
\ket{\lambda,\kappa}_{H}=\sum_{\lambda,\kappa}\,C_{\lambda,\kappa}\,e^{i\,\alpha_{\lambda, \kappa}(t)}\,\ket{\lambda, \kappa}~.
\end{align}
\newpage
\section{Application of Lewis theory to time dependent models}\label{chap-LR-HO}
In order to show the application of this theory, we aim to discuss two aspects found from this theory. First, we will establish an exact eigenstate of an explicitly time dependent Hamiltonian. For this, we consider one of the simplest time dependent models and explain the procedure for deriving the form of the Lewis invariant and Lewis phase factor, as demonstrated by Lewis \textit{et} al. in their work \cite{Lewis}. Second, to derive the geometric phase part from the Lewis phase, we will apply this method to a system of a time dependent generalized harmonic oscillator, following the approach outlined in \cite{dit}. 
\subsection{Exact eigenstate : Time dependent harmonic oscillator}
The Hamiltonian of a one dimensional simple harmonic oscillator having time dependent frequency reads
\begin{align}
H(t)&\,=\,\dfrac{p^2}{2M}\,+\,\dfrac{1}{2}\,M\,\omega^2(t)\,x^2~;\label{1HO}
\end{align}
where $M$ is time independent mass of the oscillator. It is obvious that $\omega(t)$ is an arbitrary function of time and $M$ is a real, positive constant. Since, it is a quantum mechanical model system and we have chosen to narrate this quantum theory in the natural unit ($\hbar=1$), the canonically conjugate operators satisfy the commutation relation $[x, p]\,=\,i$. In order to find out the exact eigenstate of the above Hamiltonian by using Lewis technique, the structural form of the hermitian invariant $I(t)$ is to be considered as per the standard assumption that the Lewis invariant should be of the same order and form in the canonical variables as the original Hamiltonian Eqn.(\ref{1HO}). Hence, we begin with the following homogeneous quadratic form of the invariant
\begin{align}
I(t)\,=\,\dfrac{1}{2}\left[\alpha(t)\,p^2\,+\,\beta(t)\,x^2\,+\,\gamma(t)\,\left( x\,p\,+\,px\right)\right]~;\label{1HO-inv1}
\end{align}
where $\alpha(t), \beta(t)$ and $\gamma(t)$ are real arbitrary time dependent coefficients. However, the choice of the invariant is not unique and the ansatz could have also be chosen without the factor $\dfrac{1}{2}$. But the above structure is suitable to provide its own eigensystem in a simple and convenient way. Next, in order to fulfil the property of invariance, we substitute the above ansatz in Eqn.(\ref{invprop}). It restricts the arbitrariness of the time dependent coefficients of $I(t)$ and those coefficients are constrained to obey the following differential equations,
\begin{align}
\dot{\alpha}(t)\,&=\,-\dfrac{2}{M}\,\gamma(t)\label{1HO-alph1}\\
\dot{\beta}(t)\,&=\,2\,M\,\omega^2(t)\,\gamma(t)\label{1HO-beta1}\\
\dot{\gamma}(t)\,&=\,M\omega^2(t)\alpha(t)-\dfrac{\beta(t)}{M}~.\label{1HO-gam1}
\end{align}  
In an effort to make the further procedure simpler, an auxiliary time dependent parameter $\rho(t)$ is invented such that $\alpha(t)=\rho^2(t)$. Parametrizing this variable in Eqn.(\ref{1HO-alph1}), the coefficient $\gamma(t)$ can be represented in terms of $\rho(t)$ and substituting the latest form of $\gamma(t)$ in Eqn.(\ref{1HO-gam1}) the parametrized form of $\beta(t)$ can be reproduced. Hence, the above coefficients are taken the following form,  
\begin{align}
\alpha(t)&=\rho^2(t)\label{1HO-alph2}\\
\beta(t)&=M^2\left[{{\rho}^2(t)}\,\omega^2(t)\,+\,\lbrace\,\dot{\rho}^2(t)\,+\,\rho(t)\,\ddot{\rho}(t)\,\rbrace \right]. \label{1HO-bet2}\\
\gamma(t)&=-\,M\dot{\rho}(t)\,\rho(t)~.\label{1HO-gam2}
\end{align}
However, the above three equations can be merged into a single differential equation. Substituting the earlier form of $\beta$ into Eqn.(\ref{1HO-beta1}) and multiplying both side with $\rho^2(t)$ , the following non-linear differential equation can be found as,
\begin{align}
M^2\dfrac{d}{dt}\left[\rho^3\left(\ddot{\rho}+\rho\,\omega^2(t)\right)\right]=0\nonumber\\
\Rightarrow\,\ddot{\rho}+\rho\,\omega^2(t)=\dfrac{\xi^2}{M^2\rho^3}~;\label{1HO-EP}
\end{align} 
where the constant $M$ can be absorbed into the integration constant $\xi^2$ but we prefer to keep it as a separate  
quantity as $M$ is the mass parameter of the oscillator. The non-linear differential equation in Eqn.(\ref{1HO-EP}) is known as Ermakov-Pinney (EP) \cite{Erm, Pin} equation in literature. The point is to be remembered that, this equation (Eqn.(\ref{1HO-EP})) which is emerged due to the parametrization in Eqn.(\ref{1HO-alph1}, \ref{1HO-beta1}, \ref{1HO-gam1}) actually carries the signature of the invariance property of Lewis invariant $I(t)$. Hence, the structural form of time dependent hermitian invariant in terms of $\rho(t)$ always appears with that non-linear EP equation. \\
\noindent 
Even, the parametrized form of $\beta$ in Eqn.(\ref{1HO-bet2}) can be simplified further by substituting the expression of $\ddot{\rho}$ from EP equation. Thus, $\beta$ is reduced as 
\begin{eqnarray}
\beta(t)\,=\,\dot{\rho}^2\,M^2+\dfrac{\xi^2}{\rho^2}~.
\end{eqnarray}  
With all the above expressions, we proceed with the following form of Lewis invariant
\begin{align}
I(t)&=\dfrac{1}{2}\left[\rho^2\,p^2+\left\lbrace\dot{\rho}^2\,M^2+\dfrac{\xi^2}{\rho^2}\right\rbrace\,x^2-\rho\dot{\rho}\,M\left(x\,p\,+\,p\,x\right)\right]\\
&=\dfrac{1}{2}\left[\left(\rho\,p-M\dot{\rho}\,x\right)^2+\dfrac{\xi^2x^2}{\rho^2}\right]~;\label{1HO-inv2}
\end{align}
with an additional non-linear EP equation
\begin{align}
\ddot{\rho}+\rho\,\omega^2(t)=\dfrac{\xi^2}{M^2\rho^3}~.
\end{align}
Here, we can also facilitate the advantage of the existing differential equation. Since, it is possible to solve the above differential equation \cite{Pin} exactly, a class of invariant operators can be generated corresponding to the family of real solution of EP equation.  \\
\noindent Next, we proceed to find out the eigensystem of the above invariant (Eqn.(\ref{1HO-inv2})) by eliminating the integration constant $\xi$ by making the scale transformation $\rho(t)\,\rightarrow\,\sqrt{\xi}\,\rho(t)$. This can be done in operator approach which is equivalent to the same \cite{dir} for the purpose of solving a time independent Hamiltonian of a harmonic oscillator.\\
\noindent
Defining the following unitary operators ($\hat{U^{\dagger}}\hat{U}=\hat{U}\hat{U^{\dagger}}=\textbf{I}$)
\begin{eqnarray}
\hat{U}=exp\left[\dfrac{-i\dot{\rho}\,M}{2\rho}x^2\right],\,\,\,
\label{1HO-uni}
\end{eqnarray}
the invariant is transformed as $I^{'}=\hat{U}I\hat{U}^\dagger\,$ and it takes the following structure
\begin{align}
I^{'}(t)=\rho^2\,p^2+\dfrac{{x^2}}{\rho^2}\,\,;
\label{1HO-inv-prime}   
\end{align}
which is similar to the Hamiltonian of a harmonic oscillator. Next, the ladder operators corresponding to the above form of invariant are defined as
\begin{eqnarray}
{\hat{a}}^{'}=\dfrac{1}{\sqrt{2}}\left(\dfrac{x}{\rho}+i\rho\,{p}\right)\,\,\,,\,\,\,{{\hat{a}}^{'\dagger}}=\dfrac{1}{\sqrt{2}}\left(\dfrac{x}{\rho}-i\rho{\,p}\right)~.
\label{1HO-lad1}
\end{eqnarray}
Now, applying the reverse transformation, it is very easy to bring back the ladder operators required for the purpose of solving the invariant in Eqn.(\ref{1HO-inv2}). They are obtained as 
\begin{align}
\hat{a}(t)&={\hat{U}}^\dagger{\hat{a}}^{'}\hat{U}=\dfrac{1}{\sqrt{2}}\left[\dfrac{x}{\rho}+i\rho{p}-i\dot{\rho}\,M\,x\right]\nonumber\\
\hat{{a}^{\dagger}}(t)&={\hat{U}}^\dagger{{\hat{a}}^{'\dagger}}\hat{U}=\dfrac{1}{\sqrt{2}}\left[\dfrac{x}{\rho}-i\rho{p}+i\dot{\rho}\,M\,x\right]~.
\label{1Ho-lad} 
\end{align}
It is straightforward to verify that the operators follow the canonical commutation relation $[\hat{a}, \hat{a}^{\dagger}]\,=\,1$.
Nevertheless, the multiplication of above those operators imply that
\begin{equation}
I(t)\,=\,\left(\hat{{a}^{\dagger}}\,\hat{a}\,+\,\dfrac{1}{2} \right)~.
\end{equation}
Finally, denoting $\ket{n}$ as the normalized eigenstate of $I(t)$, the eigensystem of Lewis invariant is described as
\begin{equation}
I(t)\ket{n}\,=\,\left(\,n+\dfrac{1}{2}\,\right)\,\ket{n}\,,~~n\,=\,0, 1, 2,...
\end{equation}
with the standard raising and lowering operations 
\begin{align}
\hat{a}\,\ket{n}\,=\,\sqrt{n}\,\ket{n-1}~&,~\hat{a}^{\dagger}\,\ket{n}\,=\,\sqrt{n+1}\,\ket{n+1}\nonumber\\
&\hat{a}^{\dagger}\hat{a}\,\ket{n}\,=\,n\,\ket{n}~.
\end{align}
In order to construct the eigenstate of the Hamiltonian according to Eqn.(\ref{hames}), we need to obtain the Lewis phase using Eqn.(\ref{l.phase}). For the sake of simplicity in the calculation, it is convenient to represent the diagonal matrix element in Eqn.(\ref{l.phase}) in terms of the ladder operators. Therefore, we first obtain
\begin{align}
\bra{n}\,H(t)\,\ket{n}\,&=\,\dfrac{M}{4} \left(\dot{\rho}^2+\omega^2\rho^2+\dfrac{1}{M^2\rho^2}\right)\,\bra{n}\left(\hat{a}{\hat{a}}^\dagger+{\hat{a}}^\dagger\hat{a}\right)\ket{n}\nonumber\\
&=\,\dfrac{M}{2} \left(\dot{\rho}^2+\omega^2\rho^2+\dfrac{1}{M^2\rho^2}\right)\,\left(n+\dfrac{1}{2}\right)~;\label{1HO-ham-expt}
\end{align}
and also express the diagonal matrix element $\bra{n}\dfrac{\partial}{\partial\,t}\ket{n}$ in the following manner
\begin{align}
\bra{n}\dfrac{\partial}{\partial\,t}\ket{n}\,&=\,\dfrac{1}{\sqrt{n}}\bra{n}\dfrac{\partial}{\partial\,t}\,\,{\hat{a}}^\dagger\,\ket{n-1}\nonumber\\
&=\,\dfrac{1}{\sqrt{n}}\bra{n}\dfrac{\partial\,{\hat{a}}^\dagger}{\partial\,t}\,\,\,\ket{n-1}\,+\,\,\dfrac{1}{\sqrt{n}}\bra{n}\,\,{\hat{a}}^\dagger\,\dfrac{\partial}{\partial\,t}\ket{n-1}\nonumber\\
&=\,\dfrac{1}{\sqrt{n}}\bra{n}\dfrac{\partial\,{\hat{a}}^\dagger}{\partial\,t}\,\,\,\ket{n-1}\,+\,\bra{n-1}\,\,\dfrac{\partial}{\partial\,t}\ket{n-1}~.\label{1HO-part-expt1}
\end{align}
The partial derivative of ${\hat{a}}^\dagger$ is computed as 
\begin{align}
\dfrac{\partial\,{\hat{a}}^\dagger}{\partial\,t}\,=\,\dfrac{1}{2}\left[\left\lbrace-\dfrac{2\dot{\rho}}{\rho}+i\,M\,\left(\rho\ddot{\rho}-\dot{\rho}^2\right)\right\rbrace\,\hat{a}\,+\,i\,M\,\left(\rho\ddot{\rho}-\dot{\rho}^2\right){\hat{a}}\dagger\right]~.\label{1HO-lad-derr}
\end{align}
Substituting Eqn.(\ref{1HO-lad-derr}) in Eqn.(\ref{1HO-part-expt1}), the form of the element $\bra{n}\dfrac{\partial}{\partial\,t}\ket{n}$ is simplified as
\begin{align}
\bra{n}\dfrac{\partial}{\partial\,t}\ket{n}\,&=\,\dfrac{1}{2\sqrt{n}}\bra{n}i\,M\,\left(\rho\ddot{\rho}-\dot{\rho}^2\right){\hat{a}}\dagger\,\ket{n-1}\,+\,\bra{n-1}\,\,\dfrac{\partial}{\partial\,t}\ket{n-1}\nonumber\\
&=i\,\dfrac{M}{2}\,\left(\rho\ddot{\rho}-\dot{\rho}^2\right)\,+\,\bra{n-1}\,\,\dfrac{\partial}{\partial\,t}\ket{n-1}~.
\end{align}
After $(n-1)$ times iteration 
\begin{align}
\bra{n}\dfrac{\partial}{\partial\,t}\ket{n}\,=i\,n\,\dfrac{M}{2}\,\left(\rho\ddot{\rho}-\dot{\rho}^2\right)\,\,+\,\bra{0}\,\dfrac{\partial}{\partial\,t}\ket{0}~.\label{1HO-zeropoint1}
\end{align}
The one and only information regarding the zero point contribution is that it should be purely imaginary as the operator $\dfrac{\partial}{\partial\,t}$ is non-hermitian. Taking this fact into consideration, the zero point contribution is required to be chosen in an appropriate manner. Thus, we adopt the following convenient form which vanishes while the EP parameter $\rho(t)$ reduces to a constant value,
\begin{equation}
\bra{0}\,\dfrac{\partial}{\partial\,t}\ket{0}\,=\,i\,\dfrac{M}{4}\left(\rho\ddot{\rho}-\dot{\rho}^2\right)~.\label{1HO-zeropoint2}
\end{equation}
Substituting Eqn.(\ref{1HO-zeropoint2}, \ref{1HO-zeropoint1}, \ref{1HO-ham-expt}) in Eqn.(\ref{l.phase}) and simplifying using EP equation in Eqn.(\ref{1HO-EP}), the Lewis phase factor is obtained as  
\begin{align}
\alpha_n(T)\,=\,\int_0^T\,\bra{n}\,i\,\partial_t-H(t)\,\ket{n}\,dt\,&=\,\left(n+\dfrac{1}{2}\right)\,\dfrac{M}{2}\,\int_0^T\,\left[-\rho\,\left(\ddot{\rho}\,+\,\omega^2\rho\right)-\dfrac{1}{M^2\rho^2} \right]\,dt\nonumber\\
&=\,-\dfrac{1}{M}\,\left(n+\dfrac{1}{2}\right)\,\int_0^T\,\dfrac{1}{\rho^2}\,dt~.\label{1HO-lewis}
\end{align}
Therefore, the time dependent eigenstate of the hamiltonian (Eqn.(\ref{1HO})) is found as
\begin{equation}
\ket{\psi}_n\,=\,\ket{n}\,e^{i\,\alpha_n}(t)\,;
\end{equation}
where $\alpha_n(t)$ is shown in Eqn.(\ref{1HO-lewis}).

\subsection{Berry phase from Lewis phase : Time dependent generalized harmonic oscillator (periodic Hamiltonian)}\label{sec-berry}
Here, we are about to review another nice aspect of the Lewis Riesenfeld treatment that not only builds up the eigenstate of a time dependent Hamiltonian but also provides a gentle way to derive Berry's geometric phase from the time dependent Lewis phase factor under adiabatic conditions, if the corresponding Hamiltonian structure contains a suitable term capable of forming a geometric phase when the Hamiltonian parameters evolve periodically with respect to time. Hence, we chose such a time dependent Hamiltonian containing a time reversal symmetry breaking scale invariant term which is a necessary to generate a geometric phase form and follow the methodology described in \cite{dit} for deriving Berry phase from Lewis phase. In this context, we would like to highlight other noteworthy works \cite{mor, zeng} that discuss the connections between Berry phase and Lewis phase.\\
\noindent Thus, the Hamiltonian of a time dependent harmonic oscillator interacting with an arbitrary background field reads,
\begin{align}
H(t)\,=\,\dfrac{1}{2M}\,p^2\,+\,\dfrac{M\.\omega^2(t)}{2}\,x^2+\,d(t)\,\left(x\,p\,+\,p\,x\right)~;\label{1GHO}
\end{align}
where the coefficient $d(t)$ is an arbitrary field parameter varying periodically with respect to time. That parameter, for example, can be gravitational field parameter when the oscillator interacts with gravitational wave, as also studied in a recent communication \cite{sen} for the purpose of deriving Berry phase using Lewis approach. It is noteworthy that, in the considered model, the time dependent angular frequency $\omega(t)$ also varies periodically with respect to time, and the presence of the parameter $d(t)$ with its adjoining scale invariant term is a necessary condition for the system to exhibit a geometric phase shift after evolving through a full time period on the Hamiltonian parameter space. Thus, the Hamiltonian in Eqn.(\ref{1GHO}), describes one of the simplest periodic systems in the context of the harmonic oscillator model, which allows for the presence of a geometric phase. \\
\noindent Rearranging Eqn.(\ref{1GHO}), the Hamiltonian can also be expressed as
\begin{align}
H(t)=\dfrac{1}{2M}\left(p+2\,M\,d(t)\,x\right)^2\,+\,\dfrac{M\,\omega^2_e(t)}{2}\,x^2~;
\end{align}
where $\omega_e(t)\,=\,\left[\,\omega^2(t)-4\,d^2(t)\right]^{1/2}$ acts as the effective frequency of the model. \\
\noindent In order to derive the geometric phase from the Lewis phase, it is necessary to first find the eigenstate of that system using the Lewis technique. Since this methodology is discussed in detail in the earlier section, we will only present the expressions of the Lewis invariant and Lewis phase, along with their adjoining non linear EP equation, before proceeding towards the main goal.\\
\noindent Thus, in respective manner, the structural form of the hermitian invariant $I(t)$ and Lewis phase factor are given by
\begin{equation}
I(t)=\dfrac{1}{2}\left[\rho^2\,p^2+\left[M^2(\dot{\rho}-2\rho\,d)^2+\dfrac{\xi^2}{\rho^2}\right]\,x^2-\rho\,M(\dot{\rho}-2\rho\,d)(x\,p\,+\,p\,x)\right]~;
\label{1GHO-inv}
\end{equation}
and  
\begin{equation}
\Theta_n(t)\,=\,-\dfrac{1}{M}\left(n+\dfrac{1}{2}\right)\,\int_0^T\,\dfrac{1}{\rho^2}~dt~;\label{1GHO-L.phase1}
\end{equation}
where, obviously, all the time dependent parameters adhere to the following version of non-linear EP equation,
\begin{align}
\ddot{\rho}+\rho\left[\omega_e^2-2\,\dot{d}\,\right]=\dfrac{1}{M^2\rho^3}~.\label{1GHO-EP1}
\end{align}
As we mentioned earlier that the Hamiltonian parameter $d(t)$ plays a crucial rule for the emergence of geometric phase, and the phase is going to be derived from Lewis phase, in this regard, a notable observation can be marked that the Lewis phase, unlike the Lewis invariant operator, does not occur with that coefficient. However, it can be brought inside the phase factor (Eqn.(\ref{1GHO-L.phase1})) by replacing the integrand from EP equation (Eqn.(\ref{1GHO-EP1})).
After doing so, the Lewis phase can be rewritten as 
\begin{equation}
\Theta_n(t)\,=\,-\left(n+\dfrac{1}{2}\right)\,\int_0^T\,\sqrt{\dfrac{\ddot{\rho}}{\rho}+\left(\omega_e^2-2\,\dot{d}\right)}~dt~.\label{1GHO-L.phase2}
\end{equation}
Now, if all the Hamiltonian parameters and the invented EP parameter $\rho(t)$ vary adiabatically that means they vary very slowly in comparison with the unperturbed wave function (the wave function when the Hamiltonian would not vary), the adiabatic approximation can be employed in Eqn.(\ref{1GHO-L.phase2}). \\
\noindent In the first approximation, the terms higher than the single order derivative and the terms formed due to the multiplication between the derivative terms, are considered to be negligible. Hence, we now present Eqn.(\ref{1GHO-L.phase2}) in a following manner
\begin{equation}
\Theta_n(t)\,\approx\,-\left(n+\dfrac{1}{2}\right)\,\int_0^T\,\omega_e\sqrt{\left(1-\dfrac{2\,\dot{d}}{\omega_e^2}\right)}~dt~.
\end{equation}
Next, in the above equation, we apply the binomial approximation which is consistent with the limit $\dot{d}\,<<\,\omega_e^2$ that basically implies that, at any moment, the variation of the perturbation is very less than the effective frequency of the system. Finally, the phase can be divided into two terms,
\begin{equation}
\Theta_n(t)\,\approx\,-\left(n+\dfrac{1}{2}\right)\,\left[\int_0^T\,\omega_e(t)\,~dt+\int_0^T\,\dfrac{1}{\omega_e}\dfrac{d\,d(t)}{dt}\,dt\right]~.\label{1GHO-phase-adi}
\end{equation}
Since, all the time dependent parameters are already considered to be periodic functions of time, they evolve a circular path on the parameter space $\mathcal{R}\,=\,\mathcal{R}(\omega,\,d)$ over a full time period $T$. Therefore,
\begin{align}
\mathcal{R}(0)\,=&\,\mathcal{R}(T)~~;~~\mathcal{R}\,=\,(\omega, d)\nonumber\\
&\dfrac{d}{dt}\,=\,\dfrac{d\mathcal{R}}{dt}\,\nabla\,\mathcal{R}~.
\end{align} 
Substituting the above relation in the second additive term in Eqn.(\ref{1GHO-phase-adi}) and denoting it as $\Theta_G$ we have
\begin{equation}
\Theta_G\,=\,\left(n+\dfrac{1}{2}\right)\,\oint\,\dfrac{1}{\omega_e}\left(\,\nabla_{\mathcal{R}}\,d\,\right)\,d\mathcal{R}~.
\end{equation}
This is the well known expression of Berry's geometric phase.\\
\noindent In a later chapter \ref{paper3}, we will employ this method to deduce the geometric phase for the periodic Hamiltonians of various kind of harmonic oscillators in NC space.

\chapter{Noncommutative Damped harmonic oscillator}\label{paper1}
There have been numerous studies on time-dependent classical and quantum harmonic oscillators over a long period in the literature. When considering two such oscillators, the problem becomes even more intriguing in the context of a two dimensional system. To be consistent with practical scenarios, damping must be included in the system. In practical scenarios, dissipation is attributed to the interaction between the two systems: the system we observe and the thermal reservoir. Dissipation arises as a consequence of the irreversible flow of energy into the reservoir. Dissipation arises as a consequence of the irreversible flowing of energy into the reservoir. However, an explicitly time varying Hamiltonian in a closed form provides an effective description of a dissipated system where the source of dissipation is, in principle, unknown. In the literature, there have been numerous works on damped quantum harmonic oscillator in one dimension \cite{Sebawe}-\cite{Pedrosa3}, its two dimensional equivalent is less explored~\cite{Gouba, strek1, strek2}.
A damped harmonic oscillator in two dimensional standard phase space is studied by Lawson {\it{et al.}}~\cite{Gouba}. Their solutions provide a foundation for exploring the construction of various coherent states with intriguing properties. While two dimensional quantum damped harmonic oscillators (two linearly coupled non interacting Caldirola-Kanai \cite{caldi, kanai} damping models with friction terms) in dynamic NC space were studied in \cite{strek2} to determine the exact propagator and statistical partition function of the system, our study differs from the work presented in \cite{strek2}. \\
\noindent In our work detailed in \cite{SG1}, which forms the subject of this chapter, the work by Lawson et al.~\cite{Gouba} is extended in a two dimensional dynamic NC background. To treat the system, we employ Lewis method of invariant, following the procedure given in \cite{Dey} for their two dimensional model of a simple harmonic oscillator with dynamic noncommutativity. Additionally, we also devise a fruitful procedure for obtaining a class of exact solutions for this system.



\section{Hamiltonian of the damped harmonic oscillator in noncommutative space}
Here, we start with the two dimensional time dependent Hamiltonian describing a model system consisting of a damped harmonic oscillator in a dynamic NC background. \\
\noindent Thus, within the NC framework, the Hamiltonian of the present system reads,
\begin{equation}
H(t)=\dfrac{f(t)}{2M}({P_1}^2+{P_2}^2)+\dfrac{M\omega^2(t)}{2f(t)}({X_1}^2+{X_2}^2)~;\label{1}
\end{equation}
where $f(t)$ denotes the damping factor and it is given by,
\begin{equation}
f(t)=e^{-\int_{0}^t\eta(s)ds}~,
\end{equation}
where $\eta(t)$ represents the friction coefficient. In literature, this structure of the damping factor is very well known when the friction coefficient is set to be a constant and a two dimensional Hamiltonian (placed in a standard quantum mechanical phase space) consisting such a coefficient of friction with a constant angular frequency is popularly known as Caldirola and Kanai Hamiltonian \cite{caldi, kanai}. As we further explore this model in a NC framework, later, it will be observed that this kind of damping factor leads to a certain time intervals within which the system remains physical. Apart from it, $\omega(t)$ is the oscillator's angular frequency varying with respect to time. $M$ is its constant mass parameter.\\
\noindent Since the system is considered in a dynamic NC space, the commutator brackets among the NC canonical operators existing in the Hamiltonian read (including $\hbar$)
\begin{align}
[X_j,X_k]~=i\theta(t)\epsilon_{jk}\,,\, 
&[P_j,P_k]~=i\Omega(t)\epsilon_{jk}\,,\,[X_j,P_k]\,=i\,\hbar\,\left[1+\frac{\theta(t)\Omega(t)}{4}\right]\delta_{jk};\\
&(j,k=1,2~,~\epsilon_{jk}=-\epsilon_{kj}~,~\epsilon_{12}=1)\nonumber
\end{align}
where $\theta(t)$ and $\Omega(t)$ are the dynamic NC parameters for the configuration space 
and the momentum space respectively. In the other hand, the canonical operators $(x_i,p_i)$ in standard quantum mechanical phase space follow the commutator brackets $[x_j,p_k]=i\hbar\delta_{jk}$, $[x_j,x_k]=0=[p_j,p_k]$; ($j, k=1,2$).\\
\noindent Next, the standard Bopp-shift relations \cite{mez} ($\hbar=1$) are to be implemented into the system for transforming original Hamiltonian (i.e. NC Hamiltonian) from NC space to into the standard canonical phase space. Now, the Bopp-shift relations are given by,
\begin{eqnarray}
& X_1=x_1-\dfrac{\theta(t)}{2}p_2\,\,\,;\,\,\,X_2=x_2+\dfrac{\theta(t)}{2}p_1\\
& P_1=p_1+\dfrac{\Omega(t)}{2}x_2\,\,\,;\,\,\,P_2=p_2-\dfrac{\Omega(t)}{2}x_1 \,\,.
\label{eqn1}
\end{eqnarray}
\noindent Therefore, the original Hamiltonian in terms of $(x_i,p_i)$ coordinate operators is revealed as
relation,
\begin{equation}
H=\dfrac{a(t)}{2}({p_1}^2+{p_2}^2)+\dfrac{b(t)}{2}({x_1}^2+{x_2}^2)+c(t)({p_1}{x_2}-{p_2}{x_1})\,\,\,;\label{DHO-comham}
\end{equation}
where time dependent coefficients have the following structures,
\begin{eqnarray}
a(t)&=&\dfrac{f(t)}{M}+\dfrac{M{\omega^2(t)}\theta^2(t)}{4f(t)}\label{DHO-a} \\
 b(t)&=&\dfrac{f(t){\Omega^2(t)}}{4M}+\dfrac{M{\omega^2(t)}}{f(t)}\label{DHO-b} \\
 c(t)&=&\dfrac{1}{2}\left[\dfrac{f(t)\Omega(t)}{M}+\dfrac{M\omega^2(t)\theta(t)}{f(t)} \right]. \label{DHO-c}
\end{eqnarray}
Here, the appearance of the parameter $c(t)$ can be remarked as a pure NC effect as it arises solely because of the NC parameters $\theta(t)$ and $\Omega(t)$. It should also be mentioned that although our Hamiltonian given by Eqn.(\ref{DHO-comham}) is apparently identical to that in \cite{Dey}, the time dependent coefficients in Eqn.(\ref{DHO-c}) achieve different structures. The reason is that, while the system studied in \cite{Dey} is a model of simple harmonic oscillator with dynamic noncommutativity, our study addresses the model of a time dependent damped harmonic oscillator in the same NC background as considered in \cite{Dey}. Hence, the Hamiltonian coefficients are modified by $f(t)$, the damping factor in the present system.


\section{Exact eigenstate of the model Hamiltonian}
As performed in \cite{Dey}, we also apply the Lewis technique \cite{Lewis} to solve our model system exactly. First, we construct a suitable structure of the time dependent Lewis invariant operator for the present Hamiltonian (Eqn.(\ref{DHO-comham}). Next, solving the Lewis invariant, such that    
\begin{equation}
I(t)\phi(x_1,x_2)=\epsilon \phi(x_1,x_2)
\label{DHO-eqnegn}
\end{equation}
where $\epsilon$ denotes the invariant's eigenvalue corresponding to its time dependent eigenstate $\phi(x_1,x_2)$, the eigenstate of $H(t)$ can be obtained using the following relation 
\begin{equation}
\psi(x_1,x_2,t)=e^{i\Theta(t)}\phi(x_1,x_2)~;
\label{DHO-eqnpsi}
\end{equation}
where $\psi(x_1,x_2,t)$ is the eigenfunction of the Hamiltonian (Eqn.(\ref{DHO-comham})) and $\Theta(t)$ denotes the Lewis phase factor which is real.\\



\subsection{The time dependent Lewis invariant}
We begin by referring to the portion \ref{chap-LR-HO} in the previous chapter \ref{chap-LR} where the construction procedure of Lewis invariant operator for a one dimensional system of a time dependent harmonic oscillator (actually studied by Lewis {\it et.al.}~\cite{Lewis}) is reviewed in detail. As per the methodology suggested by \cite{Lewis} and also followed in \cite{Dey}, corresponding to $H(t)$ Eqn.(\ref{DHO-comham}), we chose the following ansatz of the invariant,
\begin{equation}
I(t)=\alpha(t)({p_1}^2+{p_2}^2)+\beta(t)({x_1}^2+{x_2}^2)+\gamma(t)(x_1{p_1}+p_2{x_2}).
\label{DHO-inv1}
\end{equation}
Now, implementing the property of invariance (Eqn.(\ref{invprop})) into the form of $I(t)$ defined by 
Eqn.(\ref{DHO-inv1}) and comparing the invariant's coefficients, we have the following differential equations,
\begin{eqnarray}
\dot{\alpha}(t)&=&-a(t)\gamma(t)\label{DHO-alp1}\\
\dot{\beta}(t)&=&b(t)\gamma(t)\label{DHO-bet1}\\
\dot{\gamma}(t)&=&2\left[\,b(t)\alpha(t)-\beta(t)a(t)\,\right]
\label{DHO-gam1}
\end{eqnarray}
\noindent The subsequent task, as shown in \cite{Lewis, Dey}, is to parametrize $\alpha(t)=\rho^{2}(t)$; where $\rho(t)$ is an auxiliary parameter invented to simplify the computational procedure. In order to express the other coefficients in terms of $\rho(t)$, we substitute the above parametrization into the above relations. Hence, we find out 
\begin{eqnarray}
\gamma(t)&=&-\dfrac{2\rho\dot{\rho}}{a(t)}\label{DHO-gam2}\\
\beta(t)&=&\dfrac{1}{a(t)}\left[\dfrac{{\dot{\rho}^2}}{a(t)}+{{\rho}^2}b+\dfrac{\rho\ddot{\rho}}{a(t)}-\dfrac{\rho\dot{\rho}\dot{a}}{a^2} \right].\label{DHO-bet2}
\end{eqnarray}
Substituting the value of $\beta$ in Eqn.(\ref{DHO-bet1}), the non linear Ermakov-Pinney (EP) equation that includes a dissipative 
term~\cite{Dey, Erm, Pin}, arises in the following form,
\begin{equation}
\ddot{\rho}-\dfrac{\dot{a}}{a}\dot{\rho}+ab\rho={\xi^2}\dfrac{a^2}{\rho^3}~.
\label{DHO-EP}
\end{equation} 
where ${\xi^2}$ is an integration constant. The above form of the EP equation exactly matches with the same in \cite{Dey}. At this point, we must remember that, due to the presence of the damping factor $f(t)$, the explicit form of the time dependent 
coefficients in the Hamiltonian in Eqn.(\ref{DHO-comham}) differs from the same in the Hamiltonian studied in \cite{Dey}.\\
\noindent Now, from Eqn.(\ref{DHO-bet2}), we can gain the simplest form of $\beta$ by utilizing the EP equation. It is as follows,
\begin{eqnarray}
\beta(t)&=&\dfrac{1}{a(t)}\left[\dfrac{{\dot{\rho}^2}}{a(t)}+\dfrac{{\xi^2}{a(t)}}{\rho^2} \right].
\label{eqnew}
\end{eqnarray}  
\noindent Therefore, the hermitian invariant operator with regard to the EP parameter $\rho$ is presented as,
\begin{equation}
I(t)=\rho^2({p_1}^2+{p_2}^2)+\left(\dfrac{\dot{\rho}^2}{a^2}+\dfrac{{\xi^2}}{\rho^2}\right)({x_1}^2+{x_2}^2)-\dfrac{2\rho\dot{\rho}}{a}(x_1{p_1}+p_2{x_2}).\label{DHO-inv2}
\end{equation}
In the subsequent section, we will determine the suitable form of the ladder operators to solve the invariant system. Later, the entire system along with the ladder operators in terms of the Cartesian coordinates will be transformed into the polar coordinate system to simplify the calculation.


\subsection{The ladder operators for solving the invariant system}\label{sec-lad} 
Having established the required Hermitian invariant $I(t)$, we can now move on to calculate its eigenstates using the operator approach. This involves constructing the ladder operators and representing the invariant in respect of them. To achieve this, the invariant $I(t)$ is required to be transformed (as expressed in Eqn.(\ref{DHO-inv2})) into a more tractable form. This transformation is accomplished by applying a unitary transformation with a unitary operator $\hat{U}$ in an appropriate form. First, we begin by the following unitary operators,
\begin{eqnarray}
\hat{U}=exp\left[-\dfrac{i\dot{\rho}}{2a(t)\rho}({x_1}^2+{x_2}^2)\right],\,\,\,
\hat{U^{\dagger}}\hat{U}=\hat{U}\hat{U^{\dagger}}=\textbf{I}.
\label{DHO-uni1}
\end{eqnarray}
Next we define 
\begin{eqnarray}
\phi^{'}(x_1,x_2)=\hat{U}\phi(x_1,x_2) \,\,\,,\,\,\,
I^{'}(t)&=\hat{U}I\hat{U^\dagger}\,\,\;
\label{DHO-uni2}
\end{eqnarray}
where $\phi(x_1,x_2)$ denotes the invariant's eigenfunction (Eqn.(\ref{DHO-eqnegn})). This yields
\begin{align}
I^{'}\phi^{'}=\hat{U}I\hat{U^\dagger}\hat{U}\phi=\hat{U}I\phi=\hat{U}\epsilon\phi=\epsilon\phi^{'}.
\end{align}
With the unitary operators in Eqn.(s)(\ref{DHO-uni1}, \ref{DHO-uni2}), the invariant is transformed as, 
\begin{align}
I^{'}(t)=\rho^2({p_1}^2+{p_2}^2)+\dfrac{{\xi^2}}{\rho^2}({x_1}^2+{x_2}^2)\,\,.
\label{eqn15}   
\end{align}
The above form of the transformed invariant is structurally identical to the time dependent harmonic oscillator in two dimension. Now, it is very straightforward to have the following ladder operators,
\begin{eqnarray}
{\hat{a}_j}^{'}=\dfrac{1}{\sqrt{2\xi}}\left(\dfrac{\xi}{\rho}{x}_j+i\rho{p}_j\right)\,\,\,,\,\,\,{{\hat{a}_j}^{'\dagger}}=\dfrac{1}{\sqrt{ 2\xi}}\left(\dfrac{\xi}{\rho}{\hat{x}}_j-i\rho{p}_j\right)
\label{eqn16}
\end{eqnarray}
where $j=1,2$. Expectedly, the operators hold $[{{\hat{a}_i}^{'}},{{\hat{a}_j}^{'\dagger}}]=\delta_{ij}$.\\
\noindent Next, a reverse transformation is employed to generate the ladder operators in unprimed space. Those are found as,
\begin{eqnarray}
\hat{a_j}(t)&=&{\hat{U}}^\dagger{\hat{a}_j}^{'}\hat{U}=\dfrac{1}{\sqrt{2\xi}}\left[\dfrac{\xi}{\rho}x_j+i\rho{p_j}-\dfrac{i\dot{\rho}}{a(t)}x_j\right]\\
\label{eqn17}
\hat{{a_j}^{\dagger}}(t)&=&{\hat{U}}^\dagger{{\hat{a}_j}^{'\dagger}}\hat{U}=\dfrac{1}{\sqrt{2\xi}}\left[\dfrac{\xi}{\rho}x_j-i\rho{p_j}+\dfrac{i\dot{\rho}}{a(t)}x_j\right].
\label{eqn18} 
\end{eqnarray}
Here also, the above set of operators hold $[{{\hat{a}_i}},{{\hat{a}_j}^{\dagger}}]=\delta_{ij}$.

\noindent The above operators can be combined linearly for achieving the appropriate form of the ladder operators. Therefore, setting $\xi=1$, we define\footnote{The ladder operator $\hat{a}(t)$ should not be confused with the Hamiltonian coefficient $a(t)$},
\begin{eqnarray}
\hat{a}(t)=-\dfrac{i}{\sqrt{2}}(\hat{a}_1-i\hat{a}_2)
=\dfrac{1}{2}\left[\rho(p_1-i\,p_2)-\left(\dfrac{i}{\rho}+\dfrac{\dot{\rho}}{a(t)} \right)(x_1-i\,x_2)\right]
\label{eqn19}
\end{eqnarray}
and
\begin{eqnarray}
{\hat{a}}^\dagger(t)=\dfrac{i}{\sqrt{2}}({\hat{a}_1}^\dagger+i{\hat{a}_2}^\dagger)
=\dfrac{1}{2}\left[\rho(p_1+i\,p_2)+\left(\dfrac{i}{\rho}-\dfrac{\dot{\rho}}{a(t)} \right)(x_1+i\,x_2)\right].
\label{eqn20}
\end{eqnarray}
Once again, $[\hat{a},\hat{a}^\dagger]=1$. These are the final structure of the ladder operators for the purpose of solving the invariant.



\subsection{Polar representation of the system}\label{sec-pol}
In order to to map our two dimensional system in terms of the polar coordinate variables, we intend to discuss how the momentum operator transforms from the cartesian coordinates $(p_1, p_2)$ to the polar coordinates $(p_r, p_\theta)$. \\
\noindent Using the well known transformation of the position variables, 
\begin{eqnarray}
x_1=r\,cos\theta~,~x_2=r\,sin\theta\,,\nonumber\\
r=\sqrt{x_1^2+x_2^2}~,~\theta=tan^{-1}\dfrac{x_2}{x_1}
\end{eqnarray}
we can express the momentum vector $\vec{p}$ (in the classical framework) in its polar form, 
\begin{align}
\vec{p}&\,=\,p_1\,\hat{i}\,+\,p_2\,\hat{j}\,=\,p_r\,\hat{r}\,+\,\dfrac{p_\theta}{r}\,\hat{\theta}\\
&\,\Rightarrow\,m\dot{\vec{r}}\,=\,m\dot{r}\,\hat{r}+m\,r^2\dot{\theta}\,\hat{\theta}
\end{align}
obtaining the radial and azimuthal components of the momentum vector (in classical picture) as follows:
\begin{align}
p_r=m\dot{r}=\dfrac{x_1\,p_1+x_2\,p_2}{r}~,~p_\theta=m\,r^2\dot{\theta}=x_1\,p_2-p_1\,x_2~.
\end{align}
In quantum mechanical framework, since, the commutator brackets $[\dfrac{x_i}{r}, p_i]\,\neq\,0$, the operator $p_r$ is expressed as shown:
\begin{align}
p_r=\dfrac{1}{2}\left[\dfrac{x_1}{r}p_1+p_1\dfrac{x_1}{r}+\dfrac{x_2}{r}p_2+p_2\dfrac{x_2}{r}\right]~.\label{pr1}
\end{align}
Next, using the following commutator brackets
\begin{eqnarray}
\left[\dfrac{x_1}{r}, p_1\right]\,=\,i\,\dfrac{sin^2\,\theta}{r}~,~\left[\dfrac{x_2}{r}, p_2\right]\,=\,i\,\dfrac{cos^2\,\theta}{r}
\end{eqnarray}
and the following relation regarding the radial partial derivative 
\begin{equation}
\partial_r=\dfrac{\partial\,x_1}{\partial\,r}\partial_1+\dfrac{\partial\,x_2}{\partial\,r}\partial_2\,=\,\dfrac{x_1}{r}\partial_1+\dfrac{x_2}{r}\partial_2
\end{equation}
the polar structure of $p_r$ in Eqn.(\ref{pr1}) can be found in this manner :
\begin{align}
p_r&=\dfrac{1}{2}\left[\dfrac{x_1}{r}p_1+p_1\dfrac{x_1}{r}+\dfrac{x_2}{r}p_2+p_2\dfrac{x_2}{r}\right]\nonumber\\
&=\left[\dfrac{x_1}{r}p_1+\dfrac{x_2}{r}\,p_2\right]-\dfrac{i}{2r}\nonumber\\
&=-i\left[\dfrac{x_1\partial_1+x_2\partial_2}{r}\right]-\dfrac{i}{2r}\nonumber\\
&=-i\,\partial_r-\dfrac{i}{2r}~.\label{pr-pol}
\end{align}
Similarly, the azimuthal part in polar representation can be written as 
\begin{eqnarray}
p_\theta=x_1\,p_2-p_1\,x_2=-i\partial_{\theta}~.
\end{eqnarray}
Now, from the following form of $p_r$ and $p_\theta$ ,
\begin{eqnarray}
&p_r=\left(cos\theta\,p_1+sin\theta\,p_2\right)-\dfrac{i}{2r}\\&p_\theta=\left(-r\,sin\theta\,p_1+r\,cos\theta\,p_2\right)
\end{eqnarray}
the polar representation of the momentum variables $p_1$ and $p_2$ are derived as
\begin{eqnarray}
&p_1=-i\,cos\theta\,\partial_r+\dfrac{i}{r}\,sin\theta\,\partial_\theta\\&p_2=-i\,sin\theta\,\partial_r-\dfrac{i}{r}\,cos\theta\,\partial_\theta
\end{eqnarray}
In both the cartesian and polar coordinate frame, the position and momentum operators hold the following commutator brackets,
\begin{equation}
[r,p_r]=[\theta,p_{\theta}]=[x_1,p_1]=[x_2,p_2]=i.
\label{eqn23}
\end{equation}
Those operators hold the following anticommutator brackets,
\begin{equation}
[r, p_r]_{+}=[x_1, p_1]_{+}+[x_2, p_2]_{+}
=2(x_1 p_1+p_2 x_2)
\label{eqn24}
\end{equation}
\noindent In an effort to express the invariant $I(t)$ with regard to the polar coordinate variables, we require the following relations,
\begin{eqnarray}
({p_1}^2+{p_2}^2)&=&\left({p_r}^2+\dfrac{{p_{\theta}}^2}{r^2}-\dfrac{1}{4r^2}\right)\\
(p_1+i{p_2})&=&e^{i\theta}\left[p_r+\dfrac{i}{r}p_{\theta}+\dfrac{i}{2r} \right]      \\
(p_1-i{p_2})&=&e^{-i\theta}\left[p_r-\dfrac{i}{r}p_{\theta}+\dfrac{i}{2r} \right].
\label{eqn25}
\end{eqnarray}
Finally, polar representation of the invariant operator reads,
\begin{eqnarray}
I(t)=\dfrac{1}{\rho^2}r^2+\left(\rho{p_r}-\dfrac{\dot{\rho}}{a}r\right)^2+\left({\dfrac{\rho{p_\theta}}{r}}\right)^2-\left({\dfrac{\rho}{2r}}\right)^2
\label{DHO-inv-pol}
\end{eqnarray}
and the same for the ladder operators read,
\begin{eqnarray}
\hat{a}(t)&=&\dfrac{1}{2}\left[\left(\rho{p_r}-\dfrac{\dot{\rho}}{a(t)}r \right)-i\left(\dfrac{r}{\rho}+\dfrac{\rho{p_\theta}}{r}+\dfrac{\rho}{2r} \right)    \right]e^{-i\theta}\nonumber\\
{\hat{a}}^{\dagger}(t)&=&\dfrac{1}{2}e^{i\theta}\left[\left(\rho{p_r}-\dfrac{\dot{\rho}}{a(t)}r \right)+i\left(\dfrac{r}{\rho}+\dfrac{\rho{p_\theta}}{r}+\dfrac{\rho}{2r} \right) \right].
\label{DHO-lad-pol}
\end{eqnarray}
Both the Lewis invariant $I(t)$ 
and the corresponding ladder operators are structured similarly to those employed in \cite{Dey} for examining the undamped harmonic oscillator in NC space. However, it is important to emphasize that the time dependent coefficients in our current study are distinct because they account for the damping present in our system.


\subsection{Eigenfunction of the invariant system and Lewis phase factor}\label{sec-sol}
In the conventional method, as followed in \cite{Dey}, the initial crucial step for solving the invariant $I(t)$ is to express it using the ladder operators. It is found that
\begin{equation}
I(t)\,=\,4\left[\hat{a}^{\dagger}\hat{a}+\dfrac{1}{2}-\dfrac{p_\theta}{2}\right]~.
\end{equation}
However, another form of invariant which is more convenient for solving purpose, can be obtained from the above relation and hence, it can be defined (also done in \cite{Dey}) that
\begin{equation}
I^{'}(t)=\dfrac{I}{4}-\dfrac{p_\theta}{2}=\left(\hat{a}^{\dagger}\hat{a}+\dfrac{1}{2}\right)-p_\theta~.\label{DHO-inv-new}
\end{equation}
It is straightforward to check that the defined operator $I^{'}(t)$ also shows the property of invariance as $\dot{p_\theta}\,=\,[p_\theta, H]\,=\,0$.  Consequently, we proceed with Eqn.(\ref{DHO-inv-new}) as the Lewis invariant to construct the eigenstates of the Hamiltonian.\\ 
\noindent The commutator bracket $[I^{'}, p_\theta]\,=\,0$ indicates that both the Lewis invariant operator and the angular momentum operator can share the same eigenstate. Thus, we represent the eigenstates of $I^{'}(t)$ and $p_\theta$ as $\ket{n,l} $. Therefore, the eigensystem for both operators is found as follows:
\begin{eqnarray}
I^{'}(t)\ket{n,l}=\left(n+\dfrac{1}{2}\right)\ket{n,l}~,~p_\theta\ket{n,l}=l\ket{n,l}~,~\langle n,l|n,l \rangle =1~;\label{DHO-inv-estate}
\end{eqnarray}
where the action of the ladder operators are implemented as 
\begin{eqnarray}
\hat{a}\ket{n,l}=\sqrt{n+l}\ket{n,l-1}~,~\hat{a}^\dagger\ket{n,l}=\sqrt{n+l+1}\ket{n,l+1}~,~\hat{a}^\dagger\,\hat{a}\ket{n,l}=(n+l)\ket{n,l}~.\label{cr-an}
\end{eqnarray} 
It is evident from the above equations that the integers $n$ and $l$ can ensure that $n+l\geqslant0$.  Consequently, by introducing another positive integer $m$ such that $m=n+l$, we derive the condition $l\geqslant-n$. \\
Following the methodology outlined in \cite{Dey}, our objective is to obtain the invariant's eigenfunction with regard to the polar coordinate variables. To achieve this, we express the state vector in terms of the functional form in polar coordinate system as 
\begin{eqnarray}
\langle r,\theta | n, l \rangle = \phi_{n, l}(r, \theta) = \Phi_n(r)\,\Phi_n(\theta)~,
\end{eqnarray}
and the azimuthal part of the eigenfunction can quickly be generated as
\begin{eqnarray}
p_\theta\,\phi_n(r, \theta)=l\phi_n(r, \theta)\nonumber\\
\Rightarrow\,\phi_n(r, \theta)\,=\,\Phi_n(r)\,e^{i\,l\,\theta}~.
\end{eqnarray}
In order to find out the radial part $\Phi_n(r)$, we consider the following operation
\begin{eqnarray}
&\hat{a}\ket{n,-n}=0~~,~~\langle n,-n|n,-n\rangle=1 ; \nonumber\\
&\Rightarrow\,\hat{a}\,\,\phi_{n,-n}(r,\theta)=0~.
\end{eqnarray}
Substituting the first relation from Eqn.(\ref{DHO-lad-pol}) into the above operation and using the last relation from Eqn.(\ref{pr-pol}), we find out that
\begin{eqnarray}
\dfrac{d\Phi_n}{dr}\,=\,\dfrac{\left(n\,a\,\rho^2-a\,r^2\,+\,i\,r^2\,\rho\,\dot{\rho}\,\right)}{a\,r\,\rho^2}\Phi_n
\end{eqnarray}
Hence, the eigenfunction of the invariant $I^{'}(t)$, for the lowest value of the integer $l=-n$, can be integrated out as 
\begin{eqnarray}
\phi_{n,-n}(r, \theta)&=&\lambda_{n}\,r^{n}e^{-i\,n\theta-\dfrac{a-i\rho\dot{\rho}}{2a{\rho}^2}r^2}~;
\label{DHO-inv-efunc-grnd}
\end{eqnarray}
where $\lambda_n$ is the normalization factor which is to be calculated from the following condition 
\begin{align}
&\int_0^{\infty}\,\int_0^{2\pi}\,r\,\phi^{*}(r, \theta)\phi(r, \theta)\,dr\,d\theta\,=\,1\nonumber\\
&\Rightarrow\pi\,\lambda_n^2\int_0^{\infty}\,(r^2)^n\,e^{-r^2/\rho^2}\,d\,(r^2)\,=\,1\nonumber\\
&\Rightarrow\lambda_n^2\,=\,\dfrac{1}{\pi\,n!\rho^{2n+2}}~.\label{lamb}
\end{align}
The most general (for any arbitrary value of $m$) form of the eigenfunction of the invariant can also be deduced from the relation
\begin{equation}
{(\hat{a}^\dagger)}^{m}\ket{n,-n}=\sqrt{m!}\ket{n,m-n}~.
\end{equation}
As also found in \cite{Dey}, the eigenfunction of $I^{'}(t)$ for the higher quantum states, can be presented as 
\begin{eqnarray}
\phi_{n,m-n}(r,\theta)&=&\lambda_{n}\dfrac{{(i\rho)}^m}{\sqrt{m!}}r^{n-m}e^{i(m-n)\theta-\dfrac{a(t)-i\rho\dot{\rho}}{2a(t)
{\rho}^2}r^2}U\left(-m,1-m+n,\dfrac{r^2}{\rho^2} \right)
\label{DHO-inv-efunc1}
\end{eqnarray}
where $\lambda_n$ holds the value found in Eqn.(\ref{lamb}) and $U\left(-m,1-m+n,\dfrac{r^2}{\rho^2} \right)$ denotes the Tricomi's confluent hypergeometric function \cite{Arfken, uva}. However, the above functional form can also be written in relation to the associated Laguerre polynomial. Using the following relation
\begin{equation}
U\left(-m,1-m+n,\dfrac{r^2}{\rho^2} \right)=\dfrac{m!}{(-1)^m}\,L^{n-m}_m\left(\dfrac{r^2}{\rho^2} \right)~
\end{equation}
we can express the eigenfunction in the following manner:
\begin{align}
\phi_{n,m-n}(r,\theta, t)&=\dfrac{i^{-m}\sqrt{m!}\,\rho^{m-n-1}}{\sqrt{n!\pi}}r^{n-m}e^{i(m-n)\theta-\dfrac{a-i\rho\dot{\rho}}{2a{\rho}^2}r^2}\,L^{n-m}_m\left(\dfrac{r^2}{\rho^2} \right)~\nonumber\\
&=Q_{n,m-n}(t)\,R_{n,m-n}(r,t)\,\Phi_{n,m-n}(\theta,t)\,;\label{DHO-inv-efunc2}
\end{align}
where 
\begin{align}
R_{n,m-n}(r,t)\,=\,r^{n-m}\,e^{-\dfrac{a-i\rho\dot{\rho}}{2a{\rho}^2}r^2}\,L^{n-m}_m\left(\dfrac{r^2}{\rho^2} \right)\label{DHO-inv-efunc-R}\\
Q_{n,m-n}(t)\,=\,\dfrac{i^{-m}\sqrt{m!}\,\rho^{m-n-1}}{\sqrt{n!\pi}}~,~\Phi_{n,m-n}(\theta,t)\,=\,e^{i(m-n)\theta}\label{DHO-inv-efunc-th}~.
\end{align}
In a later section, we will find that the previously mentioned structural form of the eigenfunction is extremely useful for calculating the expectation value of the Hamiltonian.\\
\noindent The eigenfunction $\phi_{n,m-n}(r,\theta)$ adhere to the following condition of orthonormality,
\begin{equation}
\int_0^{2\pi}d\theta\int_0^{\infty}rdr\phi^{*}_{n,m-n}(r,\theta)\phi_{n^{'},m^{'}-n^{'}}(r,\theta)=\delta_{nn^{'}}\delta_{mm^{'}}.
\end{equation}

\noindent \textbf{Lewis phase~:}\vskip .20cm
\noindent The process for obtaining this phase factor from Eqn.(\ref{l.phase}) closely resembles the one outlined in the latter portion of Section \ref{chap-LR-HO}. Following the derivation presented by \cite{Dey}, we outline some key intermediate steps here. Utilizing the ladder operators' actions from Eqn.(\ref{cr-an}), the derivation method of this phase begins as follows,
\begin{align}
\dot{\Theta}_{n, l}&\,=\,\bra{n, l}\,(i\,\partial_t-H\,)\,\ket{n, l}\nonumber\\
&\,=\,\dfrac{1}{n+l}\,\bra{n, l-1}\,\hat{a}\left(i\,\partial_t-H\,\right)\,\hat{a}^{\dagger}\ket{n, l-1}\nonumber\\
&\,=\,\dfrac{1}{n+l}\,\left[\bra{n, l-1}\,[\hat{a}, i\,\partial_t-H]\,\hat{a}^{\dagger}\ket{n, l-1}+\bra{n, l-1}\,(i\,\partial_t-H)\,\hat{a}\,\hat{a}^{\dagger}\ket{n, l-1} \right]\label{ph-der-1}
\end{align}  
The value of the commutator $[\hat{a}, i,\partial_t - H]$ is essential for determining the phase factor. This calculation can be performed in either the Cartesian or polar coordinate systems since the result does not need to be expressed in terms of the coordinate operators. We have carried out this calculation in the Cartesian coordinate system, determining the value of $\partial_t \hat{a}$ as follows,
\begin{equation}
\partial_t\,\hat{a}\,=\,\dfrac{1}{2}\,\left[\dot{\rho}\,(p_1-i\,p_2)\,+\,\left(\dfrac{a}{\rho^3}-b\,\rho-\dfrac{i}{\rho^2}\dot{\rho} \right)\,(x_1-i\,x_2)\right]~,
\end{equation}
the $[\hat{a}, i\,\partial_t-H]$ can be presented as
\begin{equation}
[\hat{a}, i\,\partial_t-H]\,=\,\left(c-\dfrac{a}{\rho^2}\right)\,\hat{a}~.
\end{equation}
Again substituting the value in Eqn.(\ref{ph-der-1}), we come upto the following step,
\begin{align}
\bra{n, l}\,(i\,\partial_t-H\,)\,\ket{n, l}\,=\,\dfrac{1}{n+l}\,\bra{n, l-1}\,(i\,\partial_t-H\,)\,\ket{n, l-1}\,+\,\left(c-\dfrac{a}{\rho^2}\right)
\end{align}  
After iterating the above expression $(n+l-1)$ times, 
\begin{align}
\bra{n, l}\,(i\,\partial_t-H\,)\,\ket{n, l}\,=\,\bra{n, -n}\,(i\,\partial_t-H\,)\,\ket{n, -n}\,+\,(n+l)\,\left(c-\dfrac{a}{\rho^2}\right)
\end{align}  
Conveniently, as the vacuum state contribution, it can be chosen \cite{Dey} that $i\,\bra{n,-n}\partial_t\ket{n, -n}\,=\,\bra{n,-n}\,H\,\ket{n, -n}$. Hence, the Lewis phase factor $\Theta(t)$ is revealed as
\begin{equation}
\Theta_{\,n\,,\,l}(t)\,=\,(\,n\,+\,l\,)\,\int_0^t \left(c(T)-\dfrac{a(T)}{\rho^2(T)} \right)dT~.\label{DHO-phase1}
\end{equation}
For any provided given value of $l$ ( since $l=-n+m$), it would take the form,
\begin{equation}
\Theta_{\,n\,,\,m\,-\,n\,}(t)=m\int_0^t \left(c(T)-\dfrac{a(T)}{\rho^2(T)} \right)dT~.\label{DHO-phase2}
\end{equation}
Since the Hamiltonian coefficient $c(t)$ originates solely from the NC parameter, its presence in the above phase factor highlights a pure NC effect, as also observed in \cite{Dey}. However, due to the inclusion of a damping factor in our model, our explicit form of $c(t)$ differs from that in \cite{Dey}. Later, for various physical scenarios of damping, we will perform this integration for achieving the Lewis phase factor in a closed form.\\
\noindent Therefore, $\psi_{n, m-n}\,(r, \theta, t)$, the eigensystem of the Hamiltonian can be obtained by using  Eqn(s).(\ref{DHO-inv-efunc1}, \ref{DHO-phase2}) and it reads
\begin{eqnarray}
\psi_{n,m-n}(r,\theta,t)&=&e^{i\Theta_{n, m-n}(t)}\phi_{n, m-n}(r,\theta)\nonumber\\
&=&\lambda_{n}\dfrac{{(i\rho)}^m}{\sqrt{m!}}\exp{\left[im\int_0^t \left(c(T)-\dfrac{a(T)}{\rho^2(T)} \right)dT \right]}
\nonumber\\
&&\times~r^{n-m}e^{i(m-n)\theta-\dfrac{a(t)-i\rho\dot{\rho}}{2a(t){\rho}^2}r^2}U\left(-m,1-m+n,\dfrac{r^2}{\rho^2} \right)~;
\end{eqnarray}
where $\lambda_n$ is shown in Eqn.(\ref{lamb}).

\section{Explicit solutions for the noncommutative damped oscillator}
In this study, our primary focus is on damped oscillators in NC space. Our objective is to determine the Hamiltonian eigenfunctions in various scenarios of damping. In an effort to design various physical circumstances of damping, we consider various explicit form of the time dependent parameters existing in the structure of Hamiltonian eigenfunction. However, those must be consistent with the non linear EP equation as given in Eqn.(\ref{DHO-EP}). The method for constructing these exact analytical solutions relies on the Chiellini integrability condition \cite{man1, man2, chill}, a formalism previously utilized in \cite{Dey}, and adopted in our work as well. This means that the values of $a(t)$, $b(t)$, and $\rho(t)$ must together solve the EP equation in a manner consistent with the Chiellini integrability condition. Now, we are about to demonstrate the solving procedure of the EP equation, which will be consistent with this present study of damped oscillator in NC space. 

\subsection{Solution set of non linear Ermakov-Pinney equation}\label{sec-chil}
We begin by providing the EP equation (setting $\xi$ as $1$) found for our system
\begin{equation}
\ddot{\rho}-\dfrac{\dot{a}}{a}\dot{\rho}+ab\rho=\dfrac{a^2}{\rho^3}~.
\end{equation}
Defining the following,
\begin{eqnarray}
\dot{\rho}=\eta~,~
g(\rho)=-\dfrac{\dot{a}}{a}~,~h(\rho)=a\,b\,\rho-\dfrac{a^2}{\rho^3}~;\label{EP-comp}
\end{eqnarray}
the EP equation can be written as
\begin{equation}
\eta\dfrac{d\eta}{d\rho}+\eta\,g(\rho)+h(\rho)=0~.
\end{equation}
As per the Chilleni integrability condition \cite{chill}, if
\begin{align}
\dfrac{d}{d\rho}\left(\dfrac{h(\rho)}{g(\rho)}\right)=q\,g(\rho)~;~(q=\text{constant})\label{chill1}
\end{align}
then we consider the following as the solution of $\eta$,  
\begin{align}
\eta=\lambda_q\dfrac{h(\rho)}{g(\rho)}~~~\text{with}~~~
\lambda_q=\dfrac{-1\pm\sqrt{1-4\,q}}{2\,q}~.\label{chill2}
\end{align}
The interesting consequence emerged from the above condition is that, for a specific value of $g(\rho)$ or $h(\rho)$, the other parameters can be integrated out in the following manner: 
\begin{eqnarray}
\eta(\rho)\,=\,q\,\lambda_q\int^{\rho}\,g(\rho)\,d\rho~,~h(\rho)\,=\,q\,g(\rho)\int^{\rho}\,g(\rho)\,d\rho~;\label{chil-sol-1}
\end{eqnarray}
where the first relation can be found by substituting Eqn.(\ref{chill2}) into Eqn.(\ref{chill1}) and the second relation can directly be derived from Eqn.(\ref{chill1}). \\
\noindent Alternatively, when $h(\rho)$ is known, substituting Eqn.(\ref{chill2}) into Eqn.(\ref{chill1}) and rewriting $g(\rho)$ in terms of $h(\rho)$ and $\eta$, we have
\begin{align}
\dfrac{d\eta}{d\rho}&=q\,\lambda_q^2\,\dfrac{h(\rho)}{\eta(\rho)}~.
\end{align}
From the above relation $\eta$ can be found and substituting it in Eqn.(\ref{chill2}), $g(\rho)$ can also be obtained. They are given by
\begin{align}
\eta(\rho)\,=\,\lambda_q\sqrt{2\,q\int^{\rho}\,h(\rho)\,d
\rho}~&,~g(\rho)=\dfrac{\,h(\rho)}{\sqrt{2\,q\int^{\rho}\,h(\rho)
\,d\rho}}~.
\end{align}
In \cite{Dey}, a set of EP parameters is explicitly determined by applying various assumed values of $g(\rho)$ in Eqn.(\ref{chil-sol-1}). When $g(\rho)$ remains constant, the resulting time dependent solution varies exponentially. On the other hand, if $g(\rho)$ increases non linearly with respect to $\rho$, the resulting solution varies rationally over time. The method for obtaining these solutions is described below.
\subsubsection{Exponential EP solution}
In this context, $g(\rho)$ is assumed to be $\Gamma$, a positive, real constant. From Eqn.(\ref{EP-comp}), we can directly derive the explicit form of $a(t)$ as follows,
\begin{eqnarray}
g(\rho)=-\dfrac{\dot{a}}{a}\,=\,\Gamma\nonumber\\
\Rightarrow\,a(t)\,=\,\sigma\,e^{-\Gamma\,t}~;
\end{eqnarray}
where $\sigma$ is a constant. Subsequently, by equating the second relation in Eqn.(\ref{chill2}) with the third relation in Eqn.(\ref{EP-comp}), we obtain the following relation,
\begin{eqnarray}
h(\rho)\,=\,q\,\Gamma\int^{\rho}\,\Gamma\,d\rho\,=\,q\,\Gamma^2\,\rho\nonumber\\
\Rightarrow\,ab\rho-\dfrac{a^2}{\rho^3}\,=\,q\,\Gamma^2\,\rho~.
\end{eqnarray}
By comparing the powers of $\rho$ on both sides of the above equation, it becomes clear that $a(t),b(t),\sim,$ constant and $a(t)\sim,\rho^2(t)$. Consequently, the explicit forms of the remaining EP parameters can be derived as,
\begin{eqnarray}
b(t)\,=\,\Delta\,e^{\Gamma\,t}~,~\rho(t)\,=\,\mu\,e^{-\Gamma\,t/2}~;
\end{eqnarray}
where $\Delta, \mu$ are real, positive constants. \\
\noindent It should be noted that, the constants used in the solution are not independent, they also satisfy the following constraint relation
\begin{equation}
\mu^4=\dfrac{\xi^2{\sigma^2}}{\sigma\Delta-\dfrac{1}{4}\Gamma^2}~;
\end{equation}
which is found by substituting the EP solution set in Eqn.(\ref{DHO-EP}). 
\subsubsection{Rational EP solution}\label{DHO-rat}
Here, $g(\rho)$ is assumed to vary in a non linear manner with the function $\rho(t)$. Thus, from Eqn.(\ref{EP-comp}),
\begin{eqnarray}
g(\rho)=-\dfrac{\dot{a}}{a}\,=\,\Gamma\,\rho^{k}~;
\end{eqnarray}
where $k$ is a positive integer.
By using the second relation in Eqn.(\ref{chil-sol-1}) and the third relation in Eqn.(\ref{EP-comp}),
\begin{eqnarray}
h(\rho)\,=\,q\,\Gamma\rho^{k}\int^{\rho}\,\Gamma\rho^{k}\,d\rho\,=\,q\,\Gamma^2\,\dfrac{\rho^{2k+1}}{k+1}\nonumber\\
\Rightarrow\,ab\rho-\dfrac{a^2}{\rho^3}\,=\,q\,\Gamma^2\,\dfrac{\rho^{2k+1}}{k+1}
\end{eqnarray}
By comparing the powers of $\rho$ on both sides of the above equation, it becomes clear that $a(t)\sim\,\rho^{k+2}$ and $a(t)\,b(t)\sim\,\rho^{2k}$. The explicit form of $\rho$ can be obtained as
\begin{eqnarray}
g(\rho)=-\dfrac{\dot{a}}{a}\,=\,\Gamma\rho^{k}\nonumber\\
\Rightarrow\,-\dfrac{(k+2)\dot{\rho}}{\rho}\,=\,\Gamma\rho^{k}\nonumber\\
\Rightarrow\,\rho\,=\,\dfrac{\left(1+\dfrac{2}{k}\right)^{1/k} }{(\Gamma{t}+\chi)^{1/k}}~;
\end{eqnarray}
where $\chi$ appears as an integration constant, such that $(\Gamma{t}+\chi)~\neq~0$\\
\noindent In this context, we aim to adjust the previously derived value of $\mu$, by multiplying it by $\mu$, a real and positive constant. This adjustment will allow us to detect the presence of the parameter $\rho$ within a quantity that is explicitly computed using multiple parameters. This approach will become evident as we proceed to calculate the physical properties of the damped harmonic oscillator with noncommutativity. Thus, in our work, $\rho(t)$ holds the value  
\begin{eqnarray}
\rho\,=\,\dfrac{\mu\,\left(1+\dfrac{2}{k}\right)^{1/k} }{(\Gamma{t}+\chi)^{1/k}}~.
\end{eqnarray}
The previously found proportionality relation $a(t)\sim\,\rho^{k+2}$ and $a(t)\,b(t)\sim\,\rho^{2k}$ implies that
\begin{eqnarray}
a(t)=\dfrac{\sigma\,\left(1+\dfrac{2}{k}\right)^{\,(k+2)/k}}{(\Gamma{t}+\chi)^{\,(k+2)/k}}~,~
b(t)=\dfrac{\Delta\,\left(1+\dfrac{2}{k}\right)^{\,(k-2)/k} }{(\Gamma{t}+\chi)^{\,(k-2)/k}}~;
\end{eqnarray}
where the constant $\sigma$ and $\Delta$ have real, positive value. It is noteworthy to mention that, similar to the previous solution set, the constants adhere to the following relation
\begin{equation}
\Gamma^2\mu=(k+2)^2\,\left(\sigma\Delta\mu-\frac{\sigma^2}{\mu^3}\right)~;
\end{equation}
which is found by substituting the EP solution set in Eqn.(\ref{DHO-EP}). \\
\noindent Now, we are about to enter into the core of our work done in \cite{SG1} which explores a class of exact solution of a noncommutative damped harmonic oscillator.

\subsection{Exponential solution for model system} 
We begin by mentioning the exponentially varying EP solution set which is provided in \cite{Dey} and is also reviewed in the previous section. The exponential EP solution read, 
\begin{eqnarray}
a(t)=\sigma e^{-\Gamma{t}}\,\,\,,\,\,\,b(t)=\Delta e^{\Gamma{t}}\,\,\,,\,\,\rho(t)={\mu}e^{-\Gamma{t/2}}
\label{EP-exp}
\end{eqnarray}
with the constraint relation
\begin{equation}
\mu^4=\dfrac{{\sigma^2}}{\sigma\Delta-\dfrac{1}{4}\Gamma^2}~;
\label{EP-exp-cons}
\end{equation}
where $\sigma, \Delta, \mu$ and $\Gamma$ are the real, positive constants.\\
\noindent We will proceed with finding the eigenfunctions of the Hamiltonian using the selected time dependent coefficients (EP parameters). To accomplish this, the explicit analytical form of $f(t)$ and $\omega(t)$, the damping factor and the oscillator's frequency respectively, are to be specified. Since, we are finding the solution using Lewis technique, our solution comprises two main components : $\phi_{n, m-n}(r, \theta)$, the eigenfunction of Lewis invariant $I^{'}(t)$ and $\Theta_{n, m-n}(t)$, the Lewis phase factor. The specific form of $\phi_{n, m-n}(r, \theta)$ in Eqn.(\ref{DHO-inv-efunc1}) (also Eqn.\ref{DHO-inv-efunc2})) implies that once the EP (exact parameter) parameters are determined, the explicit form of $\phi_{n, m-n}(r, \theta)$ is fixed accordingly. However, the phase factor $\Theta_{n, m-n}(t)$ can take various explicit forms depending on the specific choices of $f(t)$ and $\omega(t)$. Thus, while $\phi_{n, m-n}(r, \theta)$ remains fixed once the EP parameter are set, a class of solution can be generated corresponding to each type of $\Theta_{n, m-n}(t)$.

\subsubsection{Eigenfunction of Lewis invariant}
The invariant's eigenfunction (Eqn.(\ref{DHO-inv-efunc1})) with regard to the exponential EP variables is derived as,
\begin{eqnarray}
\phi_{n,m-n}(r,\theta)=\lambda_{n}\dfrac{{(i{\mu}e^{-\Gamma{t/2}})}^m}{\sqrt{m!}}    r^{n-m}e^{i(m-n)\theta-\dfrac{2\sigma+i\mu^2\Gamma}{4\sigma\mu^2{e^{-\Gamma{t}}}}r^2}U\left(-m,1-m+n,\dfrac{r^2{e^{\Gamma{t}}}}{\mu^2} \right)
\label{DHO-inv-efunc-exp}
\end{eqnarray}
where $\lambda_n$ is given by
\begin{eqnarray}
\lambda_n^{\,2}\,=\,\dfrac{1}{\pi\,n!\,[\mu^2\,exp\,(-\Gamma{t})]^{1+n}}~.
\end{eqnarray}


\subsubsection{Lewis phase factors}
The calculation of the Lewis phases, influenced by diverse choices of the damping scenarios, is undoubtedly an interesting part of this work. As previously mentioned, the integral form of this phase (Eqn.(\ref{DHO-phase1})) reveals a clear NC signature through the Hamiltonian coefficient $c(t)$. Given suitable explicit forms of $f(t)$ and $\omega(t)$, the structural forms of the NC parameters $\theta(t)$ and $\Omega(t)$ can be derived. These NC parameters, in turn, define the form of $c(t)$, which can be integrated exactly according to the relation provided in Eqn.(\ref{DHO-phase1}).\vskip .20cm
\noindent{\bf $\langle A\rangle$ Solution Set-Ia}

\noindent 
Firstly, the damping factor is removed by setting $f(t)=1$. Consequently, the system's dissipation attributes to $\omega(t)$, the angular frequency decaying exponentially with respect to time. Therefore,
\begin{eqnarray}
\eta(t)=0\,\,\Rightarrow\,\,f(t)=\int_0^t\,\eta(s)\,ds\,=\,1\\
\omega(t)={\omega_0}\,exp(-\Gamma{t}/2)\,\,\,.
\label{eqn34}
\end{eqnarray}
By substituting the above relations in Eqn(s).(\ref{DHO-a}, \ref{DHO-b}), the explicit structure of the NC parameters, for the present case, are calculated to be
\begin{eqnarray}
\theta(t)&=&\dfrac{2}{M\omega_0}\, \sqrt{M\sigma\,-e^{\Gamma t}}\label{eqn35b}      \\
\Omega(t)&=&2\sqrt{\,M[\Delta\,exp\,(\Gamma{t})-M\omega_0^2\,exp\,(-\Gamma{t})]}. \label{eqn36b}
\end{eqnarray}
Substituting the above form of the dynamic NC parameters in the structural form of $c(t)$ in Eqn.(\ref{DHO-c}), we 
get,
\begin{equation}
c(t)=\sqrt{\dfrac{ \Delta\,exp\,(\Gamma{t})-M{\omega_0}^2\,exp\,(-\Gamma{t}) }{M}} 
\,+\,\omega_0\,exp\,(-\Gamma{t}/2)\sqrt{\,M\sigma\,exp\,(-\Gamma{t})-1 \,}.\label{eqn38}
\end{equation}
With the explicit form of $a(t)\,,\rho(t)$\, and $c(t)$\, in place, we perform the integration given in Eqn.(\ref{DHO-phase1}) and achieve the following closed form of the Lewis phase. The phase is explicitly found to be
\begin{eqnarray}
\Theta_{\,n\,,\,l}(t)&=&\,(\,n\,+\,l\,)\,\dfrac{\omega_0}{2\sqrt{M\sigma}\Gamma}\ \left[log_{e}\dfrac{e^{{\Gamma}t}-2M\sigma-2\sqrt{M\sigma(M\sigma-e^{{\Gamma}t})}}{1-2M\sigma-2\sqrt{M\sigma(M\sigma-1)}}\right.\nonumber\\ 
&&\left. -{\Gamma}t-2\sqrt{M\sigma(M{\sigma}e^{-{2\Gamma}t}-e^{-{\Gamma}t})}+2\sqrt{M\sigma(M\sigma-1)}\ \right]\nonumber\\
&&+\dfrac{2(n+l)}{\Gamma}\,\left[\sqrt{\frac{\Delta}{M}e^{{\Gamma}t}-{\omega_0^2}e^{-{\Gamma}t}}-\sqrt{\dfrac{\Delta}{M}-{\omega_0^2}}\right.\nonumber\\
&&\left.+2i{\omega_0}\left\{e^{-{\Gamma}t/2} {_{2}F_{1}}\left(-\frac{1}{4},\frac{1}{2},\frac{3}{4},\frac{{\Delta}e^{{2\Gamma}t}}{M\omega_0^2}\right)- {_{2}F_{1}}\left(-\frac{1}{4},\frac{1}{2},\frac{3}{4},\frac{\Delta}{M\omega_0^2}\right)\right\}\right]
-\frac{\sigma}{\mu^2}(n+l)t~.  \nonumber\\
\label{eqn39}                      
\end{eqnarray}
where $_{2}F_{1}(a,b,c;t)$ denotes the Gauss hypergeometric function. 
Interestingly, the NC parameters derived for the designed damping scenario, found to be appropriate for the calculating the Lewis phase in a closed form. Additionally, it is noteworthy that the phase includes a complex part, implying a decaying wave function over time.\\
\noindent The eigenfunction of the Hamiltonian (Eqn.(\ref{DHO-comham}), for the selected case of damping, can now be readily determined by connecting the above phase factor to Eqn.(\ref{DHO-inv-efunc-exp}), following the relation outlined in Eqn.(\ref{DHO-eqnpsi}). 
\vskip .20cm
\noindent {\bf $\langle B\rangle$ Solution Set-Ib}

\noindent Here, the system's damping $f(t)$ attributes to the exponentially decaying damping factor since the angular frequency $\omega(t)$ is set to be a constant. Therefore, 
\begin{eqnarray}
\eta(t)=\Gamma\,\,\Rightarrow\,\,f(t)=\int_0^t\,\eta(s)\,ds\,=\,exp\,(-\Gamma{t})~;~\omega(t)={\omega_0}.
\label{10x}
\end{eqnarray}
By substituting the above relations in Eqn(s).(\ref{DHO-a}, \ref{DHO-b}), the explicit structure of the NC parameters, for the present case, are calculated to be
\begin{eqnarray}
\theta(t)&=&\dfrac{2}{M\omega_0}\sqrt{M{\sigma}\,- 1}\,\,e^{-\Gamma t}\label{eqn40b}\\
\Omega(t)&=&2e^{\Gamma{t}}\sqrt{M\,[\Delta\,-M{\omega_0}^2]}.
\label{eqn41b}
\end{eqnarray}
Substituting the above form of the dynamic NC parameters in the structural form of $c(t)$ in Eqn.(\ref{DHO-c}), we 
get,
\begin{equation}
c(t)= \sqrt{\dfrac{ \Delta\,-M{\omega_0}^2\, }{M}} \,+\,\omega_0\sqrt{M\sigma-1}\,=\,constant ~. \label{eqn43}
\end{equation}
With the explicit form of $a(t)\,,\rho(t)$\, and $c(t)$\, in place, we perform the integration given in Eqn.(\ref{DHO-phase1}) and achieve the following closed form of the Lewis phase. The phase is explicitly found to be
\begin{align}
\Theta_{\,n\,,\,l}(t)\,=&(\,n\,+\,l\,)\left[-\frac{\sigma}{\mu^2}\,+\,\sqrt{\frac{ \Delta\,-M{\omega_0}^2\,}{M} } \,+\omega_0\sqrt{M\sigma-1}  \right]\,t ~.\label{eqn44}
\end{align}
Once more, we can derive an exact expression for the phase, which varies linearly with time in this scenario. It should be marked that, for being in real domain, phase is significantly influenced by the parameters $\Delta$, $M$, $\sigma$, and $\omega_0$. The phase $\Theta_{n,l}$ remains real if and only if $\Delta-M\omega_{0}^2 \geq 0$ and $M\sigma \geq 1$; otherwise, it becomes complex.\\
\noindent The eigenfunction of the Hamiltonian (Eqn.(\ref{DHO-comham}), for the selected case of damping, can now be readily determined by connecting the above phase factor to Eqn.(\ref{DHO-inv-efunc-exp}), following the relation outlined in Eqn.(\ref{DHO-eqnpsi}).


\noindent {\bf $\langle C\rangle$ Solution Set-Ic}

\noindent Here, the system's damping $f(t)$ attributes to the damping factor and the angular frequency $\omega(t)$ both decaying exponentially with respect to time. Therefore,
\begin{eqnarray}
f(t)=exp\,(-\Gamma{t})~;~\omega(t)={\omega_0}\,exp\,(-\Gamma{t}/2).
\label{eqn45}
\end{eqnarray}
By substituting the above relations in Eqn(s).(\ref{DHO-a}, \ref{DHO-b}), the explicit structure of the NC parameters, for the present case, are calculated to be
\begin{eqnarray}
\theta(t)&=&\dfrac{2}{M\omega_0}\sqrt{(M{\sigma}\,- 1)}e^{-\Gamma{t}/2}\,\,\label{eqn46}\\
\Omega(t)&=&2\sqrt{M\,[\Delta\,exp(\Gamma{t})-M{\omega_0}^2\, \,]}\,\,e^{\Gamma t/2}.\label{eqn47}
\end{eqnarray}
Substituting the above form of the dynamic NC parameters in the structural form of $c(t)$ in Eqn.(\ref{DHO-c}), we 
get,
\begin{align}
c(t)= \sqrt{\dfrac{ \Delta\,-M{\omega_0}^2\,\exp\left[-\Gamma{t}\right] }{M}} 
 +  {\omega_0}\,e^{-\Gamma t/2}\sqrt{M{\sigma}\,-1}~.\label{eqn49}
\end{align}
With the explicit form of $a(t)\,,\rho(t)$\, and $c(t)$\, in place, we perform the integration given in Eqn.(\ref{DHO-phase1}) and achieve the following closed form of the Lewis phase. The phase is explicitly found to be
\begin{eqnarray}
\Theta_{\,n\,,\,l}(t)\,&=&\,\dfrac{(\,n+l\,)}{\Gamma\,\sqrt{M}}\,\left[\,\sqrt{\Delta}\,\Gamma\,t\,+\,2\sqrt{\Delta-M\omega_0^2}\,-2\sqrt{\Delta-M\omega_0^2{\exp\,(-\Gamma{t})}}\right.\nonumber\\
&&\left.+\,2\sqrt{\Delta}\,log\,\left(\frac{\Delta+\sqrt{\Delta[\Delta-M\omega_0^2\,\exp\,(-\Gamma{t}) ]}}{\Delta+\sqrt{\Delta[\Delta-M\omega_0^2\,]}}\right)     \right]\nonumber \\
&&-\,(\,n+l\,)\left[\dfrac{\sigma\,t}{\mu^2}\,+\,\dfrac{2}{\Gamma}\,\omega_0\,\left(e^{-\Gamma t/2}-1\right)\sqrt{\,M\sigma-1} \right].\label{eqn50}
\end{eqnarray}
The eigenfunction of the Hamiltonian (Eqn.(\ref{DHO-comham}), for the selected case of damping, can now be readily determined by connecting the above phase factor to Eqn.(\ref{DHO-inv-efunc-exp}), following the relation outlined in Eqn.(\ref{DHO-eqnpsi}).


\subsection{Rational solution for model system}
We begin by mentioning the rationally varying EP solution set which is provided in \cite{Dey} and is also reviewed in the previous section \ref{DHO-rat}. The rational EP solution set read
\begin{eqnarray}
a(t)=\dfrac{\sigma\,\left(1+\dfrac{2}{k}\right)^{\,(k+2)/k}}{(\Gamma{t}+\chi)^{\,(k+2)/k}}~,~
b(t)=\dfrac{\Delta\,\left(1+\dfrac{2}{k}\right)^{\,(k-2)/k} }{(\Gamma{t}+\chi)^{\,(k-2)/k}}~,~
\rho(t)=\dfrac{\mu\left(1+\dfrac{2}{k}\right)^{1/k} }{(\Gamma{t}+\chi)^{1/k}}\label{EP-rat}
\end{eqnarray}
with the following constraint relation
\begin{equation}
\Gamma^2\mu=(k+2)^2\,\left(\sigma\Delta\mu-\frac{\sigma^2}{\mu^3}\right)~;\label{EP-rat-cons}
\end{equation}
where $k$ holds the integer values and $\sigma$, $\mu$, $\Delta$, $\chi$ and $\Gamma$ are the constants that take real, positive values and ensure that $(\Gamma{t}+\chi)~\neq~0$.


\subsubsection{Eigenfunction of Lewis invariant}
The eigenfunction of the invariant (which is given by 
Eqn.(\ref{DHO-inv-efunc1})) takes on the following form for the rational solution set of EP equation:
\begin{align}
\phi_{n\,,\,m-n}(r,\theta)=\lambda_{n}\,\dfrac{{(i\mu)}^{\,m}}{\sqrt{m!}}\left[\dfrac{k+2}{k(\Gamma{t}+\chi)}\right]^{m/k}    r^{n-m}e^{i\theta(m-n)-\dfrac{[\sigma\,(k+2)\,+\,i\mu^2\Gamma]\,\,(\Gamma{t}+\chi)^{2/k}\,\,\,k^{2/k} }{2\sigma\,(k+2)^{\,(k+2)/k}\mu^2}r^2}\nonumber \\
\times\,\,\,U\left(-m,1-m+n,\,\dfrac{r^2[k(\Gamma{t}+\chi)]^{2/k}}{\mu^2\left(k+2\right)^{2/k}}\,\right)
\label{DHO-inv-efunc-rat}
\end{align}
where $\lambda_n$ is computed as 
\begin{eqnarray}
\lambda_n^{\,2}=\dfrac{1}{\pi\,n!\mu^{2n+2}}\left[\dfrac{k(\Gamma{t}+\chi)}{k+2}\right]^{2(1+n)/k}.
\end{eqnarray}
\subsubsection{Lewis phase factor}
Once again, in order to compute the Lewis phase factor for constructing the Hamiltonian eigenfunction varying rationally with respect to time, we require the prepare the damping situation as per the rationally varying EP solution. \\
\noindent Here, the damping factor is removed by setting $f(t)=1$. Consequently, the system's dissipation attributes to $\omega(t)$, the angular frequency decaying rationally with respect to time. Therefore,
\begin{eqnarray}
\eta(t)=0\,\,\Rightarrow\,\,f(t)=1\\
\omega(t)=\dfrac{\omega_0}{(\Gamma\,t+\chi)}~.
\end{eqnarray}
By substituting the above relations in Eqn(s).(\ref{DHO-a}, \ref{DHO-b}), the explicit structure of the NC parameters, for the present case, are calculated to be
\begin{eqnarray}
\theta(t)&=& \dfrac{2\,(\Gamma\,t+\chi)}{M\,\omega_0}\,\sqrt{M\sigma\,\left[\dfrac{(k+2)}{k\,(\Gamma{t}\,+\,\chi)}\right]^{(k+2)/k}\,-\,1}     \label{eqn52}                   \\ \nonumber \\
\Omega(t)&=& \,2\,\sqrt{M\Delta\,\left[\dfrac{k+2}{k(\Gamma{t}+\chi)}\right]^{\,(k-2)/k}\,-\,\dfrac{M^{\,2}\omega_0^{\,2}}{(\Gamma\,t+\chi)^2}}~.\label{eqn53}
\end{eqnarray}
Now, for the simplification of the calculation, we chose $k=2$. This also ensures the achievement of an exact integrated form of the Lewis phase factor (Eqn.(\ref{DHO-phase1}). The EP variables are reduced to be,
\begin{eqnarray}
a(t)=\dfrac{4\sigma}{(\Gamma{t}+\chi)^{\,2}}\,\,,\,\,b(t)\,=\,\Delta\,\,,\,\,\rho(t)=\left[\dfrac{2\mu^{\,2}}{\Gamma{t}+\chi}\right]^{1/2}.
\end{eqnarray}
Substituting the above form of the dynamic NC parameters in the structural form of $c(t)$ in Eqn.(\ref{DHO-c}), we 
get,
\begin{equation}
c(t)\,=\,\dfrac{\omega_0}{(\Gamma\,t+\chi)}\,\sqrt{\dfrac{4\sigma\,M}{(\Gamma{t}+\chi)^2}\,-\,1}\,+\,\sqrt{\dfrac{\Delta}{M}\,-\,\dfrac{\omega_0^{\,2}}{(\Gamma\,t+\chi)^{\,2}}}\,\,\,\,.\label{DHO-c-rat}
\end{equation} 
With the explicit form of $a(t)\,,\rho(t)$\, and $c(t)$\, in place, we perform the integration given in Eqn.(\ref{DHO-phase1}) and achieve the following closed form of the Lewis phase. The phase is explicitly found to be
\begin{eqnarray}
\Theta_{\,n, l\,}(t)&=&\dfrac{(n+l)}{\Gamma}\,\left[\omega_0\,\,tan^{\,-1}\left(\dfrac{\omega_0}{\sqrt{\frac{\Delta}{M}{(\Gamma\,t+\chi)^2}-\omega_0^2}}\right)+\sqrt{\dfrac{\Delta\,{(\Gamma\,t+\chi)^2}}{M}-\omega_0^2}\,-\,\frac{2\sigma}{\mu^2}\,log_{e}\,\frac{(\chi+\Gamma\,t)}{\chi}\right.\nonumber\\
&&\left.-{\sqrt{\frac{\Delta}{M}{\chi^2}-\omega_0^2}}-\omega_0\,\,tan^{\,-1}\left(\dfrac{\omega_0}{\sqrt{\frac{\Delta}{M}{\chi^2}-\omega_0^2}}\right)\right] \nonumber\\
&& +\dfrac{\omega_0(n+l)}{\Gamma}\left[\dfrac{\sqrt{4\,\sigma\,M-\chi^2}}{\chi}-\dfrac{\sqrt{4\,\sigma\,M-(\chi+\Gamma\,t)^2}}{(\chi+\Gamma\,t)}
\right.\nonumber\\
&&\left. +ilog_{e}\dfrac{(\chi+\Gamma\,t)+{\sqrt{(\chi+\Gamma\,t)^2-4\,\sigma\,M}}}{\chi+\sqrt{\chi^2-4\,\sigma\,M}} \right].
\label{eqn55}
\end{eqnarray}
The eigenfunction of the Hamiltonian (Eqn.(\ref{DHO-comham})), for the selected case of damping, can now be readily determined by connecting the above phase factor to Eqn.(\ref{DHO-inv-efunc-rat}), following the relation outlined in Eqn.(\ref{DHO-eqnpsi}).



\subsection{Elementary solution for model system}
We propose a straightforward approach to solving the EP equation. The method involves the following steps.  When $\rho(t)$ is remained to be an arbitrary time varying function, if we express the rest EP parameters such that $a(t) = constant\times\dot{\rho}$ and $b(t) = constant\times\frac{a}{\rho^4}$, those variables, in the generic expressions, would always be consistent with the EP equation, provided there is a specific constraint relationship abided by the constants.    
\noindent Here we proceed with a special solution set generated from the above generic form of the EP variables. We refer to those as the phrase ``elementary solution set", which we read as
\begin{eqnarray}
a(t)={\sigma}\,\,\,\,,\,\,\,b(t)=\dfrac{{\Delta}}{{(\Gamma\,t\,+\,\chi)^4}}\,\,\,,\,\,\rho(t)=\mu(\Gamma{t}\,+\,\chi)~;\label{EP-elm}
\end{eqnarray}
$\sigma$, $\mu$, $\Delta$, $\chi$ and $\Gamma$ are the constants that take real, positive values and ensure that $(\Gamma{t}+\chi)~\neq~0$. Moreover, they adhere to the following constraint relation,
\begin{equation}
\Delta\mu^4=\sigma\,\,.\label{EP-elm-cons}
\end{equation}



\subsubsection{Eigenfunction of the Lewis invariant}
The eigenfunction of the invariant (which is given by 
Eqn.(\ref{DHO-inv-efunc1})) takes on the following form for the elementary solution set of EP equation:
\begin{eqnarray}
\phi_{n,m-n}(r,\theta)&=&\lambda_{n}\dfrac{{[i\mu(\Gamma\,t+\chi)]}^m}{\sqrt{m!}}r^{n-m}e^{i\theta(m-n)-\dfrac{\sigma-i\mu^2\Gamma(\Gamma\,t+\chi)}{2\sigma\mu^2(\Gamma\,t+\chi)^2}r^2}\nonumber \\
&&\times\,\,U\left(-m,1-m+n,\dfrac{r^2}{\mu^2(\Gamma\,t+\chi)^2} \right)
\label{DHO-inv-efunc-elm}
\end{eqnarray}
where $\lambda_n$ is given by
\begin{eqnarray}
\lambda_n^2=\dfrac{1}{\pi{n!}{\left[\mu(\Gamma\,t+\chi) \right]}^{2+2n}}~. 
\end{eqnarray}
\subsubsection{Lewis phase factors}
In order to prepare a suitable circumstances of dissipation such that it ensures the achievement of the Lewis phase in an exact form, we again consider the damping scenario chosen to compute the rationally varying eigenfunction of the Hamiltonian. Therefore,
\begin{eqnarray}
\eta(t)=0\,\,\Rightarrow\,\,f(t)=1\\
\omega(t)=\dfrac{\omega_0}{(\Gamma\,t+\chi)}~.
\end{eqnarray}
\noindent By substituting the above relations in Eqn(s).(\ref{DHO-a}, \ref{DHO-b}), the explicit structure of the NC parameters, for the present case, are calculated to be
\begin{eqnarray}
\theta(t)=\dfrac{2\,(\Gamma\,t+\chi)}{\omega_0\,M}\,\sqrt{M\,\sigma-1}\label{eqn59}\\
\Omega(t)=2\sqrt{\dfrac{M\Delta}{(\Gamma\,t+\chi)^4}-\dfrac{M^2\,\omega_0^2}{(\Gamma\,t+\chi)^{2}}}~.\label{eqn60}
\end{eqnarray}
Substituting the above form of the dynamic NC parameters in the structural form of $c(t)$ in Eqn.(\ref{DHO-c}), we 
get,
\begin{align}
c(t)=\sqrt{\dfrac{\Delta}{M(\Gamma\,t+\chi)^4}-\dfrac{\omega_0^2}{(\Gamma\,t+\chi)^2}}\,+\,\dfrac{\omega_0}{(\Gamma\,t+\chi)}\sqrt{M\,\sigma-1}.\label{DHO-c-elm}
\end{align}
With the explicit form of $a(t)\,,\rho(t)$\, and $c(t)$\, in place, we perform the integration given in Eqn.(\ref{DHO-phase1}) and achieve the following closed form of the Lewis phase. The phase is explicitly found to be
\begin{align}
\Theta_{\,n\,,\,l}(t)&=\,(n+l\,)\left[\omega_0\dfrac{\sqrt{M\,\sigma-1}}{\Gamma}\log\,\dfrac{(\Gamma\,t+\chi)}{\chi}-\frac{{\sigma}t}{\mu^2\chi(\Gamma\,t+\chi)} \right] +\,\dfrac{(\,n+l\,)}{\Gamma}\left[\sqrt{\dfrac{\Delta}{M\chi^2}-\omega_0^2}\right.\nonumber\\
&\left.-\sqrt{\dfrac{\Delta}{M(\Gamma\,t+\chi)^2}-\omega_0^2}
+\omega_0\left\{\tan^{-1}\left(\dfrac{\omega_0\chi}{\sqrt{\dfrac{\Delta}{M}-\chi^2\omega_0^2}}\right)-\tan^{-1}\left(\dfrac{\omega_0(\Gamma\,t+\chi)}{\sqrt{\dfrac{\Delta}{M}-\omega_0^2(\Gamma\,t+\chi)^2}}\right)\right\}\,\, \right].
\label{eqn62}
\end{align} 
The eigenfunction of the Hamiltonian (Eqn.(\ref{DHO-comham}), for the selected case of damping, can now be readily determined by connecting the above phase factor to Eqn.(\ref{DHO-inv-efunc-elm}), following the relation outlined in Eqn.(\ref{DHO-eqnpsi}).\\
\noindent Subsequently, an exotic model of the quantum damped harmonic oscillator, derived from the generalized solution set introduced at the start of this section, is also explored in \cite{sg1exo}.


\section{Energy profile of the model system}\label{expect}
Here, we aim to establish the expression of the energy expectation value, as obtaining an exact analytical estimation of the energy dynamics will also allow us to illustrate its behaviour graphically for various kinds of model systems, as discussed in the previous section. For doing so, we require to derive the Hamiltonian's expectation value in respect of the Hamiltonian's eigenstate. Therefore, $\langle H \rangle$, the expectation value of the Hamiltonian (Eqn.(\ref{DHO-comham})) is expressed as, 
\begin{equation}
\langle H\rangle = \dfrac{a(t)}{2}(\langle{p_1}^2\rangle+\langle{p_2}^2\rangle)+\dfrac{b(t)}{2}(\langle{x_1}^2\rangle+\langle{x_2}^2\rangle)+c(t)(\langle{p_1}{x_2}\rangle-\langle{p_2}{x_1}\rangle)\,\,\,.\label{DHO-enrg1}
\end{equation}
While the expectation values of individual canonical operators are given in \cite{Dey}, we will demonstrate the detailed calculation procedure for some of these operators in the following relation. We also describe the procedure for obtaining a generalized structure of the matrix element of the canonical coordinate operators with arbitrary finite powers. The specific value of $\langle x_i^2 \rangle$, where $i\,\in\,[1, 2]$ as provided in \cite{Dey}, can be verified as a special case of this generalized outcome.\\
\noindent This explanation aims to provide an understanding to calculate the expectation values for the remaining operators in the mentioned relation. In this regard, we denote the Hamiltonian eigenstate as $|n,l\rangle_H$. 



\subsection{Matrix elements of the canonical coordinate operators with arbitrary finite powers}
\noindent First, we calculate $_{H}\langle n,l|x^k|n,l\rangle_{H}$, the matrix elements (including $\hbar$) of the canonical coordinate operators with arbitrary finite powers. The concerned matrix element can be expressed as,
\begin{eqnarray} 
_{H}\langle n,m-n|x^k|n,m'-n\rangle_{H}&=&\int 
~_{H}\langle n,m-n |r,\theta\rangle\langle r,\theta|r^k cos^k\theta|n,m'-n\rangle_{H}\,r dr d\theta\nonumber\\
&=&\dfrac{1}{2^k}e^{i(\Theta_{n,m'-n}-\Theta_{n,m-n})}\int r^{k+1} dr d\theta~(e^{i\theta}+e^{-i\theta})^k \nonumber\\
&&~~~~~~~~~~~~~~~~~~~~~~~~~~~~~~~~\times\phi^*_{n,m-n}(r,\theta)\phi_{n,m'-n}(r,\theta)\nonumber\\ 
\label{eqn64}
\end{eqnarray} 
where $|n,l\rangle_{H}~=~e^{i\Theta_{n,l}}|n,l\rangle$, the relation that presents the Lewis theory, is substituted in the above equation.
The invariant's ($I^{'}(t)$) eigenfunction $\phi_{n,m'-n}(r,\theta)$ is expressed as $\phi_{n,m'-n}(r,\theta)$~=~$\langle r,\theta|n,m'-n\rangle$. Now, we proceed with Eqn.(\ref{eqn64}) in the following manner.
\begin{eqnarray}
_{H}\langle n,m-n|x^k|n,m'-n\rangle_{H}&=&\dfrac{\pi}{2^{k-1}} \sum_{r=0}^{k} {^{k}C_r}\delta_{m',m+2r-k}  A(n,m,m+2r-k)\nonumber\\
&\times& \int_0^\infty r~dr~r^{2(n-m-r+k)}
e^{\dfrac{-r^2}{\hbar \rho^2}} \nonumber\\
&\times& U\left(-m,1-m+n,\dfrac{r^2}{\hbar\rho^2} \right)\nonumber\\
&\times& U\left(-m-2r+k,1-m-2r+k+n,\dfrac{r^2}{\hbar\rho^2} \right)
\label{eqn67}
\end{eqnarray}
where $A(n,m,m+2r-k)=e^{i(\Theta_{n,m-n+2r-k}-\Theta_{n,m-n})}\lambda_n^2\dfrac{(-i\hbar^{1/2}\rho)^{m}(i\hbar^{1/2}\rho)^{m+2r-k}}{\sqrt{m!(m+2r-k)!}}$~. 

\noindent To simplify the calculation, we denote $w=-\dfrac{r^2}{\hbar \rho^2}$ and we have 
\begin{eqnarray}
_{H}\langle n,m-n|x^k|n,m'-n\rangle_{H}&=&\sum_{r=0}^{k}\frac{\pi}{2^k}e^{i(\Theta_{n,m-n+2r-k}-\Theta_{n,m-n})}(-1)^{k+r}i^{-k}~{^{k}C_r}\delta_{m',m+2r-k}\nonumber\\
&\times& \lambda_n^2(\hbar^{1/2} \rho)^{2n+k+2}\sqrt{m!(m+2r-k)!}\nonumber\\
&\times& \int_0^\infty dw w^{n-m-r+k}e^{-w} L_m^{(n-m)}(w) L_{m+2r-k}^{(n-m-2r+k)}(w)
\label{eqn68}
\end{eqnarray}
where we have also substituted the following expression of the hypergeometric function \cite{Arfken, uva} with regard to the associated Laguerre polynomials $L_n^{(\zeta)}(w)$, 
\begin{eqnarray}
L_n^{(\zeta)}(w)=\frac{(-1)^n}{n!}U(-n,\zeta+1,w) \label{eqn69}
\label{eqn70}
\end{eqnarray}

\noindent Now, using the Lewis phase expression in \cite{Dey},
\begin{eqnarray}
\Theta_{n,l}=(n+l)\int^t \left[c(\tau)-\frac{a(\tau)}{\rho^2 (\tau)}\right]d\tau
\label{eqn73}
\end{eqnarray} 
we deduce the following,
\begin{eqnarray}
e^{i(\Theta_{n,m-n+2r-k}-\Theta_{n,m-n})}&=&e^{i[\{(n+m-n+2r-k)-(n+m-n)\}\int^t (c(\tau)-\frac{a(\tau)}{\rho^2 (\tau)})d\tau]} \nonumber\\
&=& e^{i(0+2r-k)\int^t (c(\tau)-\frac{a(\tau)}{\rho^2 (\tau)})d\tau}\nonumber\\
&=&e^{i\Theta_{0,2r-k}}.
\label{eqn74}
\end{eqnarray}
Finally, a generic form of the matrix element of $x^k$ is deduced as,
\begin{eqnarray}
_{H}\langle n,m-n|x^k|n,m'-n\rangle_{H}&=&\sum_{r=0}^{k}\frac{\pi}{2^k}e^{i\Theta_{0,2r-k}}(-1)^{k+r}i^{-k}~{^{k}C_r}\delta_{m',m+2r-k}\nonumber\\
&\times& \lambda_n^2(\hbar^{1/2} \rho)^{2n+k+2}\sqrt{m!(m+2r-k)!}\nonumber\\
&\times& \int_0^\infty dw~w^{n-m-r+k}e^{-w} L_m^{(n-m)}(w) L_{m+2r-k}^{(n-m-2r+k)}(w).
\label{eqn75}
\end{eqnarray}
This represents a new finding in the present study and both the matrix element and the expectation value of that coordinate operator with any arbitrary power can be derived by utilizing the above result. Additionally, the above matrix element can also be expressed with respect to the invariant's eigenstate, where the Lewis phase factor would not be present. It reads,
\begin{eqnarray}
\langle n,m-n|x^k|n,m'-n\rangle&=&\sum_{r=0}^{k}\frac{\pi}{2^k}(-1)^{k+r}i^{-k}~{^{k}C_r}\delta_{m',m+2r-k}\nonumber\\
&\times& \lambda_n^2(\hbar^{1/2} \rho)^{2n+k+2}\sqrt{m!(m+2r-k)!}\nonumber\\
&\times& \int_0^\infty dw~w^{n-m-r+k}e^{-w} L_m^{(n-m)}(w) L_{m+2r-k}^{(n-m-2r+k)}(w).
\label{eqn75I}
\end{eqnarray}
\noindent Now, using our new result derived in Eqn.(\ref{eqn75}), we proceed to obtain the matrix element 
$_{H}\langle n,m-n|x|n,m'-n\rangle_{H}$ at first. It reads,
\begin{eqnarray}
_{H}\langle n,m-n|x|n,m'-n\rangle_{H}=
_{H}\langle n,m-n|x^k\mid_{\,k=1;\,r=0}|n,m'-n\rangle_{H}\nonumber\\
+ _{H}\langle n,m-n|x^k\mid_{\,k=1;\,r=1}|n,m'-n\rangle_{H}.
\label{eqn76}
\end{eqnarray}
The above matrix elements provide,
\begin{eqnarray}
_{H}\langle n,m-n|x^k\mid_{\,k=1;\,r=0}|n,m'-n\rangle_{H}= -\frac{i}{2}
{(\rho\hbar^{1/2})}\sqrt{m}e^{-i\Theta_0,1}\delta_{m,m'+1}
\label{eqn77}
\end{eqnarray}
\begin{eqnarray}
_{H}\langle n,m-n|x^k\mid_{\,k=1;\,r=1}|n,m'-n\rangle_{H}=\frac{i}{2}
{(\rho\hbar^{1/2})}\sqrt{m'}e^{i\Theta_0,1}\delta_{m',m+1}.
\label{eqn78}
\end{eqnarray}
In order to obtain Eqn(s).(\ref{eqn77}, \ref{eqn78}), we require some useful relations related to the the associated Laguerre polynomials,
\begin{eqnarray}
L_n^{(\zeta)}(w)=L_n^{(\zeta+1)}(w)-L_{n-1}^{(\zeta+1)}(w) \nonumber\\
\int_0^\infty dw~w^{\zeta} e^{-w}L_n^{(\zeta)} (w)L_m^\zeta (w) =\frac{(n+\zeta)!}{n!}\delta_{n,m}~.
\label{eqn80}
\end{eqnarray}
Combining Eqn(s).(\ref{eqn77}, \ref{eqn78}), we have
\begin{eqnarray}
_{H}\langle n,m-n|x|n,m'-n\rangle_{H}~=~\frac{i}{2}
{(\rho\hbar^{1/2})}[\sqrt{m'}e^{i\Theta_{0,1}}\delta_{m',m+1}
-\sqrt{m}e^{-i\Theta_{0,1}}\delta_{m,m'+1}].
\label{eqn79}
\end{eqnarray}
In a similar way, we can find out the expectation value of $x^2$. Therefore,
\begin{eqnarray}
_{H}\langle n,m-n|x^2|n,m'-n\rangle_{H}= _{H}\langle n,m-n|x^k\mid_{\,k=2;\,r=0}|n,m'-n\rangle_{H}\nonumber\\
+_{H}\langle n,m-n|x^k\mid_{\,k=2;\,r=1}|n,m'-n\rangle_{H}+ _{H}\langle n,m-n|x^k\mid_{\,k=2;\,r=2}|n,m'-n\rangle_{H}. 
\label{eqn81}
\end{eqnarray}
The above matrix elements provide, 
\begin{eqnarray}
_{H}\langle n,m-n|x^k\mid_{\,k=2;\,r=0}|n,m'-n\rangle_{H}&=& 
-\frac{1}{4}{(\hbar\rho^2)}e^{-i\Theta_{0,2}}\delta_{m',m-2}\sqrt{m(m-1)}\nonumber\\
_{H}\langle n,m-n|x^k\mid_{\,k=2;\,r=1}|n,m'-n\rangle_{H}&=&~\frac{1}{2}{(\hbar\rho^2)}e^{-i\Theta_{0,0}}\delta_{m,m'}(m+n+1)\nonumber\\
_{H}\langle n,m-n|x^k\mid_{\,k=2;\,r=2}|n,m'-n\rangle_{H}&=& 
-\frac{1}{4}{(\hbar\rho^2)}e^{i\Theta_{0,2}}\delta_{m',m+2}\sqrt{(m+2)(m+1)}.
\label{eqn82}
\end{eqnarray}
For computing the above results, apart from Eqn.(\ref{eqn80}), the following relation is also required.
\begin{eqnarray}
\int_{0}^{\infty} dw~w^{k+p}e^{-w}L_n^k(w)L_n^k(w)~=~\frac{(n+k)!}{n!}\times(2n+k+1)^p ~.
\label{eqn83}
\end{eqnarray}
So we have,
\begin{eqnarray}
_{H}\langle n,m-n|x^2|n,m'-n\rangle_{H}&=&\frac{(\hbar\rho^2)}{2}\delta_{m,m'}(m+n+1)\nonumber\\
&&-\frac{(\hbar\rho^2)}{4}\left[e^{-i\Theta_{0,2}}\delta_{m',m-2}\sqrt{m(m-1)}\right.\nonumber\\
&& \left. +e^{i\Theta_{0,2}}\delta_{m',m+2}\sqrt{(m+2)(m+1)}\right].
\label{eqn84}
\end{eqnarray}
It must be mentioned that the the matrix elements derived in Eqn(s).(\ref{eqn79}, \ref{eqn84}) are exactly similar to those found in \cite{Dey}. Although their results are provided in the invariant's eigenstate, our results are obtained in the Hamiltonian eigenstate.
\noindent Similarly, the matrix element of $y^k$ are deduced in the Hamiltonian eigenstate and it is found to be,
\begin{eqnarray}
_{H}\langle n,m-n|y^k|n,m'-n\rangle_{H}&=&\sum_{r=0}^{k}\frac{\pi}{2^k}e^{i\Theta_{0,2r-k}}~{^{k}C_r}\delta_{m',m+2r-k}\nonumber\\
&\times& \lambda_n^2(\hbar^{1/2} \rho)^{2n+k+2}\sqrt{m!(m+2r-k)!}\nonumber\\
&\times& \int_0^\infty dw~w^{n-m-r+k}e^{-w} L_m^{(n-m)}(w) L_{m+2r-k}^{(n-m-2r+k)}(w).
\label{eqn85}
\end{eqnarray}
In the invariant's eigenstate, the above result is expressed as,
\begin{eqnarray}
\langle n,m-n|y^k|n,m'-n\rangle&=&\sum_{r=0}^{k}\frac{\pi}{2^k}~{^{k}C_r}\delta_{m',m+2r-k}\nonumber\\
&\times& \lambda_n^2(\hbar^{1/2} \rho)^{2n+k+2}\sqrt{m!(m+2r-k)!}\nonumber\\
&\times& \int_0^\infty dw~w^{n-m-r+k}e^{-w} L_m^{(n-m)}(w) L_{m+2r-k}^{(n-m-2r+k)}(w).
\label{eqn85I}
\end{eqnarray}
Using Eqn.(\ref{eqn85}), we find out the matrix elements of $y$ and $y^2$ in the Hamiltonian eigenstate. Those are as follows,
\begin{eqnarray}
_{H}\langle n,m-n|y|n,m'-n\rangle_{H}&=&
_{H}\langle n,m-n|y^k\mid_{\,k=1;\,r=0}|n,m'-n\rangle_{H}\nonumber\\&&+ _{H}\langle n,m-n|y^k\mid_{\,k=1;\,r=1}|n,m'-n\rangle_{H}
\nonumber\\
&=&-\frac{1}{2}
{(\rho\hbar^{1/2})}[\sqrt{m}e^{-i\Theta_{0,1}}\delta_{m',m-1}
+\sqrt{m+1}e^{i\Theta_{0,1}}\delta_{m',m+1}].
\label{eqn86}
\end{eqnarray}
\begin{eqnarray}
_{H}\langle n,m-n|y^2|n,m'-n\rangle_{H}&=&_{H}\langle n,m-n|y^k\mid_{\,k=2;\,r=0}|n,m'-n\rangle_{H}\nonumber\\
&&+_{H}\langle n,m-n|y^k\mid_{\,k=2;\,r=1}|n,m'-n\rangle_{H}\nonumber\\
&&+_{H}\langle n,m-n|y^k\mid_{\,k=2;\,r=2}|n,m'-n\rangle_{H} \nonumber\\
&=&\frac{\hbar\rho^2}{4}\delta_{m',m-2}\sqrt{m(m-1)}e^{-i\Theta_{0,2}}\nonumber\\
&&+\frac{1}{2}\delta_{m,m'}{(\hbar\rho^2)}(m+n+1)+\frac{\hbar\rho^2}{4}\delta_{m',m+2}\sqrt{(m+2)(m+1)}e^{i\Theta_{0,2}}.\nonumber\\
\label{eqn87}
\end{eqnarray}
We observe once more that the structural form of the matrix element of the coordinate operator $y^k$ in the Hamiltonian eigenstate aligns with the results presented in \cite{Dey} for $k=1, 2$, However, they reported these results in the invariant's eigenstate.\\
\noindent While the results above have included $\hbar$, we use natural units ($\hbar = 1$) when applying those results to calculate the energy expectation value, as we have been working in natural units from the beginning.


\subsection{Expectation value of momentum operator raised to second power}
Here, we outline (in natural unit $\hbar\,=\,1$ )the method for determining $\langle p_i^2 \rangle$, the expectation value of the square form of the momentum operator. The procedure is significant as it yields a time dependent value related to the momentum operator and it also provides insight while obtaining the expectation value of other physical quantities, such as $\langle p_i \rangle = 0$, $\braket{x_i\,p_j}=\dfrac{1}{2}\,\epsilon_{ij}(m-n)$, where $i,j=1,2$ and $\epsilon_{ij}=-\epsilon_{ji}$ with $\epsilon_{12}=1$.  
\begin{eqnarray} 
_{H}\langle n,m-n|\,p_1^2\,|n,m-n\rangle_{H}&=&\int r dr d\theta
~_{H}\langle n,m-n |r,\theta\rangle\langle r,\theta|\left(-i\,cos\,\theta\,\partial_r+\dfrac{i}{r}sin\theta\,\partial_{\theta}\right)^2|n,m-n\rangle_{H}\nonumber\\
&=&\int r dr d\theta
~\langle n,m-n |r,\theta\rangle\langle r,\theta|\left(-i\,cos\,\theta\,\partial_r+\dfrac{i}{r}sin\theta\,\partial_{\theta}\right)^2|n,m-n\rangle\nonumber\\
&=&\int
~\phi_{n,m-n}^{*}(r,\theta,t)\left(-i\,cos\,\theta\,\partial_r+\dfrac{i}{r}sin\theta\,\partial_{\theta}\right)^2\phi_{n,m-n}(r,\theta,t)\, r dr d\theta~;
\nonumber\\
&=&\,I_1+ I_2+ I_3+ I_4+ I_5 ~.\label{DHO-psq1}
\end{eqnarray} 
We now calculate the terms $I_1, I_2, I_3, I_4$ and $I_5 $ by Using the structure of $\phi_{n,m-n}(r,\theta,t)$ in Eqn.(\ref{DHO-inv-efunc2}). 
\begin{eqnarray} 
_{H}\langle n,m-n|\,p_1^2\,|n,m-n\rangle_{H}=\,I_1+ I_2+ I_3+ I_4+ I_5 ;\label{DHO-psq2}
\end{eqnarray} 
where 
\begin{align}
I_1=&-\int\,\phi_{n,m-n}^{*}(r,\theta,t)\,cos^2\theta\,\partial_r^2\,\left[\phi_{n,m-n}(r,\theta, t)\right]\,r\,dr\,d\theta \nonumber
\\
&=-Q^{*}_{n,m-n}(t)\,Q_{n,m-n}(t)\,\int_0^\infty\,R_{n,m-n}^{*}(r,t)\,\left[\partial_r^2\,R_{n,m-n}(r,t)\right]\,r\,dr\,\nonumber\\
&\times\int_0^{2\Pi}\,\Phi_{n,m-n}^{*}(\theta,t)\,cos^2\theta\,\Phi_{n,m-n}(\theta,t)\,d\theta~,\label{DHO-I1}
\end{align}
\begin{align}
I_2&=-2\int\,\phi_{n,m-n}^{*}(r,\theta, t)\,\left(\dfrac{sin\,\theta\,cos\theta}{r^2}\right)\,\partial_\theta\,\left[\phi_{n,m-n}(r,\theta, t)\right]\,r\,dr\,d\theta 
\nonumber
\\
&=-2\,Q^{*}_{n,m-n}(t)\,Q_{n,m-n}(t)\,\int_0^\infty\,\dfrac{R_{n,m-n}^{*}(r, t)\,\,R_{n,m-n}(r, t)}{r}\,dr\,\nonumber\\
&\times\int_0^{2\Pi}\,\Phi_{n,m-n}^{*}(\theta, t)\,cos\theta\,sin\theta\,\partial_{\theta}\left[\Phi_{n,m-n}(\theta, t)\right]\,d\theta~,\label{DHO-I2}
\end{align}
\begin{align}
I_3&=2\int\,\phi_{n,m-n}^{*}(r,\theta, t)\,\left(\dfrac{sin\,\theta\,cos\theta}{r}\right)\,\partial_r\partial_\theta\,\left[\phi_{n,m-n}(r,\theta, t)\right]\,r\,dr\,d\theta 
\nonumber
\\
&=2\,Q^{*}_{n,m-n}(t)\,Q_{n,m-n}(t)\,\int_0^\infty\,R_{n,m-n}^{*}(r, t)\,\partial_r\,\left[R_{n,m-n}(r, t)\,\right]dr\,\nonumber\\
&\times\int_0^{2\Pi}\,\Phi_{n,m-n}^{*}(\theta, t)\,cos\theta\,sin\theta\,\partial_{\theta}\left[\Phi_{n,m-n}(\theta, t)\right]\,d\theta~,\label{DHO-I3}
\end{align}
\noindent
\begin{align}
I_4&=-\int\,\phi_{n,m-n}^{*}(r,\theta, t)\,\left(\dfrac{sin^2\theta}{r}\right)\,\partial_r\,\left[\phi_{n,m-n}(r,\theta, t)\right]\,r\,dr\,d\theta\nonumber\\
&=-Q^{*}_{n,m-n}(t)\,Q_{n,m-n}(t)\,\int_0^\infty\,R_{n,m-n}^{*}(r, t)\,\left[\partial_r\,R_{n,m-n}(r, t)\right]\,dr\,\nonumber\\
&\times\int_0^{2\Pi}\,\Phi_{n,m-n}^{*}(\theta, t)\,sin^2\theta\,\Phi_{n,m-n}(\theta, t)\,d\theta~, \label{DHO-I4}
\end{align}
and
\begin{align}
I_5&=-\int\,\phi_{n,m-n}^{*}(r,\theta, t)\,\left(\dfrac{sin^2\theta}{r^2}\right)\,\partial_\theta^2\,\left[\phi_{n,m-n}(r,\theta, t)\right]\,r\,dr\,d\theta\nonumber\\
&=-Q^{*}_{n,m-n}(t)\,Q_{n,m-n}(t)\,\int_0^\infty\,\dfrac{R_{n,m-n}^{*}(r, t)\,R_{n,m-n}(r, t)}{r}\,dr\,\nonumber\\
&\times\int_0^{2\Pi}\,\Phi_{n,m-n}^{*}(\theta, t)\,sin^2\theta\,\partial_\theta^2\,\left[\Phi_{n,m-n}(\theta, t)\right]\,d\theta~.\label{DHO-I5}
\end{align}
\noindent It is evident from the azimuthal components of Eqn(s).(\ref{DHO-I1}, \ref{DHO-I2}, \ref{DHO-I3}, \ref{DHO-I4}, \ref{DHO-I5}) that the terms $I_2$ and $I_3$ would provide a zero contribution. It can also be verified as,
\begin{align}
&\int_0^{2\Pi}\,\Phi_{n,m-n}^{*}(\theta, t)\,cos\theta\,sin\theta\,\partial_{\theta}\left[\Phi_{n,m-n}(\theta, t)\right]\,d\theta\nonumber\\
&=\,i\,(m-n)\,\int_0^{2\Pi}\,\Phi_{n,m-n}^{*}(\theta, t)\,cos\theta\,sin\theta\,\Phi_{n,m-n}(\theta, t)\,d\theta\nonumber\\
&=0~.\label{DHO-psq00}
\end{align}
Hence, we proceed with $I_2\,=\,0\,=I_3$.
Next, from Eqs.(\ref{DHO-inv-efunc-R}, \ref{DHO-inv-efunc-th}), we substitute the $R$, $\Phi$, and $Q$ into the remaining variables $I_1$, $I_4$, and $I_5$. Then the variable $z$ is invented and we made a parametrization $z\,=\,\dfrac{r^2}{\rho^2}$. To make the calculation simpler, we employ several useful relations related to the associated Laguerre polynomials \cite{Arfken}. First, the derivative of this polynomials is given by
\begin{align}
\partial_z\left[L^{n-m}_m\,\left(z\right)\right]&=
-\,L^{n-m+1}_{m-1}\,\left(z\right)~;~(m\,\geq\,1)\nonumber\\
&=\,0~~~~~~~~~~~~~~~~~~~;(m=0)~.\nonumber\\
\end{align}
Apart from the above relation, we also use the identity relation and the orthonormality condition for this polynomial. Those are given by 
\begin{align}
L^{n-m}_{m}(z)\,=\,L^{n-m+1}_{m}(z)-L^{n-m+1}_{m-1}(z)\nonumber\\
\int_0^\infty\,z^{n-m}\,e^{-z}\,L^{n-m}_{n}(z)\,L^{n-m}_{m}(z)
\,=\,\dfrac{\Gamma\,(2n-m+1)}{n!}\,\delta_{mn}~.
\label{DHO-psqoi}
\end{align}
We can generate the expressions of $I_1, I_4$, and $I_5$ by using the above relations (Eqn.[\ref{DHO-psqoi}]). Those are given by,
\begin{align}
I_1&=-\dfrac{m!\,(n-m)(n-m-1)}{2\,n!\,\rho^2}\,\int^{\infty}_0\,z^{n-m-1}\,e^{-z}\,L^{n-m}_{m}(z)L^{n-m}_{m}(z)\,dz\,\nonumber\\
&+\,(2n+2m+1)\,\dfrac{f}{2\rho}-\dfrac{f^2}{2}(m+n+1)+(2n-2m+1)\dfrac{m!}{\rho^2\,n!}\,\int_0^{\infty}z^{n-m}\,e^{-z}\,L^{n-m}_{m}(z)L^{n-m+1}_{m-1}(z)\,dz\nonumber\\
&-\dfrac{2\,m!}{n!\rho^2}\,\int_0^{\infty}\,z^{n-m+1}\,e^{-z}\,L^{n-m}_{m}(z)L^{n-m+2}_{m-2}(z)\,dz, 
\end{align}

\begin{align}
I_4=-\dfrac{m!\,(n-m)}{2\,n!\,\rho^2}\,\int^{\infty}_0\,z^{n-m-1}\,e^{-z}\,L^{n-m}_{m}(z)L^{n-m}_{m}(z)\,dz\,+\,\dfrac{f}{2\rho}
\nonumber\\
+\dfrac{m!}{\rho^2\,n!}\,\int_0^{\infty}z^{n-m}\,e^{-z}\,L^{n-m}_{m}(z)L^{n-m+1}_{m-1}(z)\,dz~,
\end{align}
and
\begin{align}
I_5=\dfrac{m!\,(m-n)^2}{2\,n!\,\rho^2}\,\int^{\infty}_0\,z^{n-m-1}\,e^{-z}\,L^{n-m}_{m}(z)L^{n-m}_{m}(z)\,dz~:
\end{align}
where $f=\dfrac{a-i\rho\dot{\rho}}{a\rho}$. From Eqn.(\ref{DHO-psq1}), we have

\begin{align}
_{H}\langle n,m-n|\,p_1^2\,|n,m-n\rangle_{H}&=\,\dfrac{2(n-m+1)\,m!}{n!\,\rho^2}\,\int_0^\infty\,z^{n-m}\,e^{-z}\,L^{n-m}_m(z)\,L^{n-m+1}_{m-1}(z)\,dz\nonumber\\
&-\,\dfrac{2\,m!}{n!\,\rho^2}\,\int_0^\infty\,z^{n-m+1}\,e^{-z}\,L^{n-m}_m(z)\,L^{n-m+2}_{m-2}(z)\,dz\nonumber\\
&+\dfrac{(n+m+1)}{2}\,\left[\dfrac{1}{\rho^2}+\dfrac{\dot{\rho}^2}{a^2} \right]~.\label{DHO-psqincom}
\end{align}

Now a special identity (deduced in the Appendix) is to be utilized to find out the final result. The identity is given by,
\begin{equation}
(n-m+1)\,\,\int_0^\infty\,z^{n-m}\,e^{-z}\,L^{n-m}_m(z)\,L^{n-m+1}_{m-1}(z)\,dz=\,\int_0^\infty\,z^{n-m+1}\,e^{-z}\,L^{n-m}_m(z)\,L^{n-m+2}_{m-2}(z)\,dz~.
\end{equation}
Now, Eqn.(\ref{DHO-psqincom}) yields
\begin{align}
_{H}\langle n,m-n|\,p_1^2\,|n,m-n\rangle_{H}\,=\,\dfrac{(n+m+1)}{2}\,\left[\dfrac{1}{\rho^2}+\dfrac{\dot{\rho}^2}{a^2} \right]~.\label{psq3}
\end{align}
In the similar way, it can also be obtained that
\begin{align}
_{H}\langle n,m-n|\,p_2^2\,|n,m-n\rangle_{H}\,=\,\dfrac{(n+m+1)}{2}\,\left[\dfrac{1}{\rho^2}+\dfrac{\dot{\rho}^2}{a^2} \right]
\end{align}
We should remember that, although the results are presented in \cite{Dey}, the process used to obtain them is not included. By following the detailed procedure outlined above, verifying the following expectation value will become straightforward.
\begin{eqnarray}
\braket{x_j\,p_k}=\dfrac{1}{2}\,\epsilon_{jk}(m-n)\,\,;
\end{eqnarray}
where $j,k=1,2$ and $\epsilon_{jk}=-\epsilon_{kj}$ with $\epsilon_{12}=1$.\\
Now that we are convinced we have all the necessary expectation values to calculate the energy expectation value (Eqn.(\ref{DHO-enrg1})), we will proceed directly to achieve our goal.

\subsection{Energetics of the model system}
We begin by mentioning the expectation value of the canonical operators which are needed to compute the energy expectation value Eqn.(\ref{DHO-enrg1}). Those are given by the following
relations, 
\begin{eqnarray}
\braket{x_j^2}=\dfrac{\rho^2}{2}(n+m+1)\,\,;\,\,\braket{p_j^2}=\dfrac{1}{2}\left(\dfrac{1}{\rho^2}+\dfrac{\dot{\rho}^2}{a^2} \right)\,(n+m+1)\,\,;\,\,\braket{x_j\,p_k}=\dfrac{1}{2}\,\epsilon_{jk}(m-n)\,\,;\label{eqn88}
\end{eqnarray}
where $j,k=1,2$ and $\epsilon_{jk}=-\epsilon_{kj}$ with $\epsilon_{12}=1$. So, the energy expectation value $\braket{E_{n,m-n}(t)}$ with regard to $\psi_{n,m-n}(r,\theta,t)$, the Hamiltonian eigenstate, is found to be,
\begin{align}
&\braket{E_{n,m-n}(t)}=\dfrac{1}{2}\,(n+m+1)\left[b(t)\rho^2(t)+\dfrac{a(t)}{\rho^2(t)}+\dfrac{\dot{\rho}^2(t)}{a(t)} \right]+c(t)\,(n-m)\,\,.\nonumber\\
&=\dfrac{1}{2}\left[\,(n+m+1)\left(b(t)\rho^2(t)+\dfrac{a(t)}{\rho^2(t)}+\dfrac{\dot{\rho}^2(t)}{a(t)} \right)+(n-m)\left(\dfrac{f(t)\Omega(t)}{M}+\dfrac{M\omega^2(t)\theta(t)}{f(t)}\right) \right].
\label{eqn89}
\end{align}



\noindent It is noteworthy that even as the oscillation frequency $\omega$ approaches zero, the expectation value of energy remains non zero. This occurs by reason of that $a(t)$, $b(t)$, and $c(t)$, the Hamiltonian coefficients remain finite as $\omega$ approaches zero, as evident from Eqs. (\ref{DHO-a}), (\ref{DHO-b}), and (\ref{DHO-c}). \\
\noindent We are about to explore another interesting part in this work. We will not only analytically examine the explicit form of  $\braket{E_{n,m-n}(t)}$, under various damping conditions, but also graphically visualize the time dependent behaviour of the energy persisting in the system.



\subsubsection{Energy profile for the exponential solution}
The general form of the energy expectation value, for the exponentially varying EP solution set given by Eqn.(\ref{EP-exp}), are deduced to be,
\begin{equation}
\braket{E_{n,m-n}(t)}=(n+m+1)\mu^2\Delta+c(t)\,(n-m)
\label{eqn90}
\end{equation}
where we have used the constraint relation given by Eqn.(\ref{EP-exp-cons}).\\
\noindent It is remarkable that in the general form of the energy expression mentioned above, time dependence is solely governed by the factor $c(t)$ which (Eqn.(\ref{DHO-c})) disappears in the absence of NC parameters. Thus, the observed time dependence in the energy values obtained for the exponentially varying EP solution set can be attributed exclusively to the time dependent nature of the NC space.

\vskip 0.2cm



\noindent{\bf $\langle A\rangle$ Solution Set-Ia}
\vskip 0.15cm

\noindent In this scenario of dissipation, we have chosen $f(t)=1$ and $\omega(t)=\omega_0\,e^{-\Gamma\,t/2}$. The energy expectation value in the ground state is computed to be
\begin{eqnarray}
\braket{E_{n,-n}(t)}&=&(n+1)\mu^2\Delta+\,n\,\left[\sqrt{\dfrac{ \Delta\,exp\,(\Gamma{t})-M{\omega_0}^2\,exp\,(-\Gamma{t}) }{M}} 
\,\right.\nonumber\\
&&~~~~~~~~~~~~~~~~~~~~~~~~~~\left. +\,\omega_0\,exp\,(-\Gamma{t/2})\sqrt{\,M\sigma\,exp\,(-\Gamma{t})-1 \,}\right].
\label{DHO-eqn91}
\end{eqnarray}
From Eqn.(\ref{DHO-eqn91}), it is evident that the energy expression turns complex after a specific value of time. The criterion for ensuring the energy expression remains real is given by:
\begin{eqnarray}
M\,\sigma\,e^{-\Gamma\,t}\geq 1\,\,\Rightarrow\,t\leq\,\dfrac{ln(M\,\sigma)}{\Gamma}\label{DHO-eqn92}
\end{eqnarray}
\begin{eqnarray}
e^{2\Gamma t}\geq \frac{M\omega_{0}^2}{\Delta}\,\,\Rightarrow\,
t\geq\,\frac{1}{2\Gamma}\ln{\left(\frac{M\omega_{0}^2}{\Delta}\right)}~.\label{DHO-eqn92co}
\end{eqnarray}
Thus, we have the following time range within which the energy remains physical.
\begin{eqnarray}
\frac{1}{2\Gamma}\ln{\left(\frac{M\omega_{0}^2}{\Delta}\right)}\leq t \leq \dfrac{ln(M\,\sigma)}{\Gamma}~.\label{DHO-eqn92coa}
\end{eqnarray}

\noindent It is relevant to mention that the above restriction in time axis aligns with the hermiticity of the Hamiltonian operator (Eqn.\ref{DHO-comham}). For $t>ln(M\,\sigma)/\Gamma$, as indicated by Eqn.(\ref{eqn35b}), $\theta(t)$ becomes complex, making the time dependent coefficient $c(t)$ in Eqn.(\ref{eqn38}) complex as well. Similarly, for $t<\frac{1}{2\Gamma}\ln{\left(\frac{M\omega_{0}^2}{\Delta}\right)}$, Eqn.(\ref{eqn36b}) shows that $\Omega(t)$ turns out to be complex, leading the Hamiltonian parameter $c(t)$ in Eqn.(\ref{eqn38}) to be complex again. Consequently, the Hamiltonian operator (Eqn.\ref{DHO-comham}) is no longer hermitian outside this time range. Therefore, the restriction found in time axis, for achieving a physical energy expression, is consistent with the property of hermiticity of the Hamiltonian, as it remains hermitian only within that specific time range.
\begin{figure}[t]
\centering
\includegraphics[scale=0.4]{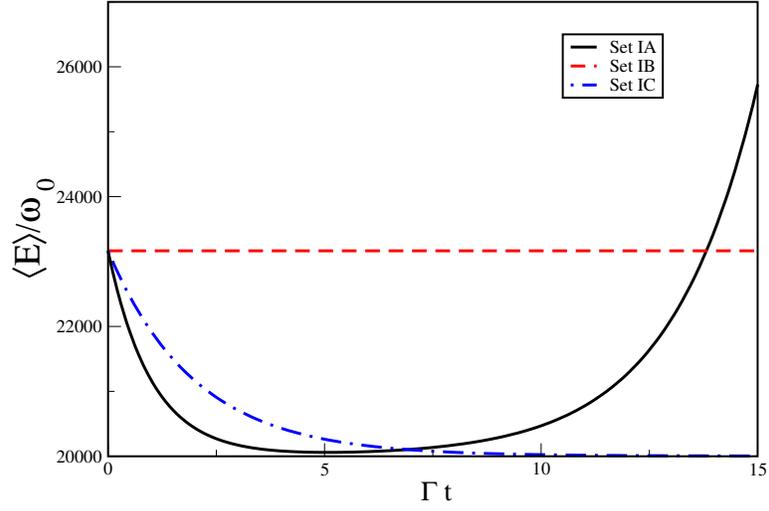}
\caption{\textit{Graph of the exponential energy expectation value (divided by $\omega_0$ to eliminate the dimension), with respect to $\Gamma$t (also a dimensionless parameter). The numerical value (in natural units) of the constants are chosen as $M,\,\Gamma,\,\mu\,=\,1$;~$\sigma,\,\Delta$=$10^7$;~$\omega_0$=$10^3$ and the quantum state is fixed as $m=0, n=1$. The energy expectation value $\langle E \rangle$ is computed when $\langle A\rangle$ Set-IA : $\omega(t)=\omega_0 e^{-{\Gamma}t/2}$ and $f(t)=1$ ; $\langle B\rangle$ Set-IB : $\omega(t)=\omega_0$ and $f(t)=e^{-{\Gamma}t}$ ; $\langle C\rangle$ 
Set-IC : $\omega(t)=\omega_0 e^{-{\Gamma}t/2}$ and $f(t)=e^{-{\Gamma}t}$. For $\langle A\rangle$, the energy initially decays and then starts to grow with respect to time. For $\langle B\rangle$, the energy is always constant over time. For $\langle C\rangle$, the energy decreases and tends to zero over time.}}  
\label{DHO-fig1}
\end{figure}


\noindent It is evident from Fig.(\ref{DHO-fig1}), that the energy exhibits an initial dissipation. However, after a certain time, it starts to grow over time. It happens in view of the fact that, at large time, $exp\,(-\Gamma{t/2})\approx\,0 $, resulting in the following approximated expression of 
energy,
\begin{equation}
E_{n,-n}(t)\approx\,(n+1)\mu^2\Delta+\,n\,\sqrt{\dfrac{ \Delta\,exp\,(\Gamma{t}) }{M}}\,\,
\label{e7}
\end{equation}
The increasing nature of the energy is reflection of the exponentially increasing Hamiltonian parameter $b(t)$ as it becomes the dominant factor in the energy expression at large time. As shown in the exponential EP solution (Eqn.(\ref{EP-exp})), while $a(t)$ decays with time, $b(t)$ grows at the same rate to satisfy the EP equation. 
\vskip 0.1cm



\noindent{{\bf{$\langle B\rangle$ Solution Set-Ib}}}

\noindent In this scenario of dissipation, we have chosen $f(t)=e^{-\Gamma\,t}$ and $\omega(t)=\omega_0$. The energy expectation value in the ground state is computed to be,
\begin{equation}
\braket{E_{n,-n}(t)}=(n+1)\mu^2\Delta+\,n\,\left[\sqrt{\dfrac{ \Delta\,-M{\omega_0}^2\, }{M}} \,+\,\omega_0\sqrt{M\sigma-1}\right].\label{e8}
\end{equation}


\noindent As shown in Fig.(\ref{DHO-fig1}), the energy expectation value notably reduces to be a constant, which is consistent with Eqn.(\ref{e8}). This constancy is caused by the balancing effect among the various decaying and increasing time dependent factors $f(t)$, $a(t)$, $b(t)$ those are varying at a same rate.   

\vskip 0.1cm
\noindent{{\bf{$\langle C\rangle$ Solution Set-Ic}}}

\noindent In this scenario of dissipation, we have chosen $f(t)=e^{-\Gamma\,t}$ and $\omega(t)=\omega_0\,e^{-\Gamma\,t/2}$. The energy expectation value in the ground state is computed to be,
\begin{equation}
\braket{E_{n,-n}(t)}=(n+1)\mu^2\Delta+\,n\,\left[\sqrt{\dfrac{ \Delta\,-M{\omega_0}^2\,exp\left[-\Gamma{t}\right] }{M}} 
 +  {\omega_0}\,exp\,(-\Gamma{t/2})\sqrt{M{\sigma}\,-1}\right].\label{e9}
\end{equation}
It is evident from Fig.(\ref{DHO-fig1}) that the energy dynamics exhibits a nice decaying nature, eventually stabilizing at a constant value at a very large time. This characteristic is also reflected in the plot depicting the variation of the expectation value of energy with time, as shown in Fig.(\ref{DHO-fig1}).

\vskip 0.1cm



\subsubsection{Energy profile for the rational solution}

The energy expectation value, for the rationally varying EP solution (at $k\,=\,2$) set given by Eqn.(\ref{EP-rat}), takes the following form,
\begin{eqnarray}
E_{n,-n}(t)&=&\dfrac{(n+1)}{2(\Gamma\,t+\chi)}\left[2\left(\dfrac{\sigma}{\mu^2}+\Delta\mu^2\right)+\dfrac{\mu^2\Gamma^2}{8\sigma} \right]\nonumber\\
&&~~~~~~~~~~~~~~~~~~~~~~~ +n\left[\dfrac{\omega_0}{\Gamma\,t+\chi}\,\sqrt{\dfrac{4\sigma\,M}{(\Gamma{t}+\chi)^2}\,-\,1}\,+\,\sqrt{\dfrac{\Delta}{M}-\dfrac{\omega_0^{\,2}}{(\Gamma\,t+\chi)^{\,2}}} \right].\label{e10}
\end{eqnarray}
where we have used the constraint relation given by Eqn.(\ref{EP-rat-cons}).

\noindent The energy dynamics exhibits a fine dissipation similar to the dissipated harmonic oscillator in the canonical phase space. However, by reason of the NC parameters, the system cannot remain physical always. The system remains physical within a certain interval of time, which is as follows
\begin{eqnarray}
\frac{1}{\Gamma}\left(\sqrt{\frac{M}{\Delta}}\omega_{0} -\chi\right)\leq t\,\leq\,\dfrac{1}{\Gamma}(2\sqrt{M\,\sigma}-\chi).
\label{e112}
\end{eqnarray}
The rationally varying energy remains physical within that particular interval of time for the similar reason as with the exponentially varying energy. The system's Hamiltonian (Eqn.\ref{DHO-comham}) is hermitian only during this period because, beyond it, $\Omega(t)$ and $\theta(t)$, the NC parameters turn imaginary. Consequently, $c(t)$, the Hamiltonian coefficient (Eqn.(\ref{DHO-c-rat})) becomes complex, causing the Hamiltonian operator to lose its hermitian property.

\vskip 0.1cm
\begin{figure}[t]
\centering
\includegraphics[scale=0.4]{ratdecay.eps}
\caption{\textit{Graph of the rational energy expectation value (divided by $\omega_0$ to eliminate the dimension), with respect to $\Gamma$t (also a dimensionless parameter). The numerical value (in natural units) of the constants are chosen as $M,\,\Gamma,\,\mu\,\chi\,=\,1$;~$\sigma,\,\Delta$=$10^7$;~$\omega_0$=$10^3$ and the quantum state is fixed as $n=1, m=0$. The expectation value of 
energy $\langle E \rangle$ is computed when $\omega(t)=\dfrac{\omega_0}{(\Gamma\,t+\chi)}$ and $f(t)=1$.}} 
\label{DHO-fig2}
\end{figure}

\noindent From Fig.(\ref{DHO-fig2}), indeed, we observe that the energy expectation value, $\langle E \rangle$, decreases over time according to a power law, consistent with the behaviour of rationally decaying solutions.


\subsubsection{Energy profile for the elementary solution}
The energy expectation value, for the elementary EP solution set given by Eqn.(\ref{EP-elm}), is deduced to be,
\begin{eqnarray}
\braket{E_{n,-n}(t)}&=&\dfrac{1}{2}(n+1)\left[\left(\Delta\mu^2+\dfrac{\sigma}{\mu^2}\right)\dfrac{1}{(\Gamma\,t+\chi)^2}+\dfrac{\mu^2\Gamma^2}{\sigma}\right]
\nonumber\\&&+n\left[\dfrac{\omega_0\sqrt{M\sigma-1}}{(\Gamma\,t+\chi)}+\dfrac{1}{(\Gamma\,t+\chi)}\sqrt{\dfrac{\Delta}{M\,(\Gamma\,t+\chi)^2}-\omega_0^2 } \right].\label{e12}
\end{eqnarray}
where we have used the constraint relation given by Eqn.(\ref{EP-elm-cons}).
Further, the constraint relation $\Delta\mu^4=\sigma$ yields the following energy expression,
\begin{align}
\braket{E_{n,-n}(t)}
&=\dfrac{1}{2}(n+1)\left[\dfrac{2\sigma}{\mu^2(\Gamma\,t+\chi)^2}+\dfrac{\mu^2\Gamma^2}{\sigma}\right]+n\,\left[\dfrac{\omega_0\sqrt{M\sigma-1}}{(\Gamma\,t+\chi)}+\dfrac{1}{(\Gamma\,t+\chi)}\sqrt{\dfrac{\Delta}{M\,(\Gamma\,t+\chi)^2}-\omega_0^2}\right].\label{e130}
\end{align}
Here also, we found a specific limit of time beyond which the Hamiltonian becomes non hermitian, resulting in a non physical system acquiring an imaginary expectation value of energy. That limit of time is found to be,
\begin{eqnarray}
\dfrac{\Delta}{M\,(\Gamma\,t+\chi)^2}\,\geq\,\omega_0^2\,\Rightarrow\,t\,\leq\,\dfrac{1}{\Gamma}\left[ \dfrac{1}{\omega_0}\sqrt{\dfrac{\Delta}{M}}-\chi \right].\label{e14}
\end{eqnarray}
As we discussed earlier, the above bound of time remains consistent with the hermiticity of the Hamiltonian (Eqn.\ref{DHO-comham}). Beyond this point in time, the Hamiltonian becomes non hermitian. This occurs because $\Omega(t)$ turns out to be complex, causing the Hamiltonian parameter $c(t)$ in Eqn.(\ref{DHO-c-elm}) to also become complex.
\begin{figure}[t]
\centering
\includegraphics[scale=0.4]{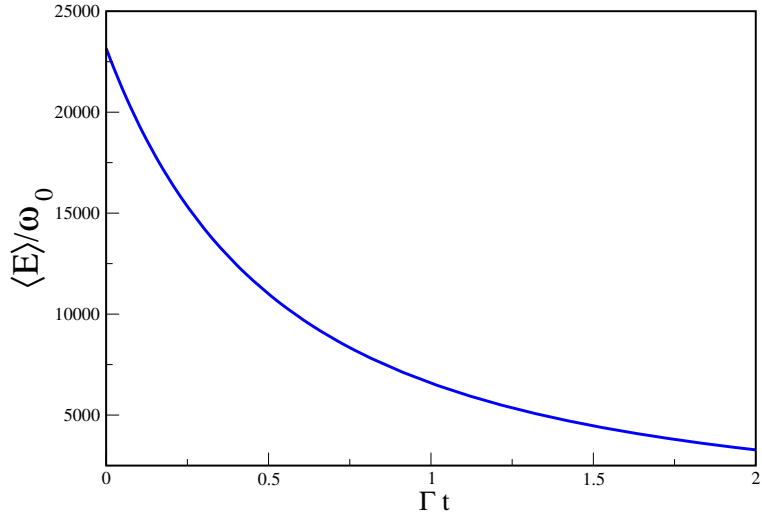}
\caption{\textit{Graph of the energy expectation value (divided by $\omega_0$ to eliminate the dimension) corresponding to the elementary EP solutions, with respect to $\Gamma$t (also a dimensionless parameter). The numerical value (in natural units) of the constants are chosen as $M,\,\Gamma,\,\mu\,\chi\,=\,1$;~$\sigma,\,\Delta$=$10^7$;~$\omega_0$=$10^3$ and the quantum state is fixed as $m=0, n=1$. The energy expectation value $\langle E \rangle$ is computed when $\omega(t)=\dfrac{\omega_0}{(\Gamma\,t+\chi)}$ and $f(t)=1$.}} 
\label{fig3}
\end{figure}
\noindent As per the Fig.(\ref{fig3}), the energy dynamics related to the elementary solution is again observed to follow a power law decay.




\section{Summary}
Here, the discussions in the present chapter is to be summarized. In this chapter, a two dimensional system of a damped harmonic oscillator with dynamic noncommutativity has been explored. The system is transformed into a canonical phase space by employing a variable shift known as the Bopp-shift, as described in existing literature. To solve the system using Lewis's method, we developed a time dependent Hermitian invariant operator. This approach resulted in a nonlinear differential equation, commonly referred to as the Ermakov-Pinney (EP) equation. We begin by obtaining the Lewis invariant in terms of the cartesian coordinate variables. Next, we transform to polar coordinates and express our results accordingly. Using the operator approach, we determine the eigenstates of the invariant and derive the Lewis phase factor in an integrated form. According to Lewis theory, the hermitian invariant and the Lewis phase factor together construct the Hamiltonian eigenfunction in an exact form. Since the eigenfunction of the system is inherently linked to the EP equation, and we aim to study the eigenfunction under various damping factors and angular frequencies, we adopt two exact analytical solution sets of the EP equation from the literature \cite{Dey}. We also provide a brief review of their derivation using the Chilleni integrability condition. The selected time dependent EP solutions from \cite{Dey} are exponential and rational in nature. Additionally, we propose a very simple method for deriving another set of EP solutions, which we refer to as elementary solutions. With these three different sets of solutions, we design various damped scenarios by selecting the angular frequency and the damping factor explicitly. For each damped model, we compute the explicit eigenfunction of the invariant and calculate the time dependent NC parameters, enabling the exact integration of the Lewis phase factor. Consequently, a class of exact eigenstates for the damping system are generated. Next, in both the Hamiltonian eigenstate and the invariant's eigenstate, we calculated the matrix elements of operators with a finite integer power. Additionally, we determined the expectation value of the momentum operator and the square form of that operator. Using these results, we computed the time dependent form of the energy expectation value for the system. With regard to the exponentially varying EP solutions, we identified three types of behaviour based on the damping factor and the oscillator frequency. When the damping factor is unity and the oscillator frequency decreases over time, the energy expectation value initially decreases before increasing. In this scenario, we also found a interval of time during which the energy dynamics behave physically. When the damping factor includes an exponentially decaying component and the oscillator frequency remains constant, the energy expectation value notably remains constant over time. When both the damping factor and the oscillator frequency decrease over time, the energy expectation value exhibits an exponential decay. With regard to the rationally decaying EP solution and elementary EP solutions, the energy dynamics are observed to follow the power law decay with a time interval within which the energy expressions remain real.






\chapter{Noncommutative Damped harmonic oscillator with magnetic field}\label{paper2}

The physicists have studied the Landau problem for a long time. The problem is regarding a dynamical charge particle affected by a magnetic field perpendicular to the particle's two dimensional plane.  
The discrete energy levels of such system are known as Landau levels which have pedagogical importance and various applications in literature. In this regard, we mention a very well known problem of a charged harmonic oscillator with a magnetic field perpendicular to the oscillator's two dimensional plane. The previous investigations into this problem are documented in works such as \cite{duval}-\cite{nair2}. Group theoretical techniques have been applied to studies in quantum Landau systems, notably in \cite{negro}, where it was significant to understand the Landau levels as local equivalence classes of the symmetry group of the configuration space. More recent studies on Landau levels are found in \cite{degh1}, \cite{degh2}. In \cite{degh1}, by using su(2) and su(1,1) Lie algebras, the authors constructed the displaced number coherent states to explore the statistical features of those. The exited coherent states for the system of a charged particle placed in a uniform magnetic field were introduced in \cite{degh2}. In a few recent communications \cite{murgan}-\cite{zhang}, the Landau problem was readdressed. Specially, in \cite{zhang}, the dynamics of the quantum states of a parametric oscillator was studied by employing the Lie transformation methodology. Additionally, we can introduce an electric field acting along the oscillator's plane. In \cite{lawson}, the charged harmonic oscillator having dynamic mass and frequency was considered. Under the influence of both magnetic field and dynamic electric field, the eigensystem of that oscillator was constructed. The problem would be more intriguing if the Landau problem is treated with the concept of noncommutativity. This scenario was previously examined in \cite{gouba}, where a dynamic magnetic field was introduced in such a NC space where the position operators hold noncommutativity.\\
\noindent In our work detailed in \cite{SG2}, which forms the subject of this chapter, we build upon this model by investigating the system with dynamic noncommutativity. Unlike \cite{gouba}, our study includes both spatial and momentum noncommutativity, and considers the NC space to be time dependent. Additionally, we introduce a damping factor to the oscillator to design a more practical scenario.
Though, several studies have examined the effects of magnetic fields and damping in physical systems \cite{satti}-\cite{nuruddeen2}, we aim to investigate such model system that incorporates both spatial and momentum noncommutativity. While the existing noncommutativity in momentum space in \cite{SG1} suggests that the system is already in a time varying magnetic field, perpendicular to the configuration plane, we still plan to introduce an external time dependent magnetic field. This is because the noncommutativity is assigned to be an inherent property of our system, and therefore, we cannot alter the field associated with it. To explore the interaction between damping and an external time dependent magnetic field, and its effect on the energy dynamics of a charged oscillator in a time dependent NC space, it is necessary to apply an external magnetic field to our previous model \cite{SG1}.\\
\noindent In our previous work \cite{SG1}, we observed that, due to the presence of damping, the energy dynamics of the oscillator shows a nice decaying nature even in a NC framework. In this study, our goal is to examine how the energy characteristics of that previous system change when the various forms of time dependent magnetic fields are applied externally on that system.

\section{Hamiltonian of the model system in NC space}
Here, we start with the two dimensional time dependent Hamiltonian describing a model system consisting of a damped harmonic oscillator influenced by a dynamic magnetic field applied externally in a dynamic NC background. \\
\noindent Thus, within the NC framework, the Hamiltonian of the present system reads,
\begin{equation}
H(t)=\dfrac{f(t)}{2M}\left[(P_1-q A_1)^2+(P_2-q A_2)^2\right]+\dfrac{M\omega^2(t)}{2f(t)}({X_1}^2+{X_2}^2)~~;\label{1}
\end{equation}
where $f(t)$ denotes the damping factor and it is given by,
\begin{equation}
f(t)=e^{-\int_{0}^t\eta(s)ds}~, 
\label{1x}
\end{equation}
where $\eta(t)$ represents the friction coefficient. In literature, this structure of the damping factor is very well known when the friction coefficient is set to be a constant and a two dimensional Hamiltonian (placed in a standard quantum mechanical phase space) consisting such a coefficient of friction with a constant angular frequency is popularly known as Caldirola and Kanai Hamiltonian \cite{caldi, kanai}. As we further explore this model in a NC framework, later, it will be observed that this kind of damping factor leads to a certain time intervals within which the system remains physical. \\
\noindent In order to introduce an external effect of a time varying magnetic field $B(t)$, the magnetic vector potential $A_i$ in our Hamiltonian is chosen in the Coulomb gauge as,
\begin{equation}
A_i=-\dfrac{B(t)}{2}\epsilon_{ij}X^{j}~~;\label{A}
\end{equation} 
where $i,j=1,2$ and $\epsilon_{ij}=-\epsilon_{ji}$ with $\epsilon_{12}=1$. Apart from it, $\omega(t)$ is the oscillator's angular frequency varying with respect to time. $M$ and $q$ implies its constant mass and charge parameter respectively. \\
\noindent Since the system is considered in a dynamic NC space, the commutator brackets among the NC canonical operators existing in the Hamiltonian read (including $\hbar$)
\begin{align}
[X_j,X_k]~=i\theta(t)\epsilon_{jk}\,,\, 
&[P_j,P_k]~=i\Omega(t)\epsilon_{jk}\,,\,[X_j,P_k]\,=i\,\hbar\,\left[1+\frac{\theta(t)\Omega(t)}{4}\right]\delta_{jk};\\
&(j,k=1,2~,~\epsilon_{jk}=-\epsilon_{kj}~,~\epsilon_{12}=1)\nonumber
\end{align}
where $\theta(t)$ and $\Omega(t)$ are the dynamic NC parameters for the configuration space 
and the momentum space respectively. In the other hand, the canonical operators $(x_i,p_i)$ in standard quantum mechanical phase space follow the commutator brackets $[x_j,p_k]=i\hbar\delta_{jk}$, $[x_j,x_k]=0=[p_j,p_k]$; ($j, k=1,2$).\\
\noindent Next, the standard Bopp-shift relations \cite{mez} ($\hbar=1$) is to be implemented into the system for reconstructing the original Hamiltonian (i.e. NC Hamiltonian) in terms of the standard commutative canonical operators explicitly. Now, the Bopp-shift relations are given by,
\begin{eqnarray}
& X_1=x_1-\dfrac{\theta(t)}{2}p_2\,\,\,;\,\,\,X_2=x_2+\dfrac{\theta(t)}{2}p_1\\
& P_1=p_1+\dfrac{\Omega(t)}{2}x_2\,\,\,;\,\,\,P_2=p_2-\dfrac{\Omega(t)}{2}x_1 \,\,.
\label{eqn1}
\end{eqnarray}
\noindent Therefore, the original Hamiltonian in terms of $(x_i,p_i)$ coordinate operators is revealed as
relation,
\begin{equation}
H=\dfrac{a(t)}{2}({p_1}^2+{p_2}^2)+\dfrac{b(t)}{2}({x_1}^2+{x_2}^2)+c(t)({x_2}{p_1}-{x_1}{p_2})\,\,\,;\label{ham}
\end{equation}
where time dependent coefficients have the following structures,
\begin{align}
&a(t)=\dfrac{f(t)}{M}+\dfrac{qB(t)f(t)\theta(t)}{2M}+\dfrac{1}{4}\left[\dfrac{q^2B^2(t)f(t)}{4M}+\dfrac{M\omega^2(t)}{f(t)}  \right]\theta^2(t),\label{a}\\
&b(t)=\dfrac{q^2B^2(t)f(t)}{4M}+\dfrac{M\omega^2(t)}{f(t)}+\dfrac{qB(t)f(t)\Omega(t)}{2M}+\dfrac{f(t)\Omega^2(t)}{4M},\label{b}\\
&c(t)=\dfrac{1}{2}\left[\dfrac{qB(t)f(t)}{M} \left( 1+\dfrac{\theta(t)\Omega(t)}{4}\right) +\dfrac{\Omega(t)f(t)}{M}+  \left(\dfrac{q^2B^2(t)f(t)}{4M}+\dfrac{M\omega^2(t)}{f(t)}  \right)\theta(t) \right]~.\label{c}
\end{align}

\noindent It should be remembered that while the generic form of the Hamiltonian (Eqn.(\ref{ham})) shares a similar form to those in \cite{Dey} and \cite{SG1} for studying a simple harmonic oscillator and a damped harmonic oscillator in dynamic NC space, the structural form of the Hamiltonian parameters (given by Eqn(s).(\ref{a})-(\ref{c})) for the present study are modified. This is because our current study examines the damped harmonic oscillator with an external time dependent magnetic field in a dynamic NC background. As a result, both $f(t)$ and $B(t)$ (the damping coefficient and the magnetic field respectively) influence and modify the Hamiltonian parameters from those studied in earlier communications \cite{Dey} and \cite{SG1}. Furthermore, the coefficient $c(t)$, unlike in our previous model \cite{SG1}, can no longer be considered as purely NC effect. Since, $b(t)$, the influence of the external magnetic field is present, the coefficient $c(t)$ would not vanish when the NC background are not considered in the scenario. Additionally, it is also important to note that these coefficients also differ from those obtained in \cite{gouba}, where the NC parameter is a constant quantity and belongs to the configuration space only.



\section{Exact eigenstate of the model Hamiltonian}
Once again, we utilize the Lewis technique \cite{Lewis} to solve the model system. A comprehensive discussion of this method can be found in the previous chapters \ref{chap-LR} and \ref{paper1}. Therefore, we will only briefly mention that by solving the Lewis invariant, such that    
\begin{equation}
I(t)\phi(x_1,x_2)=\epsilon \phi(x_1,x_2)
\label{MDHO-eqnegn}
\end{equation}
where $\epsilon$ denotes the invariant's eigenvalue corresponding to its time dependent eigenstate $\phi(x_1,x_2)$, the eigenstate of $H(t)$ can be obtained using the following relation 
\begin{equation}
\psi(x_1,x_2,t)=e^{i\Theta(t)}\phi(x_1,x_2)~;
\label{MDHO-eqnpsi}
\end{equation}
where $\psi(x_1,x_2,t)$ is the eigenfunction of the Hamiltonian (Eqn.(\ref{ham})) and $\Theta(t)$ denotes the Lewis phase factor which is real.\\
\textbf{Eigenfunction of the Lewis invariant operator and Lewis phase factor~:}
\vskip .20cm
\noindent The analytical structure of the invariant's eigenfunction is given by (for the positive integers $n$ and $m$, such that $n+l\,=m\,\geqslant\,0$) 
\begin{eqnarray}
\phi_{n,m-n}(r,\theta)&=&\braket{r,\theta|n,m-n}~\\
&=&\lambda_{n}\dfrac{{(i\rho)}^m}{\sqrt{m!}}r^{n-m}e^{i(m-n)\theta-\dfrac{a(t)-i\rho\dot{\rho}}{2a(t)
{\rho}^2}r^2}U\left(-m,1-m+n,\dfrac{r^2}{\rho^2} \right),
\label{MDHO-eqn28}
\end{eqnarray}
where $\rho$ adhere to the non linear EP equation (by setting the integration constant as zero)
\begin{equation}
\ddot{\rho}-\dfrac{\dot{a}}{a}\dot{\rho}+ab\rho=\dfrac{a^2}{\rho^3}~;
\label{MDHO-EP}
\end{equation} 
and $\lambda_n$ is given by
\begin{eqnarray}
\lambda_n^2=\dfrac{1}{\pi{n!}{(\rho^2)}^{1+n}}~. 
\label{MDHO-eqn28lam}
\end{eqnarray}
Here, $U\left(-m,1-m+n,\dfrac{r^2}{\rho^2} \right)$ denotes the
Tricomi's confluent hypergeometric function \cite{Arfken, uva} and the Hamiltonian eigenfunction $\phi_{n,m-n}(r,\theta)$ holds the orthonormality condition,
\begin{equation}
\int_0^{2\pi}d\theta\int_0^{\infty}rdr\phi^{*}_{n,m-n}(r,\theta)\phi_{n^{'},m^{'}-n^{'}}(r,\theta)=\delta_{nn^{'}}\delta_{mm^{'}}.
\label{MDHO-eqn29}
\end{equation}
The structure of $\Theta(t)$, the Lewis phase factor is given by \cite{Dey},  
\begin{equation}
\Theta_{\,n\,,\,l}(t)\,=\,(\,n\,+\,l\,)\,\int_0^t \left(c(T)-\dfrac{a(T)}{\rho^2(T)} \right)dT~.
\label{MDHO-eqn30}
\end{equation}
For any provided given value of $l$ ( since $l=-n+m$), it would take the form,
\begin{equation}
\Theta_{\,n\,,\,m\,-\,n\,}(t)=m\int_0^t \left(c(T)-\dfrac{a(T)}{\rho^2(T)} \right)dT~.
\label{MDHO-eqn31}
\end{equation}

\noindent Therefore, $\psi_{n, m-n}\,(r, \theta, t)$, the eigensystem of the Hamiltonian can be obtained by using  Eqn(s).(\ref{MDHO-eqnpsi}, \ref{MDHO-eqn28}, \ref{MDHO-eqn31}) and it reads
\begin{eqnarray}
\psi_{n,m-n}(r,\theta,t)&=&e^{i\Theta_{n, m-n}(t)}\phi_{n, m-n}(r,\theta)~\nonumber\\
&=&\lambda_{n}\dfrac{{(i\rho)}^m}{\sqrt{m!}}\exp{\left[im\int_0^t \left(c(T)-\dfrac{a(T)}{\rho^2(T)} \right)dT \right]}~
\nonumber\\
&&\times~r^{n-m}e^{i(m-n)\theta-\dfrac{a(t)-i\rho\dot{\rho}}{2a(t){\rho}^2}r^2}U\left(-m,1-m+n,\dfrac{r^2}{\rho^2} \right).
\label{eqn32}
\end{eqnarray}



\section{Explicit solutions for the model system}
In this study, we intend to investigate how the solution evolves when a time varying magnetic field is introduced into the system. Our goal is to determine the eigenfunctions of the Hamiltonian that result from the interaction between the damping factor and the external magnetic field. The different types of the damping scenarios under the influence of the external magnetic field are designed by various explicit forms of the Hamiltonian coefficients those are $a(t)$, $b(t)$ and $c(t)$. Here we should be careful because those explicit analytical forms must be consistent with the nonlinear EP equation (setting $\xi$ as $1$) specified by Eqn.(\ref{DHO-EP}). The procedure of constructing the exact analytical EP solutions follows the Chiellini integrability condition \cite{man1, man2, chill, Dey}, which was detailed in the previous chapter. In the following discussion, for our present model system, we will derive a class of exact analytical solutions which adhere to the nonlinear EP equtaion.


\subsection{Exponential solution for model system : Set-I } 
We begin by mentioning the exponentially varying EP solution set which is provided in \cite{Dey} and is also reviewed in the previous chapter \ref{paper1}. The solution set read, 
\begin{eqnarray}
a(t)=\sigma e^{-\Gamma{t}}\,\,\,,\,\,\,b(t)=\Delta e^{\Gamma{t}}\,\,\,,\,\,\rho(t)={\mu}e^{-\Gamma{t/2}}\label{MDHO-EPsoln1}
\end{eqnarray}
with the constraint relation
\begin{equation}
\mu^4=\dfrac{{\sigma^2}}{\sigma\Delta-\dfrac{1}{4}\Gamma^2}~;\label{MDHO-EPreln1}
\end{equation}
where $\sigma, \Delta, \mu$ and $\Gamma$ are the real, positive constants.\\



\subsubsection{Eigenfunctions of the invariant}
As we have already discussed in the previous chapter that once the EP (exact parameter) parameters are determined, the explicit form of $\phi_{n, m-n}(r, \theta)$, the eigenfunction of the invariant is fixed accordingly. Thus, for the exponential EP solution set (Set-I), the analytical expression of the invariant's eigenfunction is found to be
\begin{eqnarray}
\phi_{n,m-n}(r,\theta)=\lambda_{n}\dfrac{{(i{\mu}e^{-\Gamma{t/2}})}^m}{\sqrt{m!}}    r^{n-m}e^{i(m-n)\theta-\dfrac{2\sigma+i\mu^2\Gamma}{4\sigma\mu^2{e^{-\vartheta{t}}}}r^2}U\left(-m,1-m+n,\dfrac{r^2{e^{\Gamma{t}}}}{\mu^2} \right)~,
\label{MDHO-eqn33}
\end{eqnarray}
where $\lambda_n$ is given by
\begin{eqnarray}
\lambda_n^{\,2}\,=\,\dfrac{1}{\pi\,n!\,[\mu^2\,exp\,(-\Gamma{t})]^{1+n}}~.
\label{MDHO-eqn33lam}
\end{eqnarray}
\subsubsection{Lewis phase factors}
Now, we are about to examine the interesting interplay among the three parameters, $B(t)$, $f(t)$ and $\omega(t)$ by calculating the exactly integrated form of the Lewis phase factor as it changes its explicit form depending on those chosen parameters.
\vskip .20cm
\noindent {\bf $\langle a \rangle $ Set-I~,~Case I}\\
\noindent The damping effect and the external magnetic field can be chosen as 
\begin{eqnarray}
f(t)= e^{\,-\Gamma\,t}~;~\omega(t)={\omega_0}~;~B(t)=B_0~.\label{case1}
\end{eqnarray}
By substituting the above relations in Eqn(s).(\ref{a}, \ref{b}), the explicit structure of the NC parameters, for the present case, are calculated to be
\begin{align}
\theta(t)=\dfrac{8Me^{-\Gamma\,t}}{q^2B_0^2e^{-2\Gamma\,t}+4M^2\omega^2_0\,}
 &\left[\sqrt{\dfrac{q^2B_0^2\sigma e^{-2\Gamma\,t}}{4M}+\omega_0^2\,\left(M\sigma\,-1\right)}-\dfrac{qB_0e^{-\Gamma\,t}}{2M} \right]~,\nonumber\\                      
\Omega(t)&=-qB_0\,+2e^{\Gamma\,t}\sqrt{M\Delta\,-M^2\omega_0^2\,}~~~.                                                                                                           
\end{align}
In the limit as $B\rightarrow0$, we verify that the NC parameters $\theta(t)$ and $\Omega(t)$ simplify to those found in \cite{SG1}. Additionally, for a simpler case, where $t=0$ (with $B_0=0$), it is straightforward to see that $\theta(t)$ and $\Omega(t)$ reduce to the constant parameters. This corresponds to the standard time independent NC context. Furthermore, these constants can be fixed as zero, thereby recovering the results associated with the standard quantum mechanical phase space.

\noindent Substituting the above form of the dynamic NC parameters in the structural form of $c(t)$ in Eqn.(\ref{c}), we 
get,
\begin{eqnarray}
c(t)&=&\dfrac{1}{q^2B_0^2e^{-2\Gamma\,t}+4M^2\omega_0^2} \left[\left(4M^2\omega_0^2+2qB_0e^{-\Gamma\,t}\sqrt{M\Delta-M^2\omega_0^2} \right)\sqrt{\dfrac{q^2B_0^2\sigma e^{-2\Gamma\,t} }{4M}+\omega_0^2(M\sigma-1)}         \right.\nonumber \\
&&\left.          -2qB_0M\omega_0^2e^{-\Gamma\,t}-\dfrac{q^2B_0^2}{M}e^{-2\Gamma\,t}\sqrt{M\Delta-M^2\omega_0^2}              \right]+\sqrt{\dfrac{\Delta}{M}-\omega_0^2}                                                                                ~.
\end{eqnarray}
With the explicit form of $a(t)\,,\rho(t)$\, and $c(t)$\, in place, we perform the integration given in Eqn.(\ref{MDHO-eqn30}) and achieve the following closed form of the Lewis phase. The phase is explicitly found to be
\begin{align}
\small
&\Theta_{\,n\,,\,l}(t)\,=(n+l)\int_0^t \left[c(t)-\dfrac{a}{\rho^2}\right]~dT=(n+l)\left[\sqrt{\dfrac{\Delta}{M}-\omega_0^2}-\dfrac{\sigma}{\mu^2}\right]t\nonumber\\
&+\dfrac{(n+l)\omega_0\sqrt{(M\sigma-1)}}{\Gamma}
\,log\,\dfrac{\omega_0\sqrt{M\sigma-1}\,e^{\Gamma\,t}+\sqrt{\dfrac{q^2B_0^2\sigma}{4M}+\omega_0^2(M\sigma-1)e^{2\Gamma\,t}} }{\omega_0\sqrt{M\sigma-1}+\sqrt{\dfrac{q^2B_0^2\sigma}{4M}+\omega_0^2(M\sigma-1)}}
\nonumber\\
&+\dfrac{(n+l)\omega_0}{\Gamma}
\left[ tan^{-1}\dfrac{\omega_0 e^{\Gamma\,t}}{\sqrt{\dfrac{q^2B_0^2\sigma}{4M}+\omega_0^2(M\sigma-1)e^{2\Gamma\,t}}}-
tan^{-1}\dfrac{\omega_0 }{\sqrt{\dfrac{q^2B_0^2\sigma}{4M}+\omega_0^2(M\sigma-1)}}
\right]\nonumber\\
&+\dfrac{(n+l)\sqrt{M\Delta-M^2\omega_0^2}}{M\Gamma} \left[tanh^{-1}\dfrac{2M\sqrt{\dfrac{q^2B_0^2\sigma}{4M}+\omega_0^2(M\sigma-1)e^{2\Gamma\,t}}}{qB_0}-tanh^{-1}\dfrac{2M\sqrt{\dfrac{q^2B_0^2\sigma}{4M}+\omega_0^2(M\sigma-1)}}{qB_0} \right]\nonumber\\
&-\dfrac{(n+l)\sqrt{(\Delta-M\omega_0^2)\sigma}}{\Gamma}\left[tanh^{-1}\sqrt{\dfrac{q^2B_0^2\sigma+4M\omega_0^2(M\sigma-1)e^{2\Gamma\,t}}{q^2B_0^2\sigma}}-tanh^{-1}\sqrt{\dfrac{q^2B_0^2\sigma+4M\omega_0^2(M\sigma-1)}{q^2B_0^2\sigma}} \right]\nonumber\\
&-\dfrac{(n+l)\omega_0}{\Gamma}\left[tan^{-1}\dfrac{2M\omega_0 e^{\Gamma\,t}}{qB_0}-tan^{-1}\dfrac{2M\omega_0 }{qB_0}  \right]+\dfrac{(n+l)\sqrt{M\Delta-M^2\omega_0^2}}{2\Gamma\,M}\,log\,\dfrac{q^2B_0^2 e^{-2\Gamma\,t}+4M^2\omega_0^2}{q^2B_0^2 +4M^2\omega_0^2}~.
\end{align}
The eigenfunction of the Hamiltonian (Eqn.(\ref{ham}), for the selected case of damping and applied magnetic field, can now be readily determined by connecting the above phase factor to Eqn.(\ref{MDHO-eqn33}), following the relation outlined in Eqn.(\ref{MDHO-eqnpsi}).

 
\noindent {\bf $\langle b \rangle $ Set-I~,~Case II}
\vskip .10cm
\noindent The damping effect and the external magnetic field can be chosen as 
\begin{eqnarray}
f(t)= e^{-\Gamma t}~;~\omega(t)={\omega_0}~;~B(t)=B_0\,e^{\Gamma\,t}~.
\label{10X}
\end{eqnarray}
By substituting the above relations in Eqn(s).(\ref{a},\ref{b}), the explicit structure of the NC parameters, for the present case, are calculated to be
\begin{align}
\theta(t)=\dfrac{8Me^{-\Gamma\,t}}{q^2B_0^2+4M^2\omega^2_0\,}
 \left[\sqrt{\dfrac{q^2B_0^2\sigma }{4M}+\omega_0^2\,\left(M\sigma\,-1\right)}-\dfrac{qB_0}{2M} \right]~,~                       
\Omega(t)=  -qB_0\,e^{\Gamma\,t}+2e^{\Gamma\,t}\sqrt{M\Delta\,-M^2\omega_0^2\,}                                                                                                          ~. \label{exp2conNC}
\end{align}
It is important to note that in this case, the product of two dynamic NC parameters simplifies to a constant. Additionally, this constant is identical to the one found in another case, as discussed in Eqn.(\ref{-2ratdecNC}). Furthermore, it can be verified that when $B\rightarrow0$, the explicit form of $\theta(t)$ and $\Omega(t)$ match those derived in \cite{SG1}.\\
\noindent Substituting the above form of the dynamic NC parameters in the structural form of $c(t)$ in Eqn.(\ref{c}), we 
get,
\begin{eqnarray}
c(t)&=&\dfrac{1}{q^2B_0^2+4M^2\omega_0^2} \left[\left(4M^2\omega_0^2+2qB_0\sqrt{M\Delta-M^2\omega_0^2} \right)\sqrt{\dfrac{q^2B_0^2\sigma  }{4M}+\omega_0^2(M\sigma-1)}         \right.\nonumber \\
&&\left.          -2qB_0M\omega_0^2-\dfrac{q^2B_0^2}{M}\sqrt{M\Delta-M^2\omega_0^2}              \right]+\sqrt{\dfrac{\Delta}{M}-\omega_0^2}~.                                                                                \label{exp2conC}
\end{eqnarray}

\noindent At this point, for this present case, we are about to investigate the Hamiltonian coefficients in the standard quantum mechanical phase space. In the standard phase space, the Hamiltonian parameters take the following form,
\begin{align}
&a(t)=\dfrac{f(t)}{M}~,\label{a2}\\
&b(t)=\dfrac{q^2B^2(t)f(t)}{4M}+\dfrac{M\omega^2(t)}{f(t)}~,\label{b2}\\
&c(t)=\dfrac{qB(t)f(t)}{2M} ~.\label{c2}
\end{align}
For the case of exponential decay, we observe from Eqn.(\ref{MDHO-EPsoln1}) that $a(t)=\sigma e^{-\Gamma{t}}$. Additionally, Eqn.(\ref{10X}) provides $f(t)= e^{-\Gamma t}$. Substituting these parameters into Eqn.(\ref{a2}), we obtain $\sigma=\dfrac{1}{M}$.
Using this result, along with the expression $b(t)=\Delta e^{\Gamma{t}}$ for the exponentially decaying scenario in Eqn.(\ref{b2}), we derive the condition, $\Delta=\dfrac{q^2B_0^2}{4M}+M\omega_0^2$.
Furthermore, substituting $f(t)= e^{-\Gamma t}$ and $B(t)=B_0\,e^{\Gamma\,t}$ into Eqn.(\ref{c2}), we get,
\begin{align} 
c(t)=\dfrac{qB_0}{2M}~,
\label{const}
\end{align}
which is found to be a constant. Later, the energy dynamics of the above case in the standard quantum mechanical framework will also be explored.
\noindent Now we restart our discussions in the NC framework again. With the explicit form of $a(t)\,,\rho(t)$\, and $c(t)$\, in place, we perform the integration given in Eqn.(\ref{MDHO-eqn30}) and achieve the following closed form of the Lewis phase. The phase is explicitly found to be
\begin{align}
\Theta_{\,n\,,\,l}(t)\,=\dfrac{(n\,+\,l)}{q^2B_0^2+4M^2\omega_0^2} &\left[\left(4M^2\omega_0^2+2qB_0\sqrt{M\Delta-M^2\omega_0^2} \right)\sqrt{\dfrac{q^2B_0^2\sigma }{4M}+\omega_0^2(M\sigma-1)}         \right.\nonumber \\
&\left.          -2qB_0M\omega_0^2-\dfrac{q^2B_0^2}{M}\sqrt{M\Delta-M^2\omega_0^2}              \right]t+(n+l)\left[\sqrt{\dfrac{\Delta}{M}-\omega_0^2}-\dfrac{\sigma}{\mu^2}\right]t~.
\end{align}
Here, for this case, the phase is found to be linear function of time. The phase brings back its form (solution set Ib) derived in \cite{SG1} at the limit $B_0\rightarrow\,0$. \\
\noindent The eigenfunction of the Hamiltonian (Eqn.(\ref{ham}), for the selected case of damping and applied magnetic field, can now be readily determined by connecting the above phase factor to Eqn.(\ref{MDHO-eqn33}), following the relation outlined in Eqn.(\ref{MDHO-eqnpsi}).\\
\noindent Later, under these choices of damping factor and frequency, it was shown in \cite{sg2} that a purely periodic nature can be introduced to the energy dynamics of a model system of such a damped harmonic oscillator in a dynamic NC space if a time varying magnetic field can be tuned with the system in such a way that it varies sinusoidally with an amplitude that increases exponentially over time.\\
\noindent {\bf $\langle c \rangle $ Set-I~,~ Case III }
\vskip .20cm
\noindent The damping effect and the external magnetic field can be chosen as,
\begin{eqnarray}
f(t)= e^{-\Gamma\,t}~,~\omega(t)={\omega_0}~,~B(t)=B_0\,e^{-\Gamma\,t}~.
\label{10x}
\end{eqnarray}
By substituting the above relations in Eqn(s).(\ref{a},\ref{b}), the explicit structure of the NC parameters, for the present case, are calculated to be
\begin{align}
\theta(t)=\dfrac{8Me^{-\Gamma\,t}}{q^2B_0^2e^{-4\Gamma\,t}+4M^2\omega^2_0}
 &\left[\sqrt{\dfrac{q^2B_0^2\sigma\,e^{-4\Gamma\,t}}{4M}+\omega_0^2(M\sigma-1)}-\dfrac{qB_0e^{-2\Gamma\,t}}{2M} \right]~,                    
  \nonumber\\
\Omega(t)&=-qB_0e^{-\Gamma\,t}+2e^{\Gamma\,t}\sqrt{M\Delta-M^2\omega_0^2}~.                              \label{exp2decNC}
\end{align}
Substituting the above form of the dynamic NC parameters in the structural form of $c(t)$ in Eqn.(\ref{c}), we 
get,
\begin{eqnarray}	
c(t)&=&\dfrac{1}{q^2B_0^2e^{-4\Gamma\,t}+4M^2\omega_0^2} \left[\left(4M^2\omega_0^2+2qB_0e^{-2\Gamma\,t}\sqrt{M\Delta-M^2\omega_0^2} \right)\sqrt{\dfrac{q^2B_0^2\sigma e^{-4\Gamma\,t}}{4M}+\omega_0^2(M\sigma-1)}         \right.\nonumber \\
&&\left.          -2qB_0M\omega_0^2e^{-2\Gamma\,t}-\dfrac{q^2B_0^2e^{-4\Gamma\,t}}{M}\sqrt{M\Delta-M^2\omega_0^2}              \right]+\sqrt{\dfrac{\Delta}{M}-\omega_0^2} ~.
\label{exp2decC}
\end{eqnarray}
With the explicit form of $a(t)\,,\rho(t)$\, and $c(t)$\, in place, we perform the integration given in Eqn.(\ref{MDHO-eqn30}) and achieve the following closed form of the Lewis phase. The phase is explicitly found to be
\begin{align}
\small
&\Theta_{\,n\,,\,l}(t)\,=(n+l)\int_0^t \left[c(t)-\dfrac{a}{\rho^2}\right]~dT=(n+l)\left[\sqrt{\dfrac{\Delta}{M}-\omega_0^2}-\dfrac{\sigma}{\mu^2}\right]t\nonumber\\&+\dfrac{(n+l)\omega_0\sqrt{(M\sigma-1)}}{2\Gamma}
\,log\,\dfrac{\omega_0\sqrt{M\sigma-1}\,e^{2\Gamma\,t}+\sqrt{\dfrac{q^2B_0^2\sigma}{4M}+\omega_0^2(M\sigma-1)e^{4\Gamma\,t}} }{\omega_0\sqrt{M\sigma-1}+\sqrt{\dfrac{q^2B_0^2\sigma}{4M}+\omega_0^2(M\sigma-1)}}
\nonumber\\
&+\dfrac{(n+l)\omega_0}{2\Gamma}
\left[ tan^{-1}\dfrac{\omega_0 e^{2\Gamma\,t}}{\sqrt{\dfrac{q^2B_0^2\sigma}{4M}+\omega_0^2(M\sigma-1)e^{4\Gamma\,t}}}-
tan^{-1}\dfrac{\omega_0 }{\sqrt{\dfrac{q^2B_0^2\sigma}{4M}+\omega_0^2(M\sigma-1)}}
\right]\nonumber\\
&+\dfrac{(n+l)\sqrt{M\Delta-M^2\omega_0^2}}{2M\Gamma} \left[tanh^{-1}\dfrac{2M\sqrt{\dfrac{q^2B_0^2\sigma}{4M}+\omega_0^2(M\sigma-1)e^{4\Gamma\,t}}}{qB_0}-tanh^{-1}\dfrac{2M\sqrt{\dfrac{q^2B_0^2\sigma}{4M}+\omega_0^2(M\sigma-1)}}{qB_0} \right]\nonumber\\
&-\dfrac{(n+l)\sqrt{(\Delta-M\omega_0^2)\sigma}}{2\Gamma}\left[tanh^{-1}\sqrt{\dfrac{q^2B_0^2\sigma+4M\omega_0^2(M\sigma-1)e^{4\Gamma\,t}}{q^2B_0^2\sigma}}-tanh^{-1}\sqrt{\dfrac{q^2B_0^2\sigma+4M\omega_0^2(M\sigma-1)}{q^2B_0^2\sigma}} \right]\nonumber\\
&-\dfrac{(n+l)\omega_0}{2\Gamma}\left[tan^{-1}\dfrac{2M\omega_0 e^{2\Gamma\,t}}{qB_0}-tan^{-1}\dfrac{2M\omega_0 }{qB_0}  \right]+\dfrac{(n+l)\sqrt{M\Delta-M^2\omega_0^2}}{4\Gamma\,M}\,log\,\dfrac{q^2B_0^2 e^{-4\Gamma\,t}+4M^2\omega_0^2}{q^2B_0^2 +4M^2\omega_0^2}~.
\end{align}
The eigenfunction of the Hamiltonian (Eqn.(\ref{ham}), for the selected case of damping and applied magnetic field, can now be readily determined by connecting the above phase factor to Eqn.(\ref{MDHO-eqn33}), following the relation outlined in Eqn.(\ref{MDHO-eqnpsi}).

\noindent {\bf $\langle d \rangle $ Set-I~,~Case IV }
\vskip .20cm
\noindent The damping effect and the external magnetic field can be chosen as 
\begin{eqnarray}
f(t)= e^{-\Gamma\,t}~;~\omega(t)={\omega_0}e^{-\Gamma\,t/2}~;~B(t)=B_0\,e^{\Gamma\,t}~.
\end{eqnarray}
By substituting the above relations in Eqn(s).(\ref{a},\ref{b}), the explicit structure of the NC parameters, for the present case, are calculated to be
\begin{align}
\theta(t)= \dfrac{8Me^{-\Gamma\,t}}{q^2B_0^2+4M^2\omega_0^2e^{-\Gamma\,t}}\left[\sqrt{\dfrac{q^2B_0^2\sigma}{4M}+\omega_0^2e^{-\Gamma\,t}(M\sigma-1)}-\dfrac{qB_0}{2M} \right]\,,\,          
  \nonumber\\                
\Omega(t)=-qB_0e^{\Gamma\,t}+2e^{\Gamma\,t}\sqrt{M\Delta-M^2\omega_0^2e^{-\Gamma\,t}}~.\label{exp3incNC}
\end{align}
Substituting the above form of the dynamic NC parameters in the structural form of $c(t)$ in Eqn.(\ref{c}), we 
get,
\begin{eqnarray}
c(t)&=&\dfrac{1}{q^2B_0^2+4M^2\omega_0^2e^{-\Gamma\,t}} \left[\left(4M^2\omega_0^2e^{-\Gamma\,t}+2qB_0\sqrt{M\Delta-M^2\omega_0^2e^{-\Gamma\,t}} \right)\sqrt{\dfrac{q^2B_0^2\sigma}{4M}+\omega_0^2e^{-\Gamma\,t}(M\sigma-1)}         \right.\nonumber \\
&&\left.          -2qB_0M\omega_0^2e^{-\Gamma\,t}-\dfrac{q^2B_0^2}{M}\sqrt{M\Delta-M^2\omega_0^2e^{-\Gamma\,t}}              \right]+\sqrt{\dfrac{\Delta}{M}-\omega_0^2e^{-\Gamma\,t}}~.\label{exp3incC}
\end{eqnarray}
Performing the integration in Eqn.(\ref{MDHO-eqn30}), the Lewis phase in a closed form can be derived for this case also. However, we provide it in the Appendix.

\subsection{Rational solution for model system : Set-II}

We begin by mentioning the rationally varying EP solution set which is provided in \cite{Dey} and is also reviewed in the previous section. The solution set is given by the following
relations, 
\begin{eqnarray}
a(t)=\dfrac{\sigma\,\left(1+\dfrac{2}{k}\right)^{\,(k+2)/k}}{(\Gamma{t}+\chi)^{\,(k+2)/k}}~,~
b(t)=\dfrac{\Delta\,\left(\dfrac{k+2}{k} \right)^{(k-2)/k} }{(\Gamma{t}+\chi)^{\,(k-2)/k}}     ~,~\rho(t)=\dfrac{\mu\left(1+\dfrac{2}{k}\right)^{1/k} }{(\Gamma{t}+\chi)^{1/k}}~;
\label{MDHO-EPsoln2} 
\end{eqnarray}
with the constraint relation
\begin{equation}
\Gamma^2\mu=(k+2)^2\,\left(\sigma\Delta\mu-\frac{\sigma^2}{\mu^3}\right).
\label{MDHO-EPreln2}
\end{equation}
Here $k$ takes integer value and $\sigma$, $\Gamma$, $\Delta$, $\mu$ and $\chi$ are the constants and they must ensure $(\Gamma{t}+\chi)~\neq~0$.


\subsubsection{Eigenfunctions of the Lewis invariant}
For the rational EP solution set (Set-II), the analytical expression of the invariant's eigenfunction is found to be,
\begin{align}
\phi_{n\,,\,m-n}(r,\theta)=\lambda_{n}\,\dfrac{{(i\mu)}^{\,m}}{\sqrt{m!}}\left[\dfrac{k+2}{k(\Gamma{t}+\chi)}\right]^{m/k}    r^{n-m}e^{i(m-n)\theta-\dfrac{[\sigma\,(k+2)\,+\,i\mu^2\Gamma]\,\,(\Gamma{t}+\chi)^{2/k}\,\,\,k^{2/k} }{2\sigma\,(k+2)^{\,(k+2)/k}\mu^2}r^2}\nonumber \\
\times\,\,\,U\left(-m,1-m+n,\,\dfrac{r^2[k(\Gamma{t}+\chi)]^{2/k}}{\mu^2\left(k+2\right)^{2/k}}\,\right)~,
\label{MDHO-eqn51}
\end{align}
where $\lambda_n$ is given by 
\begin{eqnarray}
\lambda_n^{\,2}=\dfrac{1}{\pi\,n!\mu^{2n+2}}\left[\dfrac{k(\Gamma{t}+\chi)}{k+2}\right]^{2(1+n)/k}.
\label{MDHO-eqn51lam}
\end{eqnarray}
\subsubsection{Lewis phase factors}
As per the rational EP solution set, we must tune the oscillator's parameters $f(t)$ and $\omega(t)$, and the background field $B(t)$. To investigate the solution of $H(t)$ under the assumption of coefficients that decay rationally, we opt for a rationally decaying $\omega(t)$ and ~$B(t)$ while setting $f(t)=1$. Thus,~we have the following relations, 
\begin{eqnarray}
\eta(t)=0\,\,\Rightarrow\,\,f(t)=1~,~
\omega(t)=\dfrac{\omega_0}{(\Gamma\,t+\chi)}~,~B(t)=\dfrac{B_0}{(\Gamma\,t+\chi)}.
\label{2ratdec}
\end{eqnarray}

\noindent {\bf $\langle e \rangle $ Set-II~,~Case I }\vskip .20cm

\noindent To investigate how the characteristics of the rationally decaying solution change under the influence of a magnetic field, we choose $k=2$, in Eqn.(\ref{MDHO-EPsoln2}). Previously, this system was examined without the presence of any external field for this specific value of $k$ as reported in an earlier publication \cite{SG1}.
\noindent The simplified form of the rational EP parameters, for the value of $k=2$, are found to be
\begin{eqnarray}
a(t)=\dfrac{4\sigma}{(\Gamma{t}+\chi)^{\,2}}\,\,,\,\,b(t)\,=\,\Delta\,\,,\,\,\rho(t)=\left[\dfrac{2\mu^{\,2}}{\Gamma{t}+\chi}\right]^{1/2}.\label{2rat}
\end{eqnarray}
By substituting the above relations in Eqn(s).(\ref{a},\ref{b}), the explicit structure of the NC parameters, for the present case, are calculated to be
\begin{align}
\theta(t)=\dfrac{8M}{q^2B_0^2+4M^2\omega^2_0}&\left[\sqrt{ \dfrac{q^2B_0^2\sigma}{M}+\omega_0^2 4\sigma M-\omega_0^2(\Gamma\,t+\chi)^2}-\dfrac{qB_0}{2M}(\Gamma\,t+\chi)  \right]~,                      \nonumber \\
&\Omega(t)=- \dfrac{qB_0}{(\Gamma\,t+\chi)}+2\sqrt{M\Delta-\dfrac{M^2\omega_0^2}{(\Gamma\,t+\chi)^2}}     ~.                            \label{2ratdecNC}
\end{align}
Substituting the above form of the dynamic NC parameters in the structural form of $c(t)$ in Eqn.(\ref{c}), we 
get,
\begin{align}
c(t)&=\dfrac{1}{4M^2\omega^2_0+q^2B_0^2}\left[\left( \dfrac{4M^2\omega_0^2}{(\Gamma\,t+\chi)^2}+\dfrac{2qB_0\sqrt{M\Delta(\Gamma\,t+\chi)^2-M^2\omega_0^2}}{(\Gamma\,t+\chi)^2}\right)
\sqrt{ \dfrac{q^2B_0^2\sigma}{M}-\omega_0^2 (\Gamma\,t+\chi)^2 +4\omega_0^2\sigma M} \right.\nonumber\\ 
&\left.-\dfrac{2qB_0M\omega_0^2}{\Gamma\,t+\chi}-\dfrac{q^2B_0^2\sqrt{M\Delta(\Gamma\,t+\chi)^2-M^2\omega_0^2}}{M(\Gamma\,t+\chi)} \right]+\sqrt{\dfrac{\Delta}{M}-\dfrac{\omega_0^2}{(\Gamma\,t+\chi)^2}}~.                              \label{2ratdecC}
\end{align}
In the above expression, the extra terms arising from the magnetic field are mostly observed to be decreasing functions of time. Their impact becomes clearer when examining how the energy expectation value evolves over time in a subsequent section. The precise form of the phase factor has been determined and is shown in the Appendix.
\vskip .15cm

\noindent{\bf $\langle f \rangle $ Set-II~,~Case II }\vskip .20cm
\noindent It is observed from Eqn.(\ref{MDHO-EPsoln2}) that the rational EP solution set is not useful when $k=-2$. At this limit, $a(t)$ and $\rho(t)$ vanish and $b(t)$ diverges. Thus, we fix the following solution set 
\begin{eqnarray}
a(t)=\dfrac{\sigma}{(\Gamma{t}+\chi)^{\,(k+2)/k}}~,~
b(t)=\dfrac{\Delta}{(\Gamma{t}+\chi)^{\,(k-2)/k}}     ~,~\rho(t)=\dfrac{\mu}{(\Gamma{t}+\chi)^{1/k}}.
\label{k2} 
\end{eqnarray}
Setting $k=-2$ in the above solution set, we proceed with the following solution set,
\begin{eqnarray}
a=\sigma ~~,~~ b(t)=\dfrac{\Delta}{(\Gamma\,t+\chi)^2}~~,~~ \rho=\mu \sqrt{\Gamma\,t+\chi}\,.
\label{ratsol}
\end{eqnarray}
The constants in the above solution set are constrained to hold a relation which is found by substituting that solution set in Eqn.(\ref{MDHO-EP}). The relation is given by
\begin{equation}
-\mu^4\Gamma^2+4\sigma\Delta\mu^4=4\,\sigma^2\,;\label{rat2const}
\end{equation} 
which is similar to the constraint relation (Eqn.(\ref{MDHO-EPreln1})) found for the exponential EP solution set.
   


\noindent The analytical expression of the invariant's eigenfunction is found for the present solution set to be
\begin{eqnarray}
\phi_{n\,,\,m-n}(r,\theta)=\lambda_{n}\,\dfrac{{(i\mu\sqrt{(\Gamma\,t+\chi})}^{\,m}}{\sqrt{m!}}    r^{n-m}e^{i(m-n)\theta-\dfrac{2\sigma-i\mu^2\Gamma }{4\sigma\mu^2(\Gamma\,t+\chi)}r^2}
~U\left(-m,1-m+n,\,\dfrac{r^2}{\mu^2(\Gamma\,t+\chi)}\,\right)~,
\label{MDHO-eqn511}
\end{eqnarray}
where $\lambda_n$ is given by
\begin{eqnarray}
\lambda_n^{\,2}=\dfrac{1}{\pi\,n![\mu^2(\Gamma\,t+\chi)]^{1+n}}.
\label{eqn51lam1}
\end{eqnarray}
Again, the damping effect and the external magnetic field are chosen as,
\begin{eqnarray}
f(t)=1~,~ \omega(t)=\dfrac{\omega_0}{(\Gamma\,t+\chi)}~,~ B(t)=\dfrac{B_0}{(\Gamma\,t+\chi)}~. \label{2ratinc}
\end{eqnarray}
By substituting the above relations in Eqn(s).(\ref{a},\ref{b}), the explicit structure of the NC parameters, for the present case, are calculated to be
\begin{align}
\theta(t)= \dfrac{8M(\Gamma\,t+\chi)}{q^2B_0^2+4M^2\omega_0^2}\left[\sqrt{\dfrac{q^2B_0^2\sigma}{4M}+\omega_0^2(M\sigma-1)}-\dfrac{qB_0}{2M} \right]                  ~,~ \Omega(t)= 2\dfrac{\sqrt{M\Delta-M^2\omega_0^2}}{(\Gamma\,t+\chi)}-\dfrac{qB_0}{(\Gamma\,t+\chi)}~.                                                                                 \label{-2ratdecNC}
\end{align} 
Again, it is interesting to observe that the multiplication of the two dynamic NC parameters yields a constant value. As previously discussed in Eqn.(\ref{exp2conNC}), this multiplication results in a constant value, which remains same in this scenario as well.\\
\noindent Substituting the above form of the dynamic NC parameters in the structural form of $c(t)$ in Eqn.(\ref{c}), we get
\begin{eqnarray}
c(t)&=&\dfrac{1}{(4M^2\omega^2_0+q^2B_0^2)(\Gamma\,t+\chi)}\left[\left( 4M^2\omega_0^2+2qB_0\sqrt{M\Delta-M^2\omega_0^2}\right)
\sqrt{ \dfrac{q^2B_0^2\sigma}{4M}+\omega_0^2(M\sigma-1)  } \right.\nonumber\\ 
&&\left.-2qB_0M\omega_0^2-\dfrac{q^2B_0^2\sqrt{M\Delta-M^2\omega_0^2}}{M} \right]+\dfrac{1}{(\Gamma\,t+\chi)}\sqrt{\dfrac{\Delta}{M}-\omega_0^2 }~.
 \label{-2ratdecC}
\end{eqnarray}
With the explicit form of $a(t)\,,\rho(t)$\, and $c(t)$\, in place, we perform the integration given in Eqn.(\ref{MDHO-eqn30}) and achieve the following closed form of the Lewis phase. The phase is explicitly found to be
\begin{align}
\Theta_{\,n, l\,}(t)&=\dfrac{(n+l)}{(4M^2\omega^2_0+q^2B_0^2)\Gamma}\left[\left( 4M^2\omega_0^2+2qB_0\sqrt{M\Delta-M^2\omega_0^2}\right)
\sqrt{ \dfrac{q^2B_0^2\sigma}{4M}+\omega_0^2(M\sigma-1)  } \right.\nonumber\\ 
&\left.-2qB_0M\omega_0^2-\dfrac{q^2B_0^2\sqrt{M\Delta-M^2\omega_0^2}}{M} \right]ln\dfrac{(\Gamma\,t+\chi)}{\chi}+\dfrac{(n+l)}{\Gamma}\left[\sqrt{\dfrac{\Delta}{M}-\omega_0^2 }-\dfrac{\sigma}{\mu^2}\right]~ln\dfrac{(\Gamma\,t+\chi)}{\chi}~.\nonumber\\
\end{align}
The eigenfunction of the Hamiltonian (Eqn.(\ref{ham}), for the selected case of damping and applied magnetic field, can now be readily determined by connecting the above phase factor to Eqn.(\ref{MDHO-eqn511}), following the relation outlined in Eqn.(\ref{MDHO-eqnpsi})



\section{Energetics of the model system}
Here we aim to analyse the energy dynamics of the model under different conditions, similar to our previous research. According to \cite{SG1}, the generalized expression for the expectation value of energy $\braket{E_{n,m-n}(t)}$ in the eigenstate of Hamiltonian $\psi_{n,m-n}(r,\theta,t)$, reads
\begin{align}
\braket{E_{n,m-n}(t)}&=\dfrac{1}{2}\,(n+m+1)\left[b(t)\rho^2(t)+\dfrac{a(t)}{\rho^2(t)}+\dfrac{\dot{\rho}^2(t)}{a(t)} \right]+c(t)\,(n-m)\,\,.
\end{align}
With the structure of $c(t)$ (Eqn.(\ref{c})), the energy is also represented as 
\begin{align}
\braket{E_{n,m-n}(t)}&=\dfrac{1}{2}\,(n+m+1)\left[b(t)\rho^2(t)+\dfrac{a(t)}{\rho^2(t)}+\dfrac{\dot{\rho}^2(t)}{a(t)} \right]\nonumber \\
&+\dfrac{(n-m)}{2}\left[\dfrac{qB(t)f(t)}{M} \left( 1+\dfrac{\theta(t)\Omega(t)}{4}\right) +\dfrac{\Omega(t)f(t)}{M}+  \left(\dfrac{q^2B^2(t)f(t)}{4M}+\dfrac{M\omega^2(t)}{f(t)}  \right)\theta(t) \right]~;\label{Energy}
\end{align}
which reproduces the energy expression in \cite{SG1} while setting $B\rightarrow\,0$.\\
The energy has explicit charge dependence. It includes terms that are both linearly and quadratically dependent on the charge. Consequently, the energy is not invariant when the particle's charge changes sign. Even the energy contains a non zero, finite value when the applied field is switched off ($B\rightarrow,0$) and the oscillator's frequency ($\omega{\rightarrow}0$) is also zero. This is evident from 
Eqn.(s)(\ref{a}, \ref{b}, \ref{c}) that the Hamiltonian parameters remain non-zero while setting $B{\rightarrow}0$ and  $\omega{\rightarrow}0$.\\
\noindent Next, in various damping scenarios, we will examine the evolution of $\braket{E_{n,m-n}(t)}$ in the presence of the diverse form of the external magnetic field.

   

\subsection{Energy profile for Solution Set-I : Exponentially decaying solution}
The energy expectation value, for the exponentially varying EP solution set given by Eqn.(\ref{MDHO-EPsoln1}), takes the following form,
\begin{equation}
\braket{E_{n,m-n}(t)}=(n+m+1)\mu^2\Delta+c(t)\,(n-m)
\label{MDHO-eqn90}
\end{equation}
where we have used the constraint relation given by Eqn.(\ref{MDHO-EPreln1}).\\
\noindent Here, it should be noted that the time dependence of this general expression found for the exponential EP solution set, unlike that found in the previous work \cite{SG1}, is not a pure NC effect. This is because the Hamiltonian coefficient $c(t)$ does not vanish in the absence of NC parameters due to the presence of an externally applied magnetic field.

\vskip 0.1cm



\noindent{{\bf{$\langle A\rangle$ Set-I \,,\,Case I}}}
\vskip .20cm
\noindent Here we set $f(t)=e^{-\Gamma\,t}$ , $\omega(t)=\omega_0$ and  $B(t)=B_0$. With this the energy expression for the ground state takes the form,

\begin{align}
\braket{E_{n,-n}(t)}&=(n+1)\mu^2\Delta+\dfrac{n}{q^2B_0^2e^{-2\Gamma\,t}+4M^2\omega_0^2}\left[ -2qB_0M\omega_0^2e^{-\Gamma\,t}-\dfrac{q^2B_0^2}{M}e^{-2\Gamma\,t}\sqrt{M\Delta-M^2\omega_0^2}           \right.\nonumber\\
& \left. +\left(4M^2\omega_0^2+2qB_0e^{-\Gamma\,t}\sqrt{M\Delta-M^2\omega_0^2} \right)\sqrt{\dfrac{q^2B_0^2\sigma e^{-2\Gamma\,t} }{4M}+\omega_0^2(M\sigma-1)} \right]+n\,\sqrt{\dfrac{\Delta}{M}-\omega_0^2} .\label{exp1EN1}     
\end{align}
\noindent The nature of the energy expectation value is influenced by the values of the constants. In particular, the sign of the charge is crucial in determining this expectation value. Notably, in both the limits $t\rightarrow\infty$ and $B\rightarrow\,0$ the energy expression converges to the same constant value as found in \cite{SG1} for a damped oscillator in time dependent NC space. Additionally, incorporating a constant magnetic field into the system studied in \cite{SG1} causes the Hamiltonian to become non hermitian after a certain time limit, beyond which the energy becomes imaginary. The condition for getting the expectation value of energy to be real is as follows,

\begin{equation}
t~\leq~\dfrac{1}{2\Gamma}\,ln \dfrac{q^2B_0^2\sigma}{4M\omega_0^2(1-M\sigma)}~.\label{exp1tim1}
\end{equation}
It is noteworthy that the upper time limit, within which the energy expectation value remains real, is independent of the oscillator's charge sign. However, the actual energy value does depend on it. Figure \ref{MDHO-fig1} illustrates how the energy expectation value changes over time. Initially, there is a decrease in energy, but over time, it stabilizes and becomes constant.\\
\noindent Now, we aim to determine if the energy exhibits time variation in the absence of damping ($f(t)=1$, that is for $\Gamma=0$). In this case, 
Eqn.(\ref{exp1EN1}) has a time independent form given by, 
\begin{align}
\braket{E_{n,-n}(t)}|_{\Gamma\rightarrow\,0}&=(n+1)\mu^2\Delta+ 
\dfrac{n}{q^2B_0^2+4M^2\omega_0^2} \left[-2qB_0M\omega_0^2-\dfrac{q^2B_0^2}{M}\sqrt{M\Delta-M^2\omega_0^2}\right.\nonumber \\
&\left. +\left(4M^2\omega_0^2+2qB_0\sqrt{M\Delta-M^2\omega_0^2} \right)\sqrt{\dfrac{q^2B_0^2\sigma }{4M}+\omega_0^2(M\sigma-1)}                       \right]+n\sqrt{\dfrac{\Delta}{M}-\omega_0^2}\\
&=constant~.\nonumber
\end{align} 
So, it is seen that the energy reduces to a constant due to the removal of  the damping factor. The energetics of the system in this scenario is the same as that found in the 
next case, where damping is present, but the applied magnetic field is exponentially expanding.


\noindent{{\bf{$\langle B\rangle$ Set-I \,,\,Case II}}}\\
\noindent Here we set $f(t)=e^{-\Gamma\,t}$ , $\omega(t)=\omega_0$ and  $B(t)=B_0\,e^{\Gamma\,t}$. With this the energy expression for the ground state takes the form,
\begin{align}
\braket{E_{n,-n}(t)}&=(n+1)\mu^2\Delta+ 
\dfrac{n}{q^2B_0^2+4M^2\omega_0^2} \left[-2qB_0M\omega_0^2-\dfrac{q^2B_0^2}{M}\sqrt{M\Delta-M^2\omega_0^2}\right.\nonumber \\
&\left. +\left(4M^2\omega_0^2+2qB_0\sqrt{M\Delta-M^2\omega_0^2} \right)\sqrt{\dfrac{q^2B_0^2\sigma }{4M}+\omega_0^2(M\sigma-1)}                       \right]+n\sqrt{\dfrac{\Delta}{M}-\omega_0^2}\label{exp1EN2}
\end{align} 
Interestingly, the exponentially decaying function $f(t)$ and the exponentially increasing function $B(t)$ counterbalance each other, leading to a constant energy value. It can also observed from Fig. \ref{MDHO-fig1}. In the limit $B\rightarrow\,0$,~the constant value of energy reduces to the same obtained in \cite{SG1} for a damped oscillator in time dependent NC space. \\
\noindent Our next step involves examining the energy profile of the system in commutative space. By employing the energy expression for the system under exponentially decaying conditions [given by Eqn.(\ref{MDHO-eqn90})] and utilizing the expression for $c(t)$ [given by Eqn.(\ref{const})] for the corresponding 
commutative background, we derive the following
\begin{equation}
\braket{E_{n,-n}(t)}=(n+1)\mu^2\Delta+n\dfrac{qB_0}{2M}~;
\end{equation}
which remains a constant as we vary time.

\vskip .20cm
\noindent {\bf{$\langle C\rangle$ Set-I \,,\,Case III}}\\
\noindent Here we set $f(t)=e^{-\Gamma\,t}$ , $\omega(t)=\omega_0$ and  $B(t)=B_0e^{-\Gamma\,t}$. With this the energy expression for the ground state takes the form,
\begin{align}
\braket{E_{n,-n}(t)}&=(n+1)\mu^2\Delta+\dfrac{n}{q^2B_0^2e^{-4\Gamma\,t}+4M^2\omega_0^2} \left[   -2qB_0M\omega_0^2e^{-2\Gamma\,t}-\dfrac{q^2B_0^2e^{-4\Gamma\,t}}{M}\sqrt{M\Delta-M^2\omega_0^2}       \right.\nonumber \\
&\left.    +\left(4M^2\omega_0^2+2qB_0e^{-2\Gamma\,t}\sqrt{M\Delta-M^2\omega_0^2} \right)\sqrt{\dfrac{q^2B_0^2\sigma e^{-4\Gamma\,t}}{4M}+\omega_0^2(M\sigma-1)}                \right]+n~\sqrt{\dfrac{\Delta}{M}-\omega_0^2} ~. \label{exp1EN3}   
\end{align}
The time evolution of the energy expectation value in Case III closely resembles that of Case I, as depicted in Fig.\ref{MDHO-fig1}. However, upon closer inspection, it is evident that in Case III, the energy decays at a faster rate compared to Case I, for the same set of parameters. It must be because in Case III, unlike in Case I, the applied field is decaying as well with respect to time. Furthermore, the maximum allowable time before the system becomes non physical (where the energy becomes non real) is half of that obtained by Eqn.(\ref{exp1tim1}). Here the bound is 
found to be
\begin{equation}
t\leq\,\dfrac{1}{4\Gamma}ln \dfrac{q^2B_0^2\sigma}{4M\omega_0^2(1-M\sigma)}~.\label{exp1tim3}
\end{equation}
The above study suggests an important inference that when an external magnetic field is present, the energy of a damped oscillator typically decreases over time if the field either diminishes over time or remains constant. However, if the field grows at a rate comparable to the decay of the damping factor, the oscillator's energy tends to remain constant. Figure \ref{MDHO-fig1} illustrates this scenario when the applied field is turned off, demonstrating that the energy remains constant when $B=0$, consistent with the expressions for energy expectation values in Eqn.(s)[\ref{exp1EN1}, \ref{exp1EN2}, and \ref{exp1EN3}]. Even a constant magnetic field is able to bring about time 
variation in this energy value.
\begin{figure}[H]
\centering
\includegraphics[scale=0.4]{Exp_case.eps}
\caption{\textit{Graph of the exponential energy expectation value (divided by $\omega_0$ to eliminate the dimension), with respect to $\Gamma$t (also a dimensionless parameter). The numerical value (in natural units) of the constants are chosen as $M, q, \mu, \Gamma\,=\,1$, $B_0\,=\,10^2$, $\omega_0\,=\,10^3$, $\Delta, \sigma\,=\,10^7$ and the quantum state is fixed as $n=1, m=0$. The expectation value of 
energy $\langle E \rangle$ is computed for case I: $B(t)=B_0$, $\omega(t)=\omega_0$ and $f(t)=e^{-\Gamma\,t}$; case II: $B(t)=B_0\,e^{\Gamma\,t}$, $\omega(t)=\omega_0$ and $f(t)=e^{-\Gamma\,t}$; case III: $\omega(t)=\omega_0$, $B(t)=B_0e^{-\Gamma\,t}$ and $f(t)=e^{-\Gamma\,t}$, case IV: $B(t)=B_0e^{\Gamma\,t}$, $\omega(t)=\omega_0e^{-\Gamma\,t/2}$ and $f(t)=e^{-\Gamma\,t}$. For cases I and III, the energy starts to decay at first but ultimately becomes constant over time. For case II, the energy always stays constant over time. However, for case IV, the dynamics of energy are observed to be very interesting. Although it decays at first, it then starts to grow over time. The corresponding scenarios in the absence of the magnetic field are also plotted for cases I and IV for comparison purposes. For the cases where the angular frequency is set to be constant (cases I, II, and III), the energy expectation value also becomes constant when the external field is switched off. Thus, the external field, both in time varying and constant forms, influences the time evolution of the energy expectation value of a damped oscillator with a constant frequency. However, for case IV, where the angular frequency is a decaying function, the energy does not exhibit its expanding nature after removing the external field, which is an expanding function with time. Next, in case I, we remove damping by considering $f(t)=1$. This also results in a constant energy value. This energetics actually match those in case II, where the damping effect is nullified in the presence of an expanding magnetic field.}}  
\label{MDHO-fig1}
\end{figure}
\vskip .20cm
\noindent{{\bf{$\langle D\rangle$ Set-I, Case IV}}}\\
\noindent 
 Here we set $f(t)=e^{-\Gamma\,t}$ , $\omega(t)=\omega_0e^{-\Gamma\,t/2}$ and  $B(t)=B_0e^{\Gamma\,t}$. With this the energy expression for the ground state takes the form,
\begin{align}
\braket{E_{n,-n}(t)}&=(n+1)\mu^2\Delta+\dfrac{n}{q^2B_0^2+4M^2\omega_0^2e^{-\Gamma\,t}} \left[     -2qB_0 M\omega_0^2e^{-\Gamma\,t}-\dfrac{q^2B_0^2}{M}\sqrt{M\Delta-M^2\omega_0^2e^{-\Gamma\,t}}               \right.\nonumber \\
&\left.+\left(4M^2\omega_0^2e^{-\Gamma\,t}+2qB_0\sqrt{M\Delta-M^2\omega_0^2e^{-\Gamma\,t}} \right)\sqrt{\dfrac{q^2B_0^2\sigma}{4M}+\omega_0^2e^{-\Gamma\,t}(M\sigma-1)}  \right]\nonumber\\    &+n\sqrt{\dfrac{\Delta}{M}-\omega_0^2e^{-\Gamma\,t}}~. \label{exp2En}
\end{align}  
The energy expectation value described above follows a decaying trend in the absence of a magnetic field, as demonstrated in \cite{SG1}. For the parameters depicted in Fig.\ref{MDHO-fig1}, the energy initially decreases, then rises, and ultimately stabilizes as $t \rightarrow \infty$. Notably, if the applied magnetic field is switched off in this scenario, the energy, as depicted in Fig.\ref{MDHO-fig1}, decays over time without exhibiting the earlier observed exponential growth. Consequently, the oscillator's energy remains notably lower in the absence of this exponentially growing field within the studied time interval. The system also possess two lower bounds of time below which it becomes non-physical due to the imaginary value of energy.~The conditions are as follows,              
\begin{eqnarray}
t~ \geq~\dfrac{1}{\Gamma}ln\left[\dfrac{M\omega_0^2}{\Delta} \right]~;~
t~\geq~\dfrac{1}{\Gamma}ln\left[\dfrac{4M\omega_0^2(1-M\sigma)}{q^2B_0^2\sigma} \right]~.
\end{eqnarray}
The greater of the two bounds serves as the actual lower bound. 

\subsection{Energy profile for Solution Set-II : Rationally decaying solution}
In the preceding section, we examined two distinct solution sets derived from Eqn.(\ref{MDHO-EPsoln2}). The solution presented in Eqn.(\ref{2rat}) arises directly by substituting $k=2$ into Eqn.(\ref{MDHO-EPsoln2}), while the solution depicted in Eqn.(\ref{ratsol}) is obtained by substituting $k=-2$ into a modified version of Eqn.(\ref{MDHO-EPsoln2}).
\subsubsection{Set-II, Case I}
For the rationally decaying solution given by Eqn.(\ref{2rat}), the energy expectation value
takes the following form 
\begin{equation}
\braket{E_{n,m-n}(t)}=\dfrac{(n+m+1)}{2(\Gamma\,t+\chi)}\left[2\left(\dfrac{\sigma}{\mu^2} +\Delta\mu^2\right)+\dfrac{\mu^2\Gamma^2}{8\sigma} \right]+(n-m)c(t)~.\label{rat1EN} 
\end{equation}
where we used the constraint relation given by Eqn.(\ref{MDHO-EPreln2}).
Here we set $f(t)=1$ , $\omega(t)=\omega_0/(\Gamma\,t+\chi)$ and  $B(t)=B_0/(\Gamma\,t+\chi)$. With this the energy expression for the ground state takes the form,
\begin{align}
&\braket{E_{n,-n}(t)}=\dfrac{(n+1)}{2(\Gamma\,t+\chi)}\left[2\left(\dfrac{\sigma}{\mu^2} +\Delta\mu^2\right)+\dfrac{\mu^2\Gamma^2}{8\sigma} \right]+n\sqrt{\dfrac{\Delta}{M}-\dfrac{\omega_0^2}{(\Gamma\,t+\chi)^2}}\nonumber\\
&+\dfrac{n}{4M^2\omega^2_0+q^2B_0^2}\left[
-\dfrac{2qB_0M\omega_0^2}{\Gamma\,t+\chi}-\dfrac{q^2B_0^2\sqrt{M\Delta(\Gamma\,t+\chi)^2-M^2\omega_0^2}}{M(\Gamma\,t+\chi)}\right.
\nonumber\\
&\left.+\left( \dfrac{4M^2\omega_0^2}{(\Gamma\,t+\chi)^2}+\dfrac{2qB_0\sqrt{M\Delta(\Gamma\,t+\chi)^2-M^2\omega_0^2}}{(\Gamma\,t+\chi)^2}\right)
\sqrt{ \dfrac{q^2B_0^2\sigma}{M}-\omega_0^2 (\Gamma\,t+\chi)^2 +4\omega_0^2\sigma M   }  \right]\label{rat1EN2}~;
\end{align}
which is a decaying function of time and is seen to reduce to the same obtained in \cite{SG1} in the limit $B\rightarrow\,0$.
The time range beyond which the system becomes non-physical due to imaginary energy expectation value is as follows,
\begin{equation}
\dfrac{1}{\Gamma}\left(\omega_0\sqrt{\dfrac{M}{\Delta}}-\chi \right)\leq~t\leq~\dfrac{1}{\Gamma}\left[\sqrt{\dfrac{q^2B_0^2\sigma}{M\omega_0^2}+4\sigma M} -\chi\right]~.\label{rat1tim2}
\end{equation}
In Figure \ref{MDHO-fig2}, we conducted a comparative analysis of how the energy varies over time under two conditions: with and without an applied magnetic field. The results indicate that in both scenarios, the energy diminishes over time. However, with the magnetic field applied, the oscillator's energy is notably higher due to the additional magnetic energy present in the system.
\begin{figure}[H]
\centering
\includegraphics[scale=0.4]{Rat.eps}
\caption{\textit{Graph of the rational energy expectation value (divided by $\omega_0$ to eliminate the dimension), with respect to $\Gamma$t (also a dimensionless parameter). The numerical value (in natural units) of the constants are chosen as $M, q, \mu, \Gamma\,=\,1$, $B_0\,=\,10^{20}$, $\omega_0\,=\,10^3$, $\Delta, \sigma\,=\,10^7$ and the quantum state is fixed as $n=1, m=0$. The expectation value of 
energy $\langle E \rangle$ is computed for case I: $a(t)=\dfrac{4\sigma}{(\Gamma{t}+\chi)^{\,2}},  
\,\rho(t)=\left[\dfrac{2\mu^{\,2}}{\Gamma{t}+\chi}\right]^{1/2},\,b(t)\,=\,\Delta\,$ and $\langle B\rangle$ case II: $a=\sigma ~~,~~ \rho=\mu \sqrt{\Gamma\,t+\chi}~~,~~ b(t)=\dfrac{\Delta}{(\Gamma\,t+\chi)^2}$.  Although the energy in both cases decays rationally over time, the decay rate in case I is greater than in the other case. This is because, in case I, both parameters $a(t)$ and $\rho(t)$ are decaying functions, while in case II, only $b(t)$ is a decaying function and $\rho(t)$ increases. The corresponding scenarios in the absence of the magnetic field are plotted for both cases. The oscillator's energy also decays when the external field is switched off. Although the nature of the energy decay remains the same in both the presence and absence of the field, the energetics increase in value when the field is applied.}}  
\label{MDHO-fig2}
\end{figure}

\subsubsection{Set-II, Case II}

For the rationally decaying solution given by Eqn.(\ref{ratsol}), the energy expectation value takes the following form,

\begin{equation}
\braket{E_{n,m-n}(t)}= \dfrac{(m+n+1)\mu^2\Delta}{\Gamma\,t+\chi}+c(t)(n-m)~.\label{rat2EN}
\end{equation}
where we have used the constraint relation given by Eqn.(\ref{rat2const}).
Here we set $f(t)=1$ , $\omega(t)=\omega_0/(\Gamma\,t+\chi)$ and  $B(t)=B_0/(\Gamma\,t+\chi)$. With this the energy expression for the ground state takes the form,


\begin{align}
\braket{E_{n,-n}(t)}=\dfrac{(n+1)\mu^2\Delta}{\Gamma\,t+\chi}+\dfrac{n}{(4M^2\omega^2_0+q^2B_0^2)(\Gamma\,t+\chi)}
\left[-2qB_0M\omega_0^2-\dfrac{q^2B_0^2\sqrt{M\Delta-M^2\omega_0^2}}{M}
 \right.\nonumber\\ 
\left. +\left( 4M^2\omega_0^2+2qB_0\sqrt{M\Delta-M^2\omega_0^2}\right)
\sqrt{ \dfrac{q^2B_0^2\sigma}{4M}+\omega_0^2(M\sigma-1)  }
\right]+\dfrac{n}{(\Gamma\,t+\chi)}\sqrt{\dfrac{\Delta}{M}-\omega_0^2 }~.
\end{align}
According to Fig.\ref{MDHO-fig2}, the energy expectation value decreases over time. In fact, it approaches zero for large times, albeit at a slower rate compared to Case I. This discrepancy is anticipated because in Case II, one of the Hamiltonian parameters, $b(t)$ decreases over time while another parameter, $\rho(t)$ increases. Consequently, the overall energy decreases as $b(t)$ decreases faster (at a rate $\sim~t^{-2}$) than 
$\rho(t)$ increases (at a rate $\sim~t^{1/2}$). Upon turning off the field under these conditions, Fig.\ref{MDHO-fig2} shows that the oscillator's energy continues to decrease due to the diminishing angular frequency of the oscillator. However, in this scenario, the oscillator's energy diminishes specifically because the magnetic energy component is absent from the system.


\section{Summary}
In this chapter, our work focused on examining the impact of an external time dependent magnetic field on a two dimensional damped harmonic oscillator in NC space. Initially, we established the Hamiltonian for our system under a general magnetic field in NC space. Using Bopp-shift relations, we then transformed this Hamiltonian into commutative variables. Subsequently, we derived the exact solution for this time dependent system with an applied time-varying magnetic field by employing the well known Lewis invariant, which leads to a nonlinear differential equation known as the EP equation. To generate a class of exact solutions for this system, similar to the initial work \cite{SG1}, we consider two types of EP solutions, namely, exponentially decaying solutions and rationally decaying solutions. For each set of EP solutions, we design various model systems by selecting different explicit forms of the damping factor, angular frequency, and applied magnetic field. Consequently, we are able to obtain explicit, closed form solutions for various models. Next, we calculate the expectation value of the Hamiltonian. As expected, the expectation value of the energy changes over time. Similar to our previous work, we not only derived analytical expressions for the energy of various designed models but also graphically represented their energy profiles. In the exponentially varying scenario, an exponentially growing magnetic field can maintain the energy at a constant level despite damping, while a constant or exponentially decaying field causes the energy to decrease more rapidly due to damping. Interestingly, when an exponentially decaying frequency is present along with the damping factor, the system's behaviour under the influence of an exponentially growing field becomes even more interesting. Initially, the energy decreases over time due to the damping, but later it begins to increase under the influence of the growing field. In some cases, we identified time regions, both bounded and unbounded, where the energy cannot remain physical as the system becomes non hermitian. In the rationally varying scenario, even without the damping factor, a rationally decaying magnetic field combined with a rationally decaying angular frequency can eventually dampen the oscillator's energy. For Case I of the rational EP solution, the decaying oscillation cannot remain physical at zero energy due to the upper time bound. Conversely, for Case II, the energy physically dampens to zero over time. Additionally, we observe that at any given moment, the expectation value of energy is higher with the magnetic field present than when it is absent.

\chapter{Berry phase from Lewis phase for noncommutative periodic models}\label{paper3}
An adiabatic process involves the gradual change of external parameters \cite{born, grf}. Typically, such processes are characterized by two distinct time scales. One is the ``internal time," which relates to the motion of the system itself, while the other is the ``external time," describing the time range through which the system parameters achieve significant changes. As per the adiabatic processes, the ``external time" should be very larger than the ``internal time," indicating that  the Hamiltonian parameters vary very gradually with respect to time. In this regard, one can mention the motion of a simple harmonic oscillator (pendulum) placed on an platform oscillating gradually. The frequency of the oscillating platform is very much lesser than the same of the pendulum. For such adiabatically evolving quantum system, the quantum state of the initial Hamiltonian remains same after the Hamiltonian takes its final form after a certain time. In a simpler version, the system state does not change through out the adiabatic process. Nevertheless, a phase shift can be observed after the adiabatic evolution. As anticipated by M. V. Berry \cite{berry1, berry2}, the phase picked up by the system is not only dynamic, evolving over time, but also geometric, depending solely on the geometry of the path traversed by the system. Such geometric phase is popularly known as "Berry's phase" in literature and it can be measured physically \cite{grf}. Berry phase is relevant in a vast areas in modern physics, especially regarding the topic of topological insulators~\cite{TI1, TI2}. Although Berry was initially concerned that the path dependent Berry phase might be dominated by a bigger dynamic phase, later it was ensured that the appropriate experiments can be fabricated to identify the geometric part of the phase \cite{exp1, exp2}. In such experiments, the geometric phases acquired by the photons can be measured by the rotation of the polarization plane of light. If the plane of the polarized light rotates by an angle ``$\alpha$", the corresponding geometric phase measurement would be ``$\pm\alpha$" depending on the spin (parallel or antiparallel) of the photons. The experimental verification of the geometric phase acquisition by spins during adiabatic evolution is well documented, particularly with neutrons~\cite{exp3,exp4}. There are certain conditions of achieving a non zero Berry phase. The minimum numbers of the periodic parameters in a Hamiltonian must be two. Moreover, the eigenfunction of that Hamiltonian must be complex in nature. With these conditions in place, the periodic Hamiltonian can pick up a geometric phase if it adiabatically evolves around a closed path on its parameter space over a full time period $T$ \cite{anandan}. Nevertheless, this phase would be zero when the time reversal (TR) symmetry of the system is broken \cite{BDR}. In the theoretical perspective, after being introduced with the concept of geometric phase in adiabatically evolving systems, the physicists got interested in studying these phases for both classical and quantum systems~\cite{Shapere}. The studies regarding the generalized quantum harmonic oscillators which is capable to provide a geometric phase have been very well known over a long time \cite{Giavarini1, Giavarini2, Giavarini3}. In this regard, we wish to highlight an interesting work by Dittirich {\it{et}} al \cite{dit} where a generalized harmonic oscillator having a TR symmetry breaking scale invariant term in its Hamiltonian was studied for the purpose of deriving geometric phase from the system. To solve the Hamiltonian, they employed the Lewis-Riesenfeld technique~\cite{Lewis}. Under the adiabatic condition, they derived the form of geometric phase using that Lewis technique. Notably, the TR symmetry breaking scale invariant term can arise in the Hamiltonian when it is considered in a special NC framework which is proposed in \cite{spb} to study a geometric phase which has a pure NC origin. \\
\noindent In this chapter, we discuss our study \cite{SG3} on two periodic models of the quantum harmonic oscillator in NC space. We examine two distinct systems. System I's Hamiltonian includes an explicit term that is scale invariant and breaks TR symmetry. In contrast, system II lacks this term initially. Since, the system II is considered in the newly defined NC framework \cite{spb}, the geometric phase forming scale invariant term arises in the Hamiltonian after transforming it from the NC space to the standard quantum mechanical phase space. In system I, basically, we extend the work of \cite{dit} in a two dimensional NC space. Although system II's Hamiltonian was also studied in \cite{spb}, we solve it and deduce the Berry's geometric phase using the Lewis method of invariants, which is a very simple and elegant way to treat the time dependent Hamiltonians.\\
\noindent Our primary focus is on investigating the path dependent Berry phase in these systems. Previous studies on similar systems \cite{dit, spb} have predicted the presence of a non zero Berry phase when a scale invariant term is present in the Hamiltonian. In this discussion, by choosing various periodic form of the Hamiltonian parameters and the NC parameters, the explicit investigation of the geometric phases for such systems will be explored.

\section{Periodic model system in noncommutative space}
As previously mentioned, we analyse harmonic oscillators in a two dimensional NC space. We investigate two distinct types of systems. The first type incorporates an explicit scale invariant term in its original Hamiltonian and is considered in a time dependent NC space that adheres to the standard Bopp-shift relation. The second type, although lacking a scale invariant term in its original Hamiltonian, is considered a time dependent NC space defined by newly introduced coordinate mapping relations \cite{spb}. In both cases, the parameters of the Hamiltonians are periodic functions that depend on time.
\subsection{Hamiltonian of system-I : Generalized harmonic oscillator in noncommutative framework I}
The system we are examining consists of two dimensional generalized harmonic oscillators in a dynamic NC framework. Such a model of generalized harmonic oscillator placed in a standard quantum mechanical phase space in one dimension is a largely explored topic in literature \cite{cer}-\cite{gao}. 
\noindent Now, the system's Hamiltonian reads,
\begin{equation}
H(t)|_{NC}=P(t)({P_1}^2+{P_2}^2)+Q(t)({X_1}^2+{X_2}^2)+R(t)(X_1P_1+P_1X_1+X_2P_2+P_2X_2)~.\label{B1}
\end{equation}
Here $P(t),R(t)$ and $Q(t)$ denote the time varying Hamiltonian parameters. This system can be considered as generalized harmonic oscillators having the Hamiltonian coefficients $(P, Q, R)(t)$ those change slowly over time. The last term of the Hamiltonian mentioned above is a scale invariant term, appearing in scenarios where an external field interacts with a harmonic oscillator. For an example, when the gravitational waves interact with a harmonic oscillator \cite{GW1}, \cite{GW2}, \cite{GW3}, \cite{sen}, \cite{GW4} the system gains such a scale invariant term in its Hamiltonian. In this context, we would like to mention that, in a recent communication \cite{sen}, under Lewis-Riesenfeld framework, there has been significant research on the geometric phase arising from the periodic  model of two dimensional quantum harmonic oscillator interacting with gravitational waves. \\
\noindent Since the system is considered in a dynamic NC space, the commutator brackets among the canonical operators existing in the Hamiltonian read (including $\hbar$)
\begin{align}
[X_j,X_k]~=i\theta(t)\epsilon_{jk}\,,\, 
&[P_j,P_k]~=i\Omega(t)\epsilon_{jk}\,,\,[X_j,P_k]\,=i\,\hbar\,\left[1+\frac{\theta(t)\Omega(t)}{4}\right]\delta_{jk};\\
&(j,k=1,2~,~\epsilon_{jk}=-\epsilon_{kj}~,~\epsilon_{12}=1)\nonumber
\end{align}
where $\Omega(t)$ and $\theta(t)$, respectively, represent the dynamic NC parameters for momentum space 
and configuration space. On the other hand, the canonical operators $(x_i,p_i)$ in standard quantum mechanical phase space follow the commutator brackets $[x_j,p_k]=i\hbar\delta_{j, k}$, $[x_j,x_k]=0=[p_j,p_k]$; ($j, k= 1, 2$).\\
\noindent Next, the standard coordinate mapping relations those are the Bopp-shift relations \cite{mez} ($\hbar=1$) are to be implemented into the system for reconstructing the original Hamiltonian in terms of the standard phase space operators ($x_i, p_i$). Those coordinate mapping relations read,
\begin{eqnarray}
& X_1=x_1-\dfrac{\theta(t)}{2}p_2\,\,\,;\,\,\,X_2=x_2+\dfrac{\theta(t)}{2}p_1\\
& P_1=p_1+\dfrac{\Omega(t)}{2}x_2\,\,\,;\,\,\,P_2=p_2-\dfrac{\Omega(t)}{2}x_1 \,\,.
\label{Beqn1}
\end{eqnarray}
\noindent Therefore, the original Hamiltonian in terms of $(x_i,p_i)$ coordinate operators is revealed as
relation,
\begin{equation}
H=\dfrac{a(t)}{2}({p_1}^2+{p_2}^2)+\dfrac{b(t)}{2}({x_1}^2+{x_2}^2)+c(t)({x_2}{p_1}-{x_1}{p_2})+d(t)(x_1p_1+p_1x_1+x_2p_2+p_2x_2)\,\,\,.\label{Beqn2}
\end{equation}
The structural form of the Hamiltonian's parameters are derived as,
\begin{align}
&a(t)=2P(t)+\dfrac{\theta^2(t)}{2}\,Q(t)~,
b(t)=2\,Q(t)+\dfrac{P(t)\Omega^2(t)}{2}~;\nonumber \\
&c(t)=P(t)\Omega(t)+Q(t)\theta(t)~,
d(t)=\left(1-\dfrac{\Omega(t)\theta(t)}{4} \right)R(t)~. \label{B34a}
\end{align}
The explicit structure of the Hamiltonian parameters ($a(t), b(t),  c(t)$) align to those found in the previous communications \cite{Dey} and \cite{SG1} on the exploration of a simple and damped (respectively) harmonic oscillator with dynamic noncommutativity in two dimension. However, since our present study aims to explore the Berry's geometric phase acquired by the periodic Hamiltonian after the full time period $T$, we now assign these time dependent coefficients to have only the periodic functions as their explicit form. Additionally, the above Hamiltonian is different from the same treated in \cite{Dey} and \cite{SG1} because, unlike our present system, the earlier studies lack a scale invariant term in the Hamiltonian expressions. Our periodic Hamiltonian and the treatment performed on it are also very similar to \cite{dit} to derive the Berry's geometric phase for a generalized harmonic oscillator placed in a conventional phase space in one dimension. Consequently, $R(t)$, the Hamiltonian coefficient with the scale invariant term, leads to a latter Hamiltonian expression shown in Eqn.(\ref{Beqn2}), which gains a term associated with $d(t)$. Because of the removal of $R(t)$, the Hamiltonian reduces to the form studied in \cite{SG1}.

\subsection{Hamiltonian of system-II : Harmonic oscillator in noncommutative framework II}
In this section, we examine a slightly altered system within a different NC framework. The Hamiltonian for this system lacks the scale invariant term. Therefore, it takes a similar form to that 
considered in \cite{Dey, SG1}. Nevertheless, here, we chose a modified version of the standard NC framework (NC framework II) which is recently proposed in \cite{spb} to demonstrate that their newly proposed NC structure causes the emergence of Berry phase.

\noindent Now the system's Hamiltonian read
\begin{equation}
H(t)|_{NC}=P(t)(P_1^2+P_2^2)+Q(t)(X_1^2+X_2^2)~,\label{BHam2}
\end{equation}
where $P(t)$ and $Q(t)$ are arbitrary time dependent periodic coefficients of the Hamiltonian.\\
\noindent Since the system is considered in a dynamic NC space, the commutator brackets among the canonical operators existing in the Hamiltonian read (including $\hbar$) 
\begin{align}
[X_j,X_k]~=i\theta(t)\epsilon_{jk}\,,\, 
&[P_j,P_k]~=i\Omega(t)\epsilon_{jk}\,,\,[X_j,P_k]\,=i\,\hbar\,\delta_{jk};\\
&(j,k=1,2~,~\epsilon_{jk}=-\epsilon_{kj}~,~\epsilon_{12}=1)\nonumber
\end{align}
where $\Omega(t)$ and $\theta(t)$, respectively, represent the dynamic NC parameters for momentum space 
and configuration space. On the other hand, the canonical operators $(x_i,p_i)$ in standard quantum mechanical phase space follow the commutator brackets $[x_j,p_k]=i\hbar\delta_{j, k}$, $[x_j,x_k]=0=[p_j,p_k]$; ($j, k= 1, 2$).\\
\noindent Next, a generalized version of the standard coordinate mapping relations \cite{spb} ($\hbar=1$) are to be implemented into the system for reconstructing the original Hamiltonian in terms of the standard phase space operators ($x_i, p_i$). Those coordinate mapping relations read,
\begin{eqnarray}
&X_1=x_1-\dfrac{\theta(t)}{2}p_2+\dfrac{\sqrt{-\theta(t)\Omega(t)}}{2}x_2~,~X_2=x_2+\dfrac{\theta(t)}{2}p_1-\dfrac{\sqrt{-\theta(t)\Omega(t)}}{2}x_1~;\nonumber\\
&P_1=p_1+\dfrac{\Omega(t)}{2}x_2+\dfrac{\sqrt{-\theta(t)\Omega(t)}}{2}p_2~,~P_2=p_2-\dfrac{\Omega(t)}{2}x_1-\dfrac{\sqrt{-\theta(t)\Omega(t)}}{2}p_1~.
\end{eqnarray}
\noindent Therefore, the original Hamiltonian in terms of $(x_i,p_i)$ coordinate operators is revealed as
relation,
\begin{equation}
H=\dfrac{a(t)}{2}({p_1}^2+{p_2}^2)+\dfrac{b(t)}{2}({x_1}^2+{x_2}^2)+c(t)({x_2}{p_1}-{x_1}{p_2})+d(t)(x_1p_1+p_1x_1+x_2p_2+p_2x_2)\,\,\,.\label{Bham2}
\end{equation}
The structural form of the Hamiltonian's parameters are derived as,
\begin{align}
&a(t)=2\,P(t)\left(1-\dfrac{\theta(t)\Omega(t)}{4} \right)+Q(t)\dfrac{\theta^2(t)}{2}~,
b(t)=2\,Q(t)\left(1-\dfrac{\theta(t)\Omega(t)}{4} \right)+P(t)\dfrac{\Omega^2(t)}{2}
~,\nonumber \\
&c(t)=P(t)\Omega(t)+\theta(t)Q(t)
~,~
d(t)= \dfrac{\sqrt{-\theta(t)\Omega(t)}}{4}\left[\Omega(t)P(t)-\theta(t)Q(t) \right]                 ~.\label{B34aa}
\end{align}
It is noteworthy to mention that, $d(t)$, the Hamiltonian parameter
associated with the scale invariant term, is emerged because of the presence of noncommutativity, since, unlike our system I, it vanishes when $\Omega(t), \theta(t)$ tend to zero. Hence, it is evident that the geometric phase for the present system has a pure NC origin.\\
\noindent Before proceeding to solve the Hamiltonians (Eqn.(\ref{Beqn2}) and Eqn.(\ref{Bham2}), we also present its another convenient form which is as follows
\begin{equation}
H=\dfrac{a(t)}{2}\left[\left(\,p_1\,+\,2\,\dfrac{d}{a}x_1\right)^2\,+\,\left(\,p_2\,+\,2\,\dfrac{d}{a}x_2\right)^2\right]\,+\dfrac{\omega_p^2(t)}{2a}\left(\,x_1^2\,+\,x_2^2\,\right)\,+\,c(t)\left({p_1}{x_2}-{p_2}{x_1}\right)~;\label{Bham-anoth}
\end{equation}
where $\omega_p\,=\,\sqrt{a\,b-4\,d^2~}$ acts as the effective frequency of the system.
\section{Exact eigenstate of the model Hamiltonian}
As usual, like in our previous works \cite{SG1, SG2}, we employ the Lewis technique \cite{Lewis} to solve the Hamiltonians [Eqn.(\ref{Beqn2}) and Eqn.(\ref{Bham2})] representing both the systems (I and II). By solving the Lewis invariant, such that    
\begin{equation}
I(t)\phi(x_1,x_2)=\epsilon \phi(x_1,x_2)
\label{Beqnegn}
\end{equation}
where $\epsilon$ denotes the invariant's eigenvalue corresponding to its time dependent eigenstate $\phi(x_1,x_2)$, the eigenstate of $H(t)$ can be obtained using the following relation 
\begin{equation}
\psi(x_1,x_2,t)=e^{i\Theta(t)}\phi(x_1,x_2)~;
\label{Beqnpsi}
\end{equation}
where $\psi(x_1,x_2,t)$ is the eigenfunction of the Hamiltonian (Eqn.(\ref{ham})) and the real function $\Theta(t)$ is the Lewis phase factor, from which, as described in the previous chapter \ref{chap-LR} also, the geometric phase will be derived for our model systems.\\
\noindent Although the detailed procedure for solving a time dependent Hamiltonian using the Lewis method of invariant is covered in previous chapters \ref{chap-LR} and \ref{paper1}, we will present some important intermediate steps here before providing the final structure of the eigenfunction for the Hamiltonians considered in this study. This is because the Hamiltonian structures (Eqn.(\ref{Beqn2}) and Eqn.(\ref{Bham2})) in the present system models are more general than those in the previously discussed models.

\subsection{Ermakov-Lewis invariant system}
First, we assume an ansatz of the Lewis invariant operator which corresponds to the latter Hamiltonian [Eqn.(\ref{Beqn2}) and Eqn.(\ref{Bham2}). Conventionally, as we also discussed earlier, its structural format should be as follows,
\begin{equation}
I(t)=\alpha(t)({p_1}^2+{p_2}^2)+\beta(t)({x_1}^2+{x_2}^2)+\gamma(t)(x_1{p_1}+p_1{x_1}+x_2{p_2}+p_2{x_2}).
\label{Beqn5}
\end{equation}
Next, applying the property of invariance
\begin{equation}
\dfrac{dI}{dt}=\partial_t{I}+\dfrac{1}{i}[I,H]=0
\label{Beqn4}
\end{equation}
on the considered ansatz, we have the following set of equations hold by its time dependent coefficients.
\begin{eqnarray}
\dot{\alpha}(t)&=&4\alpha(t)\,d(t)-2a(t)\gamma(t)~,\label{Beqn6}\\
\dot{\beta}(t)&=&2b(t)\gamma(t)-4\beta(t)\,d(t)~,\label{Beqn7}\\
\dot{\gamma}(t)&=&\,b(t)\alpha(t)-\beta(t)a(t)\,~.
\label{Beqn8}
\end{eqnarray}
Parametrizing $\alpha\,=\,\rho^2$ in the above relations, we have the parametrized form of those above relations
\begin{eqnarray}
\gamma(t)&=&\dfrac{\rho}{a}(2\rho\,d-\dot{\rho})~,\label{Beqn9}\\
\beta(t)&=&\dfrac{(\dot{\rho}-2\rho\,d)^2}{a^2}+\dfrac{1}{\rho^2}~;\label{Beqn10}
\end{eqnarray}
which are linked to the following non linear differential equation
\begin{equation}
\ddot{\rho}-\dfrac{\dot{a}}{a}\dot{\rho}+\rho\,\left(ab-2\dot{d}-4d^2+2\dfrac{\dot{a}}{a}d\right)\,=\,\dfrac{a^2}{\rho^3}~,
\label{Beqn11}
\end{equation} 
known as the Ermakov-Pinney equation, which is derived by setting the integration constant to be unity. The derived EP equation resembles the form presented by \cite{dit}, \cite{cer}-\cite{gao} for exploring a time dependent generalized harmonic oscillator situated in the conventional quantum mechanical phase space. It is obvious that Eqn.(\ref{Beqn11}) simplifies to the equation obtained in \cite{Dey} and \cite{SG1,SG2} when $d(t)$, the Hamiltonian coefficient is fixed to be zero.\\
\noindent Next, with the parametrized form of $\alpha$, $\beta$ and $\gamma$, the invariant (Eqn.(\ref{Beqn5})) in terms of $\rho$ is expressed as  
\begin{equation}
I(t)=\rho^2({p_1}^2+{p_2}^2)+\left[\dfrac{(\dot{\rho}-2\rho\,d)^2}{a^2}+\dfrac{\xi^2}{\rho^2}\right]({x_1}^2+{x_2}^2)-\dfrac{\rho}{a}(\dot{\rho}-2\rho\,d)(x_1{p_1}+p_1{x_1}+x_2{p_2}+p_2{x_2})~; \label{Beqn12}
\end{equation}
which aligns with the one found in \cite{dit} for a time dependent generalized harmonic oscillator situated in the conventional quantum mechanical phase space. At the limit $d(t)\rightarrow\,0$, the above expression becomes identical to the same derived in \cite{Dey} and \cite{SG1, SG2}.




\subsection{Soultion of the Ermakov-Lewis invariant system and Lewis phase factor}
Following the method demonstrated in the portion \ref{sec-lad} from the previous chapter \ref{paper1}, we first define the unitary operators as 
\begin{eqnarray}
\hat{U}=exp\left[\dfrac{i(2\rho\,d-\dot{\rho})}{2a(t)\rho}({x_1}^2+{x_2}^2)\right],\,\,\,
\hat{U^{\dagger}}\hat{U}=\hat{U}\hat{U^{\dagger}}=\textbf{I}.
\label{Beqn13}
\end{eqnarray}
Next, we briefly describe the remaining procedure, which is already discussed in sections \ref{sec-lad}, \ref{sec-pol}, and \ref{sec-sol} under chapter \ref{paper1}, and mention the important results modified due to $d(t)$, the additional parameter that appears in the Hamiltonian (Eqn.(s) (\ref{Beqn2}) and (\ref{Bham2})) of our interest.\\
\noindent The invariant $I(t)$, when subjected to the transformation $I^{'}(t)=\hat{U}I\hat{U^\dagger}$, gains the Hamiltonian structure of a simple harmonic oscillator. Consequently, the ladder operators can be defined in the standard manner. Upon applying the reverse transformation, they have the following form,
\begin{eqnarray}
\hat{a_j}(t)&=&\dfrac{1}{\sqrt{2}}\left[\dfrac{1}{\rho}x_j+i\rho{p_j}+\dfrac{i(2\rho\,d-\dot{\rho})}{a}x_j\right]~,\\
\label{Beqn17}
\hat{{a_j}^{\dagger}}(t)&=&\dfrac{1}{\sqrt{2}}\left[\dfrac{1}{\rho}x_j-i\rho{p_j}-\dfrac{i(2\rho\,d-\dot{\rho})}{a}x_j\right]~;
\label{Beqn18} 
\end{eqnarray}
with the commutation relation $[{{\hat{a}_i}},{{\hat{a}_j}^{\dagger}}]=\delta_{ij}$. \\
\noindent We can quickly derive the final set of operators to solve the invariant in the polar coordinate system by linearly combining the above operators as shown below
\begin{align}
\hat{a}(t)&=-\dfrac{i}{\sqrt{2}}(\hat{a}_1-i\hat{a}_2)=\dfrac{1}{2}\left[\left(\rho{p_r}+\dfrac{2\rho\,d-\dot{\rho}}{a(t)}r \right)-i\left(\dfrac{r}{\rho}+\dfrac{\rho{p_\theta}}{r}+\dfrac{\rho}{2r} \right)    \right]e^{-i\theta}~,\nonumber\\
{\hat{a}}^{\dagger}(t)&=\dfrac{i}{\sqrt{2}}({\hat{a}_1}^\dagger+i{\hat{a}_2}^\dagger)=\dfrac{1}{2}e^{i\theta}\left[\left(\rho{p_r}+\dfrac{2\rho\,d-\dot{\rho}}{a(t)}r \right)+i\left(\dfrac{r}{\rho}+\dfrac{\rho{p_\theta}}{r}+\dfrac{\rho}{2r} \right) \right]~.
\label{Beqn27}
\end{align}
It is evident from the above equation that the ladder operators prepared in this study have been modified by the parameter $d(t)$. When $d(t)$ is set to zero, these operators revert to those derived in the previous model \cite{SG1}.\\
The Lewis invariant $I(t)$ can also be presented in terms of the polar coordinates as
\begin{eqnarray}
I(t)=\dfrac{1}{\rho^2}r^2+\left(\rho{p_r}-\dfrac{\dot{\rho}-2\rho\,d}{a}r\right)^2+\left({\dfrac{\rho{p_\theta}}{r}}\right)^2-\left({\dfrac{\rho}{2r}}\right)^2
~,
\label{Beqn26}
\end{eqnarray}
\noindent Again, starting from the existing form of the invariant $I(t)$ in Eqn.(\ref{Beqn26}), a new and appropriate form of the invariant $I^{'}(t)$ is to be derived, as done in \cite{Dey, SG1}, in terms of the ladder operator, such that, 
\begin{equation}
I^{'}(t)=\dfrac{I}{4}-\dfrac{p_\theta}{2}=\left(\hat{a}^{\dagger}\hat{a}+\dfrac{1}{2}\right)-p_\theta~.
\end{equation}
Next, we redo the procedure demonstrated in the portion \ref{sec-sol} from the chapter \ref{paper1} and the eigenfunction of the invariant $I^{'}(t)$ is found to be
\begin{eqnarray}
\phi_{n,m-n}(r,\theta)=\lambda_{n}(t)\dfrac{{(i\rho)}^m}{\sqrt{m!}}r^{n-m}e^{i(m-n)\theta-\dfrac{a-i\rho(\dot{\rho}-2\rho\,d)}{2a
{\rho}^2}r^2}U\left(-m,1-m+n,\dfrac{r^2}{\rho^2} \right)~;
\label{Beqn28}
\end{eqnarray}
where $n$ and $m$ are positive integers constrained by $n+l\,=m\,\geqslant\,0$ and the normalization factor $\lambda_n(t)$ is given by
\begin{eqnarray}
\lambda_n^2(t)=\dfrac{1}{\pi{n!}{(\rho^2)}^{1+n}}~;
\label{eqn28lam}
\end{eqnarray}
and $U\left(-m,1-m+n,\dfrac{r^2}{\rho^2} \right)$ presents
Tricomi's confluent hypergeometric function \cite{Arfken, uva}. It is noteworthy to state that the exact analytical form of the eigenfunction of the Lewis invariant corresponding to our Hamiltonian (Eqn.(\ref{Beqn2}) and Eqn.(\ref{Bham2}), which have a more general form than those considered in \cite{Dey, SG1, SG2}), appears with an additional Hamiltonian coefficient $d(t)$, and reduces to the one found for the models in \cite{Dey, SG1, SG2} when $d(t)$ equals zero.\\
\noindent As usual, the Hamiltonian eigenfunction holds the following orthonormal relation,
\begin{equation}
\int_0^{2\pi}d\theta\int_0^{\infty}rdr\phi^{*}_{n,m-n}(r,\theta)\phi_{n^{'},m^{'}-n^{'}}(r,\theta)=\delta_{nn^{'}}\delta_{mm^{'}}.
\label{Beqn29}
\end{equation}
Now, we proceed to have the Lewis phase which plays very crucial role in this work. It is crucial not only for obtaining the eigensystem of the Hamiltonian, as demonstrated in our earlier works~\cite{SG1, SG2}, but also for determining the geometric phase form that a system acquires after a full time period $T$. \\
\noindent The Lewis phase form matches that found in \cite{Dey, SG1, SG2}. It is presented as,
\begin{equation}
\Theta_{n,l}(t)=(n+l)\int_0^{t}{\left[c(\tau)-\dfrac{a(\tau)}{\rho^2(\tau)}\right] d\tau}~.\label{Beqn31}
\end{equation}
It is important to note that, as we discussed in chapter \ref{chap-LR} while deriving the geometric phase for a one dimensional generalized harmonic oscillator model, the parameter $d(t)$ which is crucial for forming a Berry phase is not explicitly present in the phase factor. However, by utilizing the EP equation, we can incorporate this factor into the phase (Eqn.(\ref{Beqn31})). This allows us to proceed to the essential part of our work, deriving Berry's geometric phase from the adiabatically approximated form of the Lewis phase.

\section{Berry's geometric phase from Lewis phase}
First, we revisit the portion \ref{sec-berry} (in the chapter \ref{chap-LR}), where we derived the Berry phase from the Lewis phase of a one dimensional generalized harmonic oscillator. Given that our Hamiltonian is more generalized than the one previously considered, this section will detail the procedure specific to our current study.\\
We start with the rearranged form of the EP equation as given in Equation (\ref{Beqn11}). We rearrange the equation as
\begin{align}
\dfrac{1}{\rho}\dfrac{d}{dt}\left(\dfrac{\dot{\rho}}{a} \right)-\left[2\dfrac{d}{dt}\left(\dfrac{d}{a} \right)-\left(\dfrac{ab-4d^2}{a} \right)+\dfrac{a}{\rho^4} \right]=0~.\label{BEP2}
\end{align}
Substituting $\dfrac{a}{\rho^2}$ from the above equation into the Lewis phase form, it is possible to bring the factor $d(t)$ inside it. But for the sake of simplification of the calculation, we shall do it after enforcing adiabatic approximation on the system.
\subsection{Lewis Phase Factor under adiabatic approximation}
We begin by assuming that Hamiltonian is varying adiabatically, meaning the system parameters are changing gradually. Next, the Lewis phase factor is to be investigated under the condition while varying the Hamiltonian adiabatically, adopting the procedure demonstrated in \cite{dit}.\\
In order to include the adiabatic effect, we assume that, in the parameter space, both the Hamiltonian parameters and the EP parameter evolve gradually over a full time period $T$, such that $[a,b,c,d, \rho]\,(0)=[a,b,c,d, \rho]\,(T)$.
As an adiabatic effect, it is valid to ignore the variation rate of second order and the product of the variations having first order. Hence, we drop out the term including $\dot{\rho}$ from the equation mentioned above.
The EP equation (Eqn.(\ref{BEP2})) is now approximated as,
\begin{align}
\left[2\dfrac{d}{dt}\left(\dfrac{d}{a} \right)-\left(\dfrac
{ab-4d^2}{a} \right)+\dfrac{a}{\rho^4} \right]=0~.\label{BEPapprox}
\end{align}
At this moment, the value of $a/\rho^2$ stands as
\begin{align}
\dfrac{a}{\rho^2}~\approx~\omega_p\sqrt{ 1-\dfrac{2\,a}{{\omega_p}^2}\dfrac{d}{dt}\left(\dfrac{d}{a} \right) }~;
\end{align}
where $\omega_p$ is given by,
\begin{eqnarray}
\omega_p=\sqrt{ab-4d^2}.
\label{omg}
\end{eqnarray}
Next, we impose the binomial approximation under the following condition,
\begin{equation}
\dfrac{d}{dt}\left(\dfrac{d}{a}\right)\,<<\,\dfrac{\omega_p^2}{2a}~;
\end{equation}
which is consistent with the adiabatic condition, as shown by the Hamiltonian structure in Eqn.(\ref{Bham-anoth}), indicating that the rate of change of the external influence is negligible compared to the effective frequency of the system.
\noindent Under 
binomial approximation, $a/\rho^2$ takes on the following form,
\begin{align}
\dfrac{a}{\rho^2}~\approx~\omega_p\left[ 1-\dfrac{a}{{\omega_p}^2}\dfrac{d}{dt}\left(\dfrac{d}{a} \right) \right]\label{Bapprox}~;
\end{align}
So, the adiabatically approximated expression of the Lewis phase is given by, 
\begin{align}
\Theta_{n,l} (t)=&(n+l)\int_0^t \left[c-\dfrac{a}{\rho^2}\right]~d\tau\nonumber\\
\approx&(n+l)\int_0^t\left[c-\omega_p\left\lbrace 1-\dfrac{a}{{\omega_p}^2}\dfrac{d}{d\tau}\left(\dfrac{d}{a} \right) \right\rbrace\right]~d\tau~.\label{Blewisad}
\end{align}
The structure of the approximated phase factor resembles that derived in \cite{dit}. Nevertheless, due to our system being situated in a NC space, there is an extra component that involves the Hamiltonian coefficient $c(t)$. This component of the phase is purely dynamic and continues to evolve over time.

\subsection{Derivation of the geometric Phase}
We assume that the Hamiltonian which evolves adiabatically in its parameter space $\mathcal{R}$, undergoes a cycle through a time period $T$. Therefore,
\begin{eqnarray}
{\mathcal{R}} (0) = \mathcal{R} (T)~~~;~~~\mathcal{R}=(a,b,c,d)~;
\end{eqnarray}
For deriving the geometric phase expression , we substitute the following relation in Eqn.(\ref{Blewisad}),
\begin{equation}
\dfrac{d}{dt}=\dfrac{d\mathcal{R}}{dt}\nabla_{\mathcal{R}}.
\label{Re} 
\end{equation}
Substituting it in Eqn.(\ref{Blewisad}), we get the geometric part of the phase, i.e. the Berry 
phase as,
\begin{align}
\Theta^G_{n,l}=(n+l)\oint^{\mathcal{R}} \dfrac{a}{\omega_p}\,\overrightarrow{\nabla}_{\mathcal{R}}\left(\dfrac{d}{a} \right).d \overrightarrow{\mathcal{R}}~.\label{berry}
\end{align}
It is important to note that the derived phase is exactly similar to that presented in \cite{dit}, which examines a time dependent generalized harmonic oscillator situated in a standard quantum mechanical phase space in one dimension. Additionally, \cite{spb} provides another consistent expression for the geometric phase, achieved through the application of a specific gauge transformation.
\begin{equation}
\dfrac{a}{\omega_p}d\left(\dfrac{d}{a}\right)=-\dfrac{b}{\omega_p}d\left(\dfrac{d}{b}\right)-d\left[tan^{-1}\sqrt{\dfrac{ab}{4d^2}-1} \right]=-\dfrac{b}{\omega_p}d\left(\dfrac{d}{b}\right)-d\left[tan^{-1}\left(\dfrac{\omega_p}{2d}\right) \right]~.\label{gauge}
\end{equation}
Therefore, another form of Eqn.(\ref{berry}) is given by ,
\begin{align}
\Theta^G_{n,l}=-(n+l)\oint^{\mathcal{R}} \dfrac{b}{\omega_p}\,\overrightarrow{\nabla}_{\mathcal{R}}\left(\dfrac{d}{b} \right).d \overrightarrow{\mathcal{R}}~.\label{berryb}
\end{align}



\section{Berry phases for various choices of Hamiltonian parameters }
We now aim to investigate the explicit existence of the Berry phase by choosing various periodic functional form of the Hamiltonian parameters. Most of the earlier communications investigating the Berry phase for the harmonic oscillators have only derived a generic form of the geometric phase, without calculating the exact phase value for the system in question. Integrating Eqn.[\ref{berry}] and computing the exact phase value for a periodic set of Hamiltonian parameters is crucial to verify whether a non trivial geometric phase is acquired by the periodic system as it undergoes a cycle on the parameter space. In order to have the Berry phase in an explicit form, the periodic functional form of the original Hamiltonian (before employing the coordinate mapping relations) parameters $P(t)$, $Q(t)$, $R(t)$ and those of the NC parameters are to be chosen at first. 
Next, by using Eqn.(s)(\ref{B34a} and \ref{B34aa}), $a$, $b$, $c$, and $d$, the coefficients of the latter (after employing the coordinate mapping relations) Hamiltonians for both the systems (I and II) are to be determined explicitly. Finally, as per Eqn.(\ref{berry}), the calculated periodic parameters ($a$, $b$, $c$, $d$) corresponding to the chosen periodic parameters ($P$, $Q$, $R$, $\theta$, $\Omega$ ) result in an explicit value of the Berry's geometric phase.\\
\noindent It is also important to note that, unlike in our previous works \cite{SG1, SG2}, the chosen forms of the Hamiltonian parameters do not need to satisfy the exact form of the EP equation in Eqn.(\ref{Beqn11}), as that equation is now approximated in Eqn.(\ref{Bapprox}).
\subsection{Explicit investigation of Berry phase : system I}
Here, for our first system which includes a scale invariant term in the Hamiltonian from the beginning and hence the presence of the considered NC (standard) framework is not essential to generate a Berry phase, we intend to investigate a finite , non zero value of Berry phase explicitly. 
\subsubsection{Case I : Non trivial geometric phase without noncommutativity}
In this scenario, the noncommutativity is removed by assigning the explicit value of the NC parameters as zero. Hence, the parameter space in which the periodic system evolves exists in a conventional quantum phase space. The coefficients of the original Hamiltonian Eqn.(\ref{B1}) are set to hold the following forms ,
\begin{eqnarray}
P(t)=\dfrac{1}{M}=constant~,~Q(t)=\dfrac{1}{2}M\omega_0^2=constant~,~ R(t)=\dfrac{\omega_0}{\sqrt{2}}\,cos\,(ft)~.\label{BCaseI1}
\end{eqnarray}
The NC parameters are
\begin{eqnarray}
\theta(t)=0,~\Omega(t)=0~.\label{BCaseI2}
\end{eqnarray}
It demonstrates a model persisting a periodically varying coefficient of the scale invariant term in its Hamiltonian. Since, this is a standard quantum mechanical scenario, it can be designed in the laboratory.\\
\noindent 
By using Eqn.(\ref{B34a}), the above chosen values lead to calculate the following (latter) Hamiltonian parameters as
\begin{eqnarray}
a(t)=\dfrac{2}{M}   ~,~ b(t)=M\omega_0^2 ~,~
c(t)=0 ~,~ d(t)=\dfrac{\omega_0}{\sqrt{2}}\,cos(ft)~.\label{BCaseI3}
\end{eqnarray}
Finally, by using Eqn.(\ref{berry}), the above calculated periodic forms result in the following value of Berry phase as
\begin{eqnarray}
\Theta^G_{n,l}&=&(n+l)\oint^{\mathcal{R}} \dfrac{a}{\omega_p}\,\overrightarrow{\nabla}_{\mathcal{R}}\left(\dfrac{d}{a} \right).d \overrightarrow{\mathcal{R}} \\
&=&(n+l)\int^{2\pi/f}_0 \dfrac{a}{\omega_p}\dfrac{d}{dt}\left(\dfrac{d}{a} \right) dt \\
&=& -(n+l)\int^{2\pi/f}_0 \dfrac{f}{2} dt \\
&=& -(n+l)\pi~.
\end{eqnarray}
Again, by using Eqn.(\ref{berryb}), the phase emerges to be
\begin{align}
\Theta^G_{n,l}=-(n+l)\oint^{\mathcal{R}} \dfrac{b}{\omega_p}\,\overrightarrow{\nabla}_{\mathcal{R}}\left(\dfrac{d}{b} \right).d \overrightarrow{\mathcal{R}}=(n+l)\pi~.
\end{align}
However, the equivalence between both the values of $\Theta^G_{n,l}$ can easily be verified in the following manner,
\begin{align}
e^{i\Theta^G_{n,l}}=e^{-i(n+l)\pi}=e^{i(n+l)\pi}=cos[(n+l)\pi].
\end{align}
Therefore, because of the presence of a scale invariant term in the original Hamiltonian, a non trivial Berry phase is achieved  in the standard quantum mechanical framework. \\
\noindent Additionally, the emergence of a non zero Berry phase can also be analysed through a graphical representation. We aim to visualize how the finite phase is picked up by the system. To illustrate it, we require the integral expression of the above phase, which is found to be
\begin{eqnarray}
I_{\Theta^G_{n,l}}(t)&=&(n+l)\int^{\mathcal{R}} \dfrac{a}{\omega_p}\overrightarrow{\nabla}_{\mathcal{R}}\left(\dfrac{d}{a} \right).d \overrightarrow{\mathcal{R}} \label{ind}\\
&=&(n+l)\int^t_0 \dfrac{a}{\omega_p}\dfrac{d}{dt}\left(\dfrac{d}{a} \right) dt \\
&=&-(n+l)\dfrac{ft}{2}~.
\label{g1}
\end{eqnarray}
As seen from Fig.(\ref{Bfig1}), the dynamic nature of $I_{\Theta^G_{n,l}}$ is aperiodic. This causes the Hamiltonian to acquire a non zero geometric phase after a time period $T$. As expected, after the ending of each cycle, the phase remains constant.

\subsubsection{Case II : Non trivial geometric phase with momentum  noncommutativity}
\begin{figure}[t]
\centering
\includegraphics[scale=0.4]{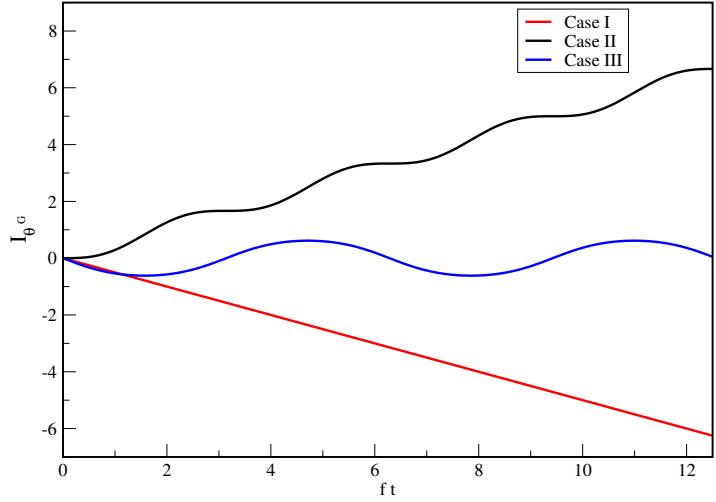}
\caption{\textit{Plot of the evolution of the integral expression ($I_{\Theta^G_{n,l}}$) associated with the derived geometric phase, with respect to $ft$ (dimensionless quantity), where $f$ denotes the frequency of the periodic parameters. The value of the constants (in natural units) related to the periodic parameters are chosen as $l\,=\,0$; $n\,=\,1$; $M, \omega_0\,=\,1$; $\theta_0, \Omega_0\,=\,2$. Case I deals with the system I situated in the standard quantum mechanical phase space and  case II, III deal with the same system with dynamic noncommutativity. While case I and Case II show an aperiodic nature of   
$I_{\Theta^G_{n,l}}$ with respect to time, in case III, the nature of $I_{\Theta^G_{n,l}}$ is purely periodic which is consistent with the vanishing of the geometric phase.}}  
\label{Bfig1}
\end{figure}

Here we remove the spatial noncommutativity and consider the system in such a framework where the momentum operators do not commute. The coefficients of the original Hamiltonian Eqn.(\ref{B1}) are set to hold the following forms ,
\begin{eqnarray}
P(t)=\dfrac{1}{M}=constant~,~Q(t)=\dfrac{1}{2}M[\omega_0\,sin^3(ft)]^2~,~ R(t)=\dfrac{\omega_0}{\sqrt{2}}\,sin^3(ft)~.
\end{eqnarray}
The NC parameters are
\begin{eqnarray}
\theta(t)=0~,~\Omega(t)=\Omega_0\,cos(ft)~.
\end{eqnarray}
By using Eqn.(\ref{B34a}), the above chosen values lead to calculate the following (latter) Hamiltonian parameters as,
\begin{eqnarray}
a(t)=\dfrac{2}{M} ~,~ b(t)=M\,\omega_0^2\,sin^6(ft)+\dfrac{\Omega_0^2}{2M}\,cos^2(ft)    \\
c(t)=\dfrac{\Omega_0}{M}\,cos(ft)~,~ d(t)=\dfrac{\omega_0}{\sqrt{2}}\,sin^3\,(ft)~.
\end{eqnarray}
\noindent Finally, by using Eqn.(\ref{berry}), the above calculated periodic forms result in the following value of Berry phase as
\begin{eqnarray}
\Theta^G_{n,l}&=&(n+l)\oint^{\mathcal{R}} \dfrac{a}{\omega_p}\,\overrightarrow{\nabla}_{\mathcal{R}}\left(\dfrac{d}{a} \right).d \overrightarrow{\mathcal{R}} \nonumber\\
&=&(n+l)\int^{2\pi/f}_0 \dfrac{a}{\omega_p}\dfrac{d}{dt}\left(\dfrac{d}{a} \right) dt \nonumber\\
&=& (n+l)\int^{2\pi/f}_0 \dfrac{3\,M\,\omega_0 f}{\sqrt{2}\,\Omega_0}\dfrac{\left[1- cos(2ft)\right]}{2} dt \nonumber\\
&=&(n+l)\dfrac{3\,M\,\omega_0\,\pi}{\sqrt{2}\,\Omega_0}~.
\end{eqnarray}
Once again, like the previous case, we analysis the above non zero result graphically. Hence, $I_{\Theta^G_{n,l}}$, the integral expression corresponding to the above phase is given by  
\begin{align}
I_{\Theta^G_{n,l}}
=&(n+l)\int^{\mathcal{R}} \dfrac{a}{\omega_p}\overrightarrow{\nabla}_{\mathcal{R}}\left(\dfrac{d}{a} \right).d \overrightarrow{\mathcal{R}} \\
=&(n+l)\int^t_0 \dfrac{a}{\omega_p}\dfrac{d}{dt}\left(\dfrac{d}{a} \right) dt \\
=&(n+l)\dfrac{3\,f\,M\,\omega_0}{2\sqrt{2}\,\Omega_0}\left[t-\dfrac{1}{2f}\,sin(2ft)\right].
\label{g5} 
\end{align}
Here also, as observed from Fig. (\ref{Bfig1}), the nature of $I_{\Theta^G_{n,l}}$ is found to be aperiodic and is consistent with the non zero finite value of the Berry phase found for this case.


\subsubsection{Case III : Vanishing of Berry phase in a noncommutative framework}

Here we consider the NC framework in which our original Hamiltonian (Eqn.(\ref{B1})) was considered at first. The coefficients of the original Hamiltonian Eqn.(\ref{B1}) are set to hold the following forms,
\begin{eqnarray}
P(t)=\dfrac{1}{M}=constant~,~Q(t)=\dfrac{1}{2}M\omega_0^2  ~,~ R(t)=\dfrac{\omega_0}{\sqrt{2}}~.
\end{eqnarray}
The NC parameters are
\begin{eqnarray}
\theta(t)=\theta_0 sin(ft)~,~\Omega(t)=\Omega_0 sin(ft)~.
\end{eqnarray}
By using Eqn.(\ref{B34a}), the above chosen values lead to calculate the following (latter) Hamiltonian parameters as,
\begin{eqnarray}
a(t)=\dfrac{2}{M}+\dfrac{M\,\omega_0^2\,\theta_0^2}{4}\,sin^2(f t)   ~,~ b(t)=M\,\omega_0^2+\dfrac{\Omega_0^2 sin^2(ft)}{2M}    \\
c(t)= \left(\dfrac{\Omega_0}{M}+\dfrac{M\omega_0^2\theta_0}{2}\right) sin(f t) ~,~ d(t)=\dfrac{\omega_0}{\sqrt{2}}\left(1-\dfrac{\theta_0 \Omega_0 sin^2(ft)}{4}\right)~.
\end{eqnarray}
\noindent Finally, by using Eqn.(\ref{berry}), the above calculated periodic forms result in the following value of Berry phase as
\begin{eqnarray}
\Theta^G_{n,l}&=&(n+l)\oint^{\mathcal{R}} \dfrac{a}{\omega_p}\,\overrightarrow{\nabla}_{\mathcal{R}}\left(\dfrac{d}{a} \right).d \overrightarrow{\mathcal{R}} \nonumber\\
&=&(n+l)\int^{2\pi/f}_0 \dfrac{a}{\omega_p}\dfrac{d}{dt}\left(\dfrac{d}{a} \right) dt \nonumber\\
&=&0~.
\end{eqnarray}
Surprisingly, for the above chosen parameters, the phase is revealed to be vanished, although there was a scale invariant term (which is necessary to generate a geometric phase) in the Hamiltonian from the beginning. \\
\noindent In this scenario as well, we proceed to analyse $I_{\Theta^G_{n,l}}$, the integral expression of the derived phase. Our objective is to visually explore why the system does not acquire a finite phase graphically, unlike in the earlier cases. Hence, $I_{\Theta^G_{n,l}}$, the integral expression corresponding to the above phase is given by
\begin{align}
I_{\Theta^G_{n,l}}
=&(n+l)\int^{\mathcal{R}} \dfrac{a}{\omega_p}\overrightarrow{\nabla}_{\mathcal{R}}\left(\dfrac{d}{a} \right).d \overrightarrow{\mathcal{R}} \label{ind}\\
=&(n+l)\int^t_0 \dfrac{a}{\omega_p}\dfrac{d}{dt}\left(\dfrac{d}{a} \right) dt \\
=&-(n+l)\, tan^{-1} \left(\dfrac{M\theta_0\omega_0}{2\sqrt{2}} sin(f t)   \right)~.
\label{g7}
\end{align}
As seen from Fig.(\ref{Bfig1}), the dynamic nature of $I_{\Theta^G_{n,l}}$ is purely periodic. This causes the Hamiltonian to acquire a null geometric phase after a time period $T$. 

\subsection{Explicit investigation of Berry phase : system II}
In the second system (II) lacking an explicit scale invariant interaction term in its original Hamiltonian, the geometric phase has a pure NC origin since the modified version \cite{spb} of the standard noncommutativity provides a scale invariant term that is a geometric phase forming term, in the latter Hamiltonian. As evident from Eqn.(\ref{berry}) and Eqn.(\ref{berryb}), to achieve a non zero finite value of Berry phase, $d(t)$, the coefficient of that scale invariant term must not vanish. Now, in system II, $d(t)$ remains non zero until both the NC parameters ($\theta(t)$, $\Omega(t)$) are present. Therefore, as also discussed in \cite{spb}, it indicates an apparent existence of a non zero geometric phase in the NC framework where both the NC parameters have non zero finite explicit forms. \\
\noindent Now, that crucial role of noncommutativity to provide a Berry phase is to be verified explicitly and for doing so, we consider the NC framework in which our original Hamiltonian of the second system (Eqn.(\ref{BHam2})) was considered.    

\subsubsection{Case I: Vanishing of Berry phase in noncommutative framework}
The coefficients of the original Hamiltonian Eqn.(\ref{BHam2}) are set to hold the following forms,
\begin{eqnarray}
P(t)=\dfrac{1}{M}=constant~,~Q(t)=\dfrac{1}{2}M\omega_0^2 ~. 
\end{eqnarray}
The NC parameters are
\begin{eqnarray}
\theta(t)=\theta_0\,cos(ft)~,~\Omega(t)=-\Omega_0\,cos(ft)~.
\end{eqnarray}
By using Eqn.(\ref{B34aa}), the above chosen values lead to calculate the following (latter) Hamiltonian parameters as,
\begin{align}
a(t)= \dfrac{2}{M}\left(1+\dfrac{\theta_0\Omega_0 cos^2(ft)}{4}\right)+\dfrac{1}{4}M\omega_0^2\theta_0^2 cos^2(ft)\,,\,b(t)= \left[M\omega_0^2 \left(1+\dfrac{\theta_0\Omega_0 cos^2(ft)}{4}\right)+\dfrac{\Omega_0^2 cos^2(ft)}{2M}\right]  \nonumber\\
c(t)=-\dfrac{\Omega_0 cos(ft)}{M}+\dfrac{M\omega_0^2\theta_0 cos(ft)}{2}\,,\, 
d(t)= \dfrac{\sqrt{\Omega_0\theta_0}cos(ft)}{4}\left(\dfrac{-\Omega_0 cos(ft)}{M}-\dfrac{1}{2}M\omega_0^2\theta_0 cos(ft)\right)~.
\end{align}
\noindent Finally, by using Eqn.(\ref{berry}), the above calculated periodic forms result in the following value of Berry phase as
\begin{eqnarray}
\Theta^G_{n,l}&=&(n+l)\oint^{\mathcal{R}} \dfrac{a}{\omega_p}\,\overrightarrow{\nabla}_{\mathcal{R}}\left(\dfrac{d}{a} \right).d \overrightarrow{\mathcal{R}} \nonumber\\
&=&(n+l)\int^{2\pi/f}_0 \dfrac{a}{\omega_p}\dfrac{d}{dt}\left(\dfrac{d}{a} \right) dt \nonumber\\
&=&0~.
\end{eqnarray}
Despite fixing a non zero finite value for the NC parameters which result in a non zero coefficient of the scale invariant term in the latter Hamiltonian, the geometric phase vanishes surprisingly.
\subsubsection{Case II: Vanishing of Berry phase in noncommutative framework}
The coefficients of the original Hamiltonian Eqn.(\ref{BHam2}) are set to hold the following forms,
\begin{eqnarray}
P(t)=\dfrac{sin^2(ft)}{M},~Q(t)=\dfrac{1}{2}M\omega_0^2 sin^2(ft) ~. 
\end{eqnarray}
The NC parameters are,
\begin{eqnarray}
\theta(t)=\theta_0~,~\Omega(t)=-\Omega_0~.
\end{eqnarray}
By using Eqn.(\ref{B34aa}), the above chosen values lead to calculate the following (latter) Hamiltonian parameters as,
\begin{eqnarray}
a(t)= \left[\dfrac{2}{M}\left(1+\dfrac{\theta_0\Omega_0}{4}\right)+\dfrac{M\omega_0^2\theta_0^2}{4}\right]sin^2(ft) ~,~ 
b(t)= \left[M\omega_0^2\left(1+\dfrac{\theta_0\Omega_0}{4}\right)+\dfrac{\Omega_0^2}{2M}\right]sin^2(ft) \nonumber\\
c(t)= -\dfrac{\Omega_0 sin^2(ft)}{M}+\dfrac{\theta_0 M\omega_0^2 sin^2(ft)}{2} ~,~ 
d(t)= -\dfrac{\sqrt{\Omega_0\theta_0}}{4}\left(\dfrac{\Omega_0}{M}+\dfrac{\theta_0}{2}M\omega_0^2\right)sin^2(ft)~.
\end{eqnarray}
\noindent Finally, by using Eqn.(\ref{berry}), the above calculated periodic forms result in the following value of Berry phase as
\begin{eqnarray}
\Theta^G_{n,l}&=&(n+l)\oint^{\mathcal{R}} \dfrac{a}{\omega_p}\,\overrightarrow{\nabla}_{\mathcal{R}}\left(\dfrac{d}{a} \right).d \overrightarrow{\mathcal{R}} \nonumber\\
&=&(n+l)\int^{2\pi/f}_0 \dfrac{a}{\omega_p}\dfrac{d}{dt}\left(\dfrac{d}{a} \right) dt \nonumber \\
&=&0~.
\end{eqnarray}
Again, the phase is revealed to be zero.\\
\noindent 
In this context, $\dfrac{d(t)}{a(t)}$ remains constant, leading to the result that despite non zero NC parameters, the Berry phase is still zero in both cases examined. Therefore, the presence of the non zero NC parameters do not necessarily result in a non zero geometric phase. This is a significant conclusion of this work.

\section{Summary} 

In this chapter, we explore whether a periodic Hamiltonian consisting a geometric phase forming scale invariant term can ensure the actual existence of a geometric phase. Two distinct types of systems are examined under two different kinds of NC framework. While the first system which already includes a scale invariant term in its original Hamiltonian is considered in a standard NC framework, the second one which lacks the geometric phase forming term in the original Hamiltonian gains that term in its latter form of the Hamiltonian after being transformed by a modified version of the standard coordinate mapping relations. Both the systems are treated with Lewis-Riesenfeld approach for determining their exact eigenstate. Under the adiabatic approximation, a generalized expression for the Berry phase is derived. We then, under various framework of space (both configuration and momentum space), find out the Berry phases explicitly for both systems by using different periodic functional forms of the Hamiltonian parameters. For the first system, we could achieve an actual existence of Berry phase when the NC parameters were not present in the place, which is expected because of the presence of that scale invariant term in the original Hamiltonian. Interestingly, in the presence of the NC parameters, the phase is revealed to be vanished for a specific set of periodic parameters. Those phenomena were also explained graphically by plotting the integral expressions corresponding those derived phases with respect to time. Expectedly, the non zero geometric phases the aperiodic dynamics of their integral expressions over time and the vanishing Berry phase showed a pure periodic nature corresponding to its integral expression over time. In the second system, the Berry phase is found as a pure NC effect in literature. Although the structure of that newly proposed NC framework \cite{spb} suggested an apparent existence of Berry phase due to its ability of inserting a scale invariant term in the latter Hamiltonian of the periodic system, for most choices of the periodic parameters, we have observed the geometric phases to be vanished even in the presence of the non zero, finite NC parameters.   

\chapter{Simple harmonic oscillator with modified noncommutativity}\label{paper4}
In the realm of noncommutativity, using the standard Bopp-shift relations~\cite{mez} as a coordinate mapping relations to transform a system from NC space to ordinary commutative space has proven convenient. Nevertheless, as discussed in the previous chapter, a recent study~\cite{spb} proposed a generalized version of these standard Bopp-shift relations. In the proposed relations, while the commutator brackets among configuration operators and those among momentum operators match those in standard noncommutativity, the commutator brackets between configuration and momentum coordinates are the same as in the ordinary phase space of quantum mechanics, $[X_k, P_l] = i \delta_{kl}$, for $k, l = 1, 2$. This is in contrast to the standard Bopp-shift relations, where the commutator $[X_k, P_l]$ depends on the NC parameters $\theta$ and $\Omega$. Another distinction is that while both the standard and modified versions involve real constants for the NC parameters $\theta$ and $\Omega$, the modified version imposes an additional condition $\theta \Omega < 0$ to prevent the appearance of imaginary numbers in the coordinate mapping structure. Furthermore, an intriguing aspect highlighted in \cite{spb} and discussed previously is its ability to include a time reversal (TR) symmetry breaking scale invariant term in the latter form of the Hamiltonian, expressed in terms of commutative variables. This scale invariant term suggests the possibility of a non trivial Berry's geometric phase emerging under adiabatic conditions when the Hamiltonian evolves periodically over time. These unique characteristics make this modified NC structure a compelling area for further investigation, especially in the exploration of simple prototype models within its framework.\\
\noindent In our study detailed in \cite{SG4}, which forms the subject of this chapter, we investigate a time independent harmonic oscillator placed in dynamic NC background by using the modified version of the standard coordinate mapping relations introduced by \cite{spb}. Our methodology follows the approach pioneered by \cite{Dey}, who employed the standard Bopp-shift relations to transform the Hamiltonian of a two dimensional time independent simple harmonic oscillator from NC space into the ordinary quantum mechanical phase space.\\
\noindent The extended form of the Bopp shift connecting NC variables with commutative ones has led to the observation of a geometric phase shift in planar NC quantum mechanics \cite{spb}. Our earlier research, detailed in \cite{SG3}, demonstrated that such transformations enable exact analytical solutions for quantum systems. This addresses an intriguing query whether we can establish a quantum theory in a time varying NC framework governed by the NC algebra introduced by \cite{spb}. Exploring this question is crucial as it promises deeper insights into quantum theory within dynamic NC backgrounds.

\section{Hamiltonian of harmonic oscillator model in modified noncommutative space} 
Here, we consider the time independent Hamiltonian of the model of a system which is a combination of two non interacting simple harmonic oscillators in two dimensional dynamic NC background. The model of such time independent harmonic oscillators which have equal time independent 
frequencies, and equal mass, was also studied in \cite{Dey} in time dependent NC space. In our work, that model is extended by placing the system in a NC space, incorporating the recent modification introduced in \cite{spb}.\\
\noindent Thus, the present Hamiltonian has the following structure in NC space as,
\begin{equation}
H(t)|_{NC}=\dfrac{1}{2\,M}(P_1^2+P_2^2)+\dfrac{1}{2}\,M\,\omega^2(X_1^2+X_2^2)~;\label{Ham2}
\end{equation}
where $M$ and $\omega$ are constants which denote the mass and the angular frequency of the oscillator respectively. The commutation relations between the position and momentum operators in NC space are given by (including $\hbar$) 
\begin{align}
[X_1,X_2]=i\theta(t)~,~[P_1,P_2]=i\Omega(t)~,~
[X_1,P_1]=\,i\hbar\,=[X_2,P_2]~;
\end{align}
where $\theta(t)$ and $\Omega(t)$, in respective manner, are the configuration and momentum NC parameters. The commutator brackets among the canonical operators $(x_i,p_i)$ in ordinary quantum mechanical phase space are
\begin{equation}
[x_j,x_k]=0=[p_j,p_k]~,~[x_j,p_k]=i\hbar\,\delta_{jk}; (j, k=1,2).
\end{equation}
In order to transform the Hamiltonian [Eqn.(\ref{Ham2})] into the ordinary quantum mechanical phase space, we apply the modified form of the standard Bopp-shift relations \cite{spb}. The relations are given ($\hbar\,=\,1$) by  
\begin{eqnarray}
&X_1=x_1-\dfrac{\theta(t)}{2}p_2+\dfrac{\sqrt{-\theta(t)\Omega(t)}}{2}x_2~,~X_2=x_2+\dfrac{\theta(t)}{2}p_1-\dfrac{\sqrt{-\theta(t)\Omega(t)}}{2}x_1~;\label{Xbop}\\
&P_1=p_1+\dfrac{\Omega(t)}{2}x_2+\dfrac{\sqrt{-\theta(t)\Omega(t)}}{2}p_2~,~P_2=p_2-\dfrac{\Omega(t)}{2}x_1-\dfrac{\sqrt{-\theta(t)\Omega(t)}}{2}p_1~;\label{Pbop}
\end{eqnarray}
where $\Omega(t)\theta(t)<\,0$.

\noindent The Hamiltonian in terms of the commutative variables $(x_i,p_i)$ then takes the form
\begin{equation}
H(t)=\dfrac{a(t)}{2}\left({p_1}^2+{p_2}^2\right)+\dfrac{b(t)}{2}\left({x_1}^2+{x_2}^2\right)+c(t)\left({p_1}{x_2}-{p_2}{x_1}\right)+d(t)\left(p_1x_1+x_1p_1+p_2x_2+x_2p_2\right)\,\,\,;\label{ham2}
\end{equation}
where the structural form of the coefficients $a(t), b(t), c(t), d(t)$ are given by
\begin{align}
&a(t)=\dfrac{1}{M}\left(1-\dfrac{\Omega(t)\theta(t)}{4} \right)+\dfrac{M\,\omega^2\theta^2(t)}{4}~,
b(t)=M\,\omega^2\left(1-\dfrac{\Omega(t)\theta(t)}{4} \right)+\dfrac{\Omega^2(t)}{4\,M}
~,\nonumber \\
&c(t)=\dfrac{\Omega(t)}{2M}+\dfrac{M\omega^2\theta(t)}{2}
~,~
d(t)= \dfrac{\sqrt{-\Omega(t)\theta(t)}}{4}\left[\dfrac{\Omega(t)}{2M}-\dfrac{M\omega^2\theta(t)}{2} \right]~.   \label{ham2co}              
\end{align}
The structure of the Hamiltonian [Eqn.(\ref{ham2})] and the Hamiltonian parameters $a(t)$, $b(t)$, $c(t)$, and $d(t)$ [Eqn.(\ref{ham2co})] resemble those employed in \cite{SG3} (from the previous chapter) for exploring explicit forms of geometric phases in NC space. However, it's important to clarify that our present investigation has different objectives compared to those addressed in \cite{SG3}.

\section{Exact eigenstate of the Hamiltonian}
We employ the Lewis technique \cite{Lewis} to address all our model systems, with a detailed exposition available in earlier chapters \ref{chap-LR} and in \ref{paper1}. Consequently, we will briefly mention that by solving the Lewis invariant $I(t)$, such that    
\begin{equation}
I(t)\phi(x_1,x_2)=\epsilon \phi(x_1,x_2)~;
\end{equation}
where $\epsilon$ denotes the invariant's eigenvalue corresponding to its time dependent eigenstate $\phi(x_1,x_2)$, the eigenstate of $H(t)$ can be obtained using the following relation 
\begin{equation}
\psi(x_1,x_2,t)=e^{i\Theta(t)}\phi(x_1,x_2)~;\label{L.eqn}
\end{equation}
where $\psi(x_1,x_2,t)$ is the eigenfunction of the Hamiltonian (Eqn.(\ref{ham2})) and the real function $\Theta(t)$ is the Lewis phase factor.\vskip .20cm
\noindent \textbf{Eigenfunction of the Lewis invariant and Lewis phase factor~:}
\vskip .20cm
\noindent Since the Hamiltonian (Eqn.(\ref{ham2})) in this study matches that of our earlier work \cite{SG3}, as detailed in the previous chapter, we adopt the findings from there and express them using polar coordinate variables.\\
\noindent The eigenfunction of the invariant can be expressed as (for the positive integers $n$ and $m$, such that $n+l\,=m\,\geqslant\,0$)
\begin{eqnarray}
\phi_{n,m-n}(r,\theta, t)=\dfrac{i^m\,\rho^{m-n-1}}{\sqrt{m!n!\pi}}r^{n-m}e^{i(m-n)\theta-\dfrac{a-i\rho(\dot{\rho}-2\rho\,d)}{2a{\rho}^2}r^2}U\left(-m,1-m+n,\dfrac{r^2}{\rho^2} \right)~;\label{efU}
\end{eqnarray}
where $\rho$ adhere to the non linear EP equation (by setting the integration constant as zero)
\begin{equation}
\ddot{\rho}-\dfrac{\dot{a}}{a}\dot{\rho}+\rho\,\left(ab-2\dot{d}-4d^2+2\dfrac{\dot{a}}{a}d\right)=\dfrac{a^2}{\rho^3}~.
\label{EP}
\end{equation}
As we also discusses in the earlier chapters, $U\left(-m,1-m+n,\dfrac{r^2}{\rho^2} \right)$ is known as
Tricomi's confluent hypergeometric function \cite{Arfken, uva} which  can also be presented in terms of the 
associated Laguerre polynomial. Thus,
\begin{equation}
U\left(-m,1-m+n,\dfrac{r^2}{\rho^2} \right)=\dfrac{m!}{(-1)^m}\,L^{n-m}_m\left(\dfrac{r^2}{\rho^2} \right)~.
\end{equation}
Thus, the eigenfunction of the invariant operator can also be expressed as,
\begin{align}
\phi_{n,m-n}(r,\theta, t)&=\dfrac{i^{-m}\sqrt{m!}\,\rho^{m-n-1}}{\sqrt{n!\pi}}r^{n-m}e^{i(m-n)\theta-\dfrac{a-i\rho(\dot{\rho}-2\rho\,d)}{2a{\rho}^2}r^2}\,L^{n-m}_m\left(\dfrac{r^2}{\rho^2} \right)~\nonumber\\
&=Q_{n,m-n}(t)\,R_{n,m-n}(r,t)\,\Phi_{n,m-n}(\theta,t)\,;\label{efL}
\end{align}
where the components $R_{n, m-n}(r,t), \Phi_{n, m-n}(\theta,t)$ and $Q_{n, m-n}(t)$ denote the following expressions,
\begin{align}
&R_{n,m-n}(r,t)=r^{n-m}e^{-\dfrac{a-i\rho(\dot{\rho}-2\rho\,d)}{2a{\rho}^2}r^2}\,L^{n-m}_m\left(\dfrac{r^2}{\rho^2} \right)
\,,\nonumber\\
&Q_{n,m-n}(t)=\dfrac{i^{-m}\sqrt{m!}\,\rho^{m-n-1}}{\sqrt{n!\pi}}~,\Phi_{n,m-n}(\theta,t)=e^{i(m-n)\theta}.\label{efLparts}
\end{align}
It is obvious that $\phi_{n,m-n}(r,\theta)$, the Hamiltonian eigenfunction holds the following 
orthonormality condition,
\begin{equation}
\int_0^{2\pi}d\theta\int_0^{\infty}rdr\phi^{*}_{n,m-n}(r,\theta)\phi_{n^{'},m^{'}-n^{'}}(r,\theta)=\delta_{nn^{'}}\delta_{mm^{'}}.
\label{orth}
\end{equation}
The structure of $\Theta(t)$, the Lewis phase factor is given by \cite{Dey},  
\begin{equation}
\Theta_{n, l}(t)=\,(n+l)\,\int_0^{t}{\left[c(\tau)-\dfrac{a(\tau)}{\rho^2(\tau)}\right] d\tau}.
\end{equation}
For any provided given value of $l$ ( since $l=-n+m$), it would take the form,
\begin{equation}
\Theta_{n,m-n}(t)=\,m\,\int_0^{t}{\left[c(\tau)-\dfrac{a(\tau)}{\rho^2(\tau)}\right] d\tau}.\label{phase2}
\end{equation}
Therefore, $\psi_{n, m-n}\,(r, \theta, t)$, the eigensystem of the Hamiltonian can be obtained by using Eqn.(\ref{L.eqn}).
\begin{equation}
\psi_{n, m-n}\,(r, \theta, t)\,=\,e^{i\,\Theta_{n,m-n}(t)}\,\phi_{n, m-n}\,(r, \theta, t)
\end{equation}
In our previous chapter to discuss \cite{SG3}, we dealt with a similar EP equation (Eqn.(\ref{EP})), albeit in its approximated form, to determine the explicit value of the Berry phase. However, for our current study, we must proceed with the exact form of this equation. The EP equation has been altered by the parameter $d(t)$, introduced through a modified version of noncommutativity \cite{spb}. Therefore, we require the precise analytical functional expression of the additional parameter $d(t)$ corresponding to the other EP variables $a(t), \rho(t)$, and $b(t)$, which were already found in \cite{Dey} and utilized in our prior investigations \cite{SG1, SG2}.
\section{Solution set of the Ermakov-Pinney equation}

In order to derive the explicit solution sets of the EP equation, we proceed as outlined in \cite{Dey}, where the solutions for the EP equation (Eqn.(\ref{EP})) under the condition $d(t),=,0$ were constructed using the Chiellini integrability condition \cite{chill}.

\noindent Here, we expand upon that solution set by deducing the analytical expression of the fourth EP parameter $d(t)$ associated with the derived values of $a(t)$, $\rho(t)$, and $b(t)$ in \cite{Dey}. Moreover, we will also verify that all the EP parameters including the additional one, are still consistent with the Chiellini integrability condition.

\subsection{Exponential solution set of EP equation} 
The exponentially varying EP solution which is already obtained in \cite{Dey} is given by the following
relations, 
\begin{eqnarray} 
a(t)=\sigma e^{-\Gamma{t}}\,\,\,,\,\,\,b(t)=\Delta e^{\Gamma{t}}\,\,\,,\,\,\rho(t)={\mu}e^{-\Gamma/2}\,\,\,\,\,;\label{exp1}
\end{eqnarray}
where $\Delta, \mu, \Gamma$ and $\sigma$ are real, positive constants which are constrained to follow the relation
\begin{equation}
4\sigma\Delta\mu^4-\mu^4\Gamma^2-4\sigma^2=0~;\label{exp1c}
\end{equation}
which is formed after directly substituting the above EP solution set [Eqn.(\ref{exp1})] into Eqn.(\ref{EP}) with $d\,=\,0$.\\
\noindent Substituting the analytical form of $a(t)$, $b(t)$ and $\rho(t)$ into the EP equation [Eqn.(\ref{EP})], we have the following differential equation for the unknown coefficient $d(t)$,
\begin{equation}
\dot{d}+2d^2+\Gamma\,d=\dfrac{4\mu^4\sigma\Delta-4\sigma^2-\mu^4\Gamma^2}{8\mu^4}\equiv\Bbbk\,\text{(constant)}~.\label{expk}
\end{equation}
The analytical form of $d(t)$ is calculated from the above differential equation and it is found to be,
\begin{align}
d(t)\,=\,\dfrac{1}{4}\left[\sqrt{8\Bbbk+\Gamma^2}~~\dfrac{\complement\,e^{\sqrt{8\Bbbk+\Gamma^2}\,\,t}+1}{\complement\,e^{\sqrt{8\Bbbk+\Gamma^2}\,\,t}-1}-\Gamma \right]~~;\label{expd1}
\end{align}
where $\complement$, the constant emerged out from the integration, is chosen such that it exceeds unity, ensuring the solution remains non divergent for all times $t$. To elaborate, we first determine the critical time $t_0$ at which the above parameter would diverge. The criterion for divergence in this context is specified as follows,

\begin{equation}
\complement\,e^{\sqrt{\Gamma^2+8\Bbbk}\,\,t_0}\,=\,1
\end{equation}  
and it provides the critical time $t_0$ as
\begin{equation}
t_0\,=\,-\dfrac{log\,\complement}{\sqrt{8\Bbbk+\Gamma^2}}~.
\end{equation}
Therefore, selecting $\complement$ to exceed $1$ allows us to render $t_0$ negative. Hence, $t$, the physical time cannot tend to $t_0$ and the possibility of the divergence can be avoided in a nice manner.

\subsubsection*{A simple form of exponential EP parameter\,:}
\noindent It is important to highlight that Eqn.(\ref{expd1}) can provide different kind of special solutions depending on the choice of the constant $\Bbbk$ in Eqn.(\ref{expk}). To illustrate this, we examine the simplest case by setting $\Bbbk$ to zero. This leads to the following relation, derived from Eqn.(\ref{expk}),

\begin{align}
\mu^4\sigma\Delta-\sigma^2=\dfrac{\mu^4\Gamma^2}{4}~;\label{expk0}
\end{align}  
which is exactly similar to the relation shown in Eqn.(\ref{exp1c}).
\noindent Thus, we find out a very convenient form of the EP parameter $d(t)$ which is as follows,
\begin{equation}
d(t)|_{\Bbbk=0}=\dfrac{\Gamma}{2\left(\complement\,e^{\Gamma\,t}-1\right)}~~;~~\complement\,>1~.\label{expd0}
\end{equation}

\noindent At the limit $\complement\rightarrow\infty$, $d(t)$ approaches to zero. 

\subsubsection{Chilleni integrability condition for the solution}
Our goal is to show that the solution sets derived earlier are also consistent with the Chiellini integrability condition \cite{chill}, which was utilized in \cite{Dey} to obtain the explicit analytical form of the first three EP parameters. For a detailed discussion on this integrability condition, we refer to portion \ref{sec-chil} in chapter \ref{paper1}. Here, we will briefly restate the condition.

\noindent Defining the components of the EP equation (Eqn. \ref{EP}) in the following manner
\begin{eqnarray}
\dot{\rho}=\eta~,~
g(\rho)=-\dfrac{\dot{a}}{a}~,~h(\rho)=\rho\left(ab-2\dot{d}-4\,d^2+2\dfrac{\dot{a}}{a}d \right)-\dfrac{a^2}{\rho^3}~;\label{chil1}
\end{eqnarray}
we are able to transform Eqn.(\ref{EP}) into the following first order differential equation 
\begin{equation}
\eta\dfrac{d\eta}{d\rho}+\eta\,g(\rho)+h(\rho)=0~.\label{chil2}
\end{equation}
Now as per the the Chilleni integrability condition, if
\begin{align}
\dfrac{d}{d\rho}\left(\dfrac{h(\rho)}{g(\rho)}\right)=q\,g(\rho)~;~(q=\text{constant})\label{chil3}
\end{align}
then the solution of Eqn.(\ref{chil2}) is
\begin{align}
\eta=\lambda_q\dfrac{h(\rho)}{g(\rho)}~~~\text{with}~~~
\lambda_q=\dfrac{\pm\sqrt{1-4\,q}-1}{2\,q}~.\label{chil4}
\end{align}
Here we show that the exponentially varying EP solution sets should be consistent with the Chiellini integrability condition. To achieve this, we substitute the analytical form for the said EP solution set, along with their constraint relation [Eqn.(\ref{expk})], into Eqn.(\ref{chil1}). This substitution yields explicit analytical form for $\eta(\rho)$, $g(\rho)$, and $h(\rho)$,
\begin{align}
\eta(\rho)=-\dfrac{\Gamma}{2}\rho~~,~~g(\rho)=\Gamma~~,~~
h(\rho)=\dfrac{\Gamma^2}{4}\rho~.\label{chil5}
\end{align}
With these values at hand, it is straightforward to show that the Chiellini integrability conditions stated in Eqns.(\ref{chil3}, \ref{chil4}) are satisfied and the constants ($q$ and $\lambda_q$) associated with this condition takes the following value as,
\begin{eqnarray}
q=\dfrac{1}{4}~~,~~\lambda_q=-2~;
\end{eqnarray}
which also follows $\lambda_q=(-1-\sqrt{1-4\,q})\,/2\,q$.

\subsection{Rational solution set of EP equation}
The rationally varying EP solution which is already obtained in \cite{Dey} is given by the following
relations, 
\begin{eqnarray}
a(t)=\dfrac{\sigma\,\left(1+\dfrac{2}{k}\right)^{\,(k+2)/k}}{(\Gamma{t}+\chi)^{\,(k+2)/k}}~,~
b(t)=\dfrac{\Delta\,\left(1-\dfrac{2}{k} \right)^{(k-2)/k} }{(\Gamma{t}+\chi)^{\,(k-2)/k}}~,~\rho(t)=\dfrac{\mu\left(1+\dfrac{2}{k}\right)^{1/k} }{(\Gamma{t}+\chi)^{1/k}}~;\label{rat1}
\end{eqnarray}
where $k\,\,\in\,\,N$ and $\sigma, \Delta, \mu, \chi$ and $\Gamma$ are real, positive constants which are constrained to follow the relation
\begin{equation}
\left(\sigma\Delta\mu^4-\sigma^2 \right)\left(k+2\right)^2=\mu^4\Gamma^2~;\label{ratc1}
\end{equation}
which is formed after directly substituting the above EP solution set [Eqn.(\ref{rat1})] into Eqn.(\ref{EP}) with $d\,=\,0$.\\
\noindent Substituting the analytical form of $a(t)$, $b(t)$ and $\rho(t)$ into the EP equation [Eqn.(\ref{EP})]. This results in the following differential equation for the unknown coefficient $d(t)$,
\begin{align}
\dfrac{1}{k^2\left(\Gamma{t}+\chi\right)^{2}}\left[\left(\Delta\sigma\mu-\dfrac{\sigma^2}{\mu^3}\right)\left(k+2\right)^2-\Gamma^2\mu\right]=2\mu\left[\dot{d}(t)+2d^2(t)+d(t)\dfrac{\Gamma(k+2)}{k(\Gamma\,t+\chi)} \right]~.\label{ratdiff}
\end{align}
It is evident that the above dynamic equation would reduce to be a time independent relation if $d(t)$ takes the following analytical form,
\begin{equation}
d(t)=\dfrac{\delta}{\left(\Gamma\,t+\chi\right)}~;\label{ratd}
\end{equation}  
where $\delta$ is a real, positive constant. Substituting the assumed form of $d(t)$ into Eqn.(\ref{ratdiff}), we have the following time independent constraint relation, 
\begin{equation}
4\delta\,\mu^4\,k^2\left(\dfrac{\Gamma}{k}+\delta \right)=\left(\sigma\mu^4\Delta-\sigma^2 \right)\left(k+2\right)^2-\Gamma^2\mu^4~.\label{ratc2}
\end{equation}
Thus, the rationally varying EP solution set is given by Eqns.(\ref{rat1}, \ref{ratd}) with the constraint relation Eqn.(\ref{ratc2}). Notably, while the variables $b(t)$, $a(t)$ and $\rho(t)$ can provide various explicit forms based on the numerical values of $k$, the variable $d(t)$ does not depend on the value of $k$.

\subsubsection{Chilleni integrability condition for the solution}
Once again, we demonstrate that the rationally varying EP solution sets should be consistent with the Chiellini integrability condition. To achieve this, we substitute the analytical form for the said EP solution set, along with their constraint relation [Eqn.(\ref{ratc2})], into Eqn.(\ref{chil1}). This substitution yields explicit analytical form for $\eta(\rho)$, $g(\rho)$, and $h(\rho)$,
\begin{align}
\eta(\rho)=-\dfrac{\rho^{k+1}\Gamma}{\mu^k(k+2)}~~;~~g(\rho)=\Gamma\dfrac{\rho^k}{\mu^k}~~;~~
h(\rho)=\dfrac{\rho^{2k+1}\Gamma^2}{\mu^{2k}\left(k+2\right)^2}~.\label{chil5rat}
\end{align}
With these values at hand, it is straightforward to show that the Chiellini integrability conditions stated in Eqns.(\ref{chil3}, \ref{chil4}) are satisfied and the constants ($q$ and $\lambda_q$) associated with this condition takes the following value as,
\begin{eqnarray}
q=\dfrac{k+1}{\left(k+2\right)^2}~~,~~\lambda_q=-\left(k+2\right)~;
\end{eqnarray}
which also follows $\lambda_q=(-1-\sqrt{1-4\,q})\,/2\,q$.

\section{Expectation values and uncertainty relations}
We now fix our goal as to determine the energy expectation value, and the generalized uncertainty relations among the canonical operators. As per Eqn[\ref{ham2}], $\langle H \rangle$ is to be derived from the following relation which is given as,
\begin{align}
\langle H\rangle &= \dfrac{a(t)}{2}\left(\langle{p_1}^2\rangle+\langle{p_2}^2\rangle\right)+\dfrac{b(t)}{2}\left(\langle{x_1}^2\rangle+\langle{x_2}^2\rangle\right)+c(t)\left(\langle{p_1}{x_2}-{p_2}{x_1}\rangle\right)\nonumber\\
&+\,d(t)\left(\langle{p_1}{x_1}+{x_1}{p_1}\rangle+\langle{p_2}{x_2}+{x_2}{p_2}\rangle\right)\,\,\,.\label{enrgexp}
\end{align}
It should be noted that, $|n,l\rangle_H$ denotes the eigenstates of the Hamiltonian $H(t)$. 

\subsection{Expectation values of various operators} 
We begin by presenting the configuration space operators and the momentum operators in terms of the polar coordinate variables. They reads,
\begin{align}
x_1\,=\,r\,cos\,\theta~&,~x_2\,=\,r\,sin\,\theta\nonumber\\
p_1\,=\,\left[-i\,cos\,\theta\,\partial_r+\dfrac{i}{r}sin\theta\,\partial_{\theta}\right]~&,~p_2\,=\,\left[-i\,sin\,\theta\,\partial_r-\dfrac{i}{r}cos\theta\,\partial_{\theta}\right]~.
\end{align}
We need the expectation values of the configuration space operators and the momentum operators $x_1$, $x_2$, $p_1$, and $p_2$, as well as their respective squared forms. For this purpose, we refer to section \ref{expect} from chapter \ref{paper1}, where our previous work \cite{SG1} is discussed in detail. The method to compute these quantities is thoroughly explained there. By following the same methodology used for the model \cite{SG1} under standard noncommutativity \cite{mez}, we have obtained results for the current model system \cite{SG4} under modified noncommutativity \cite{spb}.\\
\noindent The expectation values regarding the position operators have been found to be
\begin{align}
_{H}\langle n,m-n|\,x_i\,|n,m-n\rangle_{H}\,&=\,0~,\nonumber\\
_{H}\langle n,m-n|\,x_i^2\,|n,m-n\rangle_{H}\,&=\,\dfrac{(n+m+1)}{2}\,\rho^2~;\label{xsq}
\end{align}
where $i=1, 2$. \\
\noindent Our results, derived using a modified version of the standard coordinate mapping relations, matches with the same found in \cite{Dey, SG1}, where a dynamic NC framework was considered using the standard coordinate mapping relations. It is obvious because, in our generalized eigenfunction (Eqn.(\ref{efL})) of the invariant, the additional parameter $d(t)$ introduced through the modification in the coordinate mapping relations appears inside a phase factor. Consequently, the $d(t)$ factor does not influence the expectation value calculations as it is eliminated in the process.\\
\noindent The expectation values regarding the momentum operators have been found to be ,
\begin{align}
_{H}\langle n,m-n|\,p_i\,|n,m-n\rangle_{H}\,&=\,0~,\nonumber\\
_{H}\langle n,m-n|\,p_1^2\,|n,m-n\rangle_{H}\,&=\,\dfrac{(m+n+1)}{2}\,\left[\dfrac{1}{\rho^2}+\dfrac{(\dot{\rho}-2\rho\,d)^2}{a^2} \right]~;\label{psq3}
\end{align}
where $i=1, 2$. \\
\noindent The inclusion of the coefficient $d(t)$ in the above result allows us to observe the impact of using modified NC background instead of the standard NC background. When $d(t)$ is set to zero, our result reverts to the one previously obtained in \cite{Dey, SG1}. Additionally, in the present system, the emergence of $d(t)$ is exclusively influenced by the NC parameters, making its presence a distinct NC effect in the expression.\\
\noindent Next, we intend to derive the expectation value of $\left(x_1\,p_1+p_1\,x_1\right)$. The polar representation of the quantity reads,
\begin{align}
\left(p_1\,x_1+x_1\,p_1\right)=\,-i\,+\,2\,x_1\,p_1&\nonumber\\
=\,-i-2i\,\left(\,r\,cos^2\theta\,\partial_r-sin\theta\,cos\theta\,\partial_\theta\,\right)&~.
\end{align}
Now, we shall utilize the same technique followed in calculating $\braket{p_i^2}$ in chapter \ref{paper1}. It provides,
\begin{align} 
&_{H}\langle n,m-n|\left(p_1\,x_1+x_1\,p_1\right)|n,m-n\rangle_{H}\nonumber\\
&=\int 
~_{H}\langle n,m-n |r,\theta\rangle\langle r,\theta|\left[\,-2i\,\left(\,r\,cos^2\theta\,\partial_r-sin\theta\,cos\theta\,\partial_\theta\,\right)-i\right]|n,m-n\rangle_{H}~r dr d\theta\nonumber\\
&=\int ~\langle n,m-n |r,\theta\rangle\langle r,\theta|\,\left[\,-2i\,\left(\,r\,cos^2\theta\,\partial_r-sin\theta\,cos\theta\,\partial_\theta\,\right)-i\right]|n,m-n\rangle~r dr d\theta
\nonumber\\
&=-i-2i\int~(r dr d\theta)~
~\phi_{n,m-n}^{*}(r,\theta, t)\,\left(\,r\,cos^2\theta\,\partial_r-sin\theta\,cos\theta\,\partial_\theta\,\right)\phi_{n,m-n}(r,\theta, t)\, \nonumber\\
&=\,I_1^{'}+I_2^{'}-i~;
\end{align} 
where 
\begin{align}
I_1^{'}=-2i\,Q^{*}_{n,m-n}\,Q_{n,m-n}\,\int_0^\infty\,R_{n,m-n}^{*}(r, t)\,r^2\,\left[\partial_r\,R_{n,m-n}(r, t)\right]\,dr\,\int_0^{2\Pi}\,\Phi_{n,m-n}^{*}(\theta, t)\,cos^2\theta\,\Phi_{n,m-n}(\theta, t)\,d\theta
\end{align}
and
\begin{align}
I_2^{'}&=2i\,Q_{n,m-n}(t)\,Q^{*}_{n,m-n}(t)\,\int_0^\infty\,R_{n,m-n}(r, t)\,R_{n,m-n}^{*}(r, t)\,r\,dr\,\nonumber\\
&\times\int_0^{2\Pi}\,\Phi_{n,m-n}^{*}(\theta, t)\,sin\theta\,cos\theta\,\partial_\theta\left[\Phi_{n,m-n}(\theta, t)\right]\,d\theta~.
\end{align}
The azimuthal portions of the integral $I_2^{'}$ indicates that the terms $I_2^{'}$ would provide zero contribution. Because
\begin{align}
&\int_0^{2\Pi}\,\Phi_{n,m-n}^{*}(\theta, t)\,sin\theta\,cos\theta\,\partial_{\theta}\left[\Phi_{n,m-n}(\theta, t)\right]\,d\theta\nonumber\\
&=\,i\,(m-n)\,\int_0^{2\Pi}\,\Phi_{n,m-n}^{*}(\theta, t)\,sin\theta\,cos\theta\,\Phi_{n,m-n}(\theta, t)\,d\theta\nonumber\\
&=0~.\label{psq00}
\end{align}
Next, using Eqn.[\ref{efLparts}], we substitute $R$, $\Phi$, and $Q$ into the present expression of $I_1^{'}$. To perform the integration, we employ several useful and specific relations regarding the associated Laguerre polynomials. The derivative of the associated Laguerre polynomial reads \cite{Arfken},
\begin{align}
\partial_z\left[L^{n-m}_m\,\left(z\right)\right]&=
-\,L^{n-m+1}_{m-1}\,\left(z\right)~;~(m\,\geq\,1)\nonumber\\
&=\,0~~~~~~~~~~~~~~~~~~~;(m=0)~.\nonumber\\
\end{align}
Moreover, the associated Laguerre polynomials have the following identity relation and the orthonormality condition \cite{Arfken},
\begin{align}
L^{n-m}_{m}(z)\,=\,L^{n-m+1}_{m}(z)-L^{n-m+1}_{m-1}(z)\nonumber\\
\int_0^\infty\,e^{-z}\,z^{n-m}\,L^{n-m}_{n}(z)\,L^{n-m}_{m}(z)
\,=\,\dfrac{\Gamma\,(2n-m+1)}{n!}\,\delta_{mn}~.\label{psqoi}
\end{align}
We proceed to evaluate the integral $I_1^{'}$ using the above relations related to the associated Laguerre polynomial, which results in the following outcome
\begin{align}
_{H}\langle n,m-n|\left(p_1\,x_1+x_1\,p_1\right)|n,m-n\rangle_{H}=(n+m+1)\,\dfrac{\rho\,\left(\dot{\rho}-2\rho\,d\right)}{a}~.\label{si1}
\end{align}
\noindent Similarly, we can also derive
\begin{align}
_{H}\langle n,m-n|\left(p_2\,x_2+x_2\,p_2\right)|n,m-n\rangle_{H}=(n+m+1)\,\dfrac{\rho\,\left(\dot{\rho}-2\rho\,d\right)}{a}~,\label{si2}
\end{align}
and hence, express both the quantities as
\begin{align}
_{H}\langle n,m-n|\left(p_i\,x_i+x_i\,p_i\right)|n,m-n\rangle_{H}=(n+m+1)\,\dfrac{\rho\,\left(\dot{\rho}-2\rho\,d\right)}{a}~;\,i=1, 2\,.\label{si}
\end{align}

\noindent We are now left with calculating $\braket{H(t)}$ is $\braket{p_1\,x_2-x_1\,p_2}$. The polar representation of this quantity
\begin{equation}
\left(p_1\,x_2-x_1\,p_2\right)\,=\,i\partial_{\theta}~.
\end{equation} 
The expectation value of the above quantity can be calculated very easily and it is given by,  
\begin{align}
_{H}\langle n,m-n|\left(p_1\,x_2-x_1\,p_2\right)|n,m-n\rangle_{H}=(n-m)~;\label{zee}
\end{align}
where the individual terms in $\braket{p_1\,x_2-x_1\,p_2}$ reads
\begin{align}
_{H}\langle n,m-n|\left(x_1\,p_2\right)|n,m-n\rangle_{H}=\dfrac{(m-n)}{2}~;~_{H}\langle n,m-n|\left(x_2\,p_1\right)|n,m-n\rangle_{H}=\dfrac{(n-m)}{2}~.\label{zee1}
\end{align} 

\noindent With all the important quantities for deriving $\braket{H(t)}$, we now compute the expectation values of two additional bilinear operators, $p_2p_1$ and $x_2x_1$. These expectation values are essential for determining the generalized form of the uncertainty equalities for the NC coordinate operators, which we will demonstrate later.
\begin{align}
x_2x_1\,=&\,r^2\,cos\theta\,sin\theta~,\nonumber\\
p_2\,p_1\,=\,-cos\theta\,sin\theta\,\partial_r^2+\dfrac{cos\,2\theta}{r^2}\,\partial_{\theta}-&\dfrac{cos\,2\theta}{r}\partial_r\partial_{\theta}+\dfrac{sin\theta\,\cos\theta}{r}\partial_r+\dfrac{sin\theta\,cos\theta}{r^2}\partial_{\theta}^2~.
\end{align}
By carefully examining the azimuthal components of these quantities and drawing on insights from previous calculations, it becomes evident that none of these quantities contributes to their expectation values in the eigenstate of $H(t)$. Thus, we present the following results,
\begin{align}
\braket{p_2\,p_1}=0~,~\braket{x_2\,x_1}=0 ~.\label{bi}
\end{align}

\subsection{Energetics of the model system}
This section focuses on exploring the time evolution of the system's energy expectation value both analytically and graphically. We have conducted the calculations needed to determine the expectation value of $H(t)$ with respect to its own time varying eigenfunction $\psi_{n,m-n}(r, \theta)$. Consequently, by using Eqn.(\ref{enrgexp}) , $\braket{E_{n,m-n}(t)}$, the generic form of the energy expectation value is established as 
\begin{align}
\braket{E_{n,m-n}(t)}&=\dfrac{1}{2}\,(n+m+1)\left[b(t)\rho^2(t)+\dfrac{a(t)}{\rho^2(t)}+\dfrac{\dot{\rho}^2(t)-4\rho^2(t)\,d^2(t)}{a(t)} \right]+c(t)\,(n-m)~;\label{enr}
\end{align}
which provides the same found in \cite{SG1} when $d(t)$ is absent.

\noindent Previously, we found the explicit analytical structure of $d(t)$, the Hamiltonian coefficient that follows the EP equation, corresponding to the other form of the EP parameters $a(t)$, $\rho(t)$, and $b(t)$ generated in \cite{Dey}. Our current goal is to explicitly analyse the energy expression in Eqn.(\ref{enr}) by using those explicit EP solution sets. It is important to remember that $c(t)$, the Hamiltonian coefficient in the generic form of the energy expression, does not follow the EP equation. However, following the methodology outlined in \cite{SG1}, it can be determined explicitly. To achieve this, we need to calculate the explicit forms of the NC parameters $\Omega(t)$ and $\theta(t)$ by substituting $b(t)$, $a(t)$, and $d(t)$ into Eqn.(\ref{ham2co}). This yields the precise form of $c(t)$. For the simplicity of our calculations, we examine the energy dynamics using Eqn.(\ref{enr}) with $n=m$. This gives 
\begin{align}
\braket{E_{m, 0}(t)}&=\dfrac{1}{2}\,(2\,m+1)\left[\dfrac{a(t)}{\rho^2(t)}+b(t)\rho^2(t)+\dfrac{\dot{\rho}^2(t)-4\rho^2(t)\,d^2(t)}{a(t)} \right]~.\label{enr1}
\end{align}

\subsubsection{Exponentially varying energy profiles}
\begin{figure}[H]
\centering
\includegraphics[scale=1]{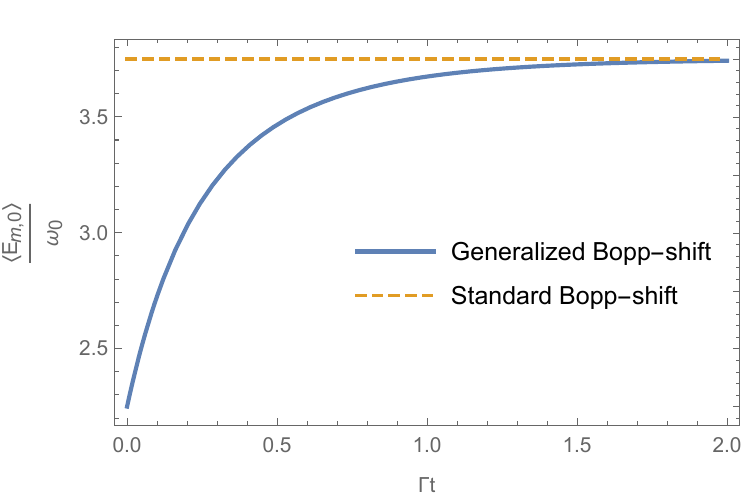}
\caption{\textit{Graph of the exponential energy expectation value (divided by $\omega_0$ to eliminate the dimension), with respect to $\Gamma$t. The expectation value of 
energy $\langle E \rangle$ is computed when $\langle A\rangle$ the system belongs to the modified noncommutativity and the value of the constants  are (in accordance with Eqn.(\ref{exp1c})) chosen as $m, n, \mu, \Gamma=1$ and $\Delta=\dfrac{5}{4},\,\complement=2$; $\langle B\rangle$ the system belongs to the standard noncommutativity and the value of the constants  are (as per Eqn.(\ref{exp1c})) as $m, \mu, \omega_0, \sigma, \Gamma=1$, $\Delta=\dfrac{5}{4}$ and $\complement\,\rightarrow\,\infty$. For $\langle A\rangle$ the energy dynamics initially increases and saturates ultimately at a large time. However, for $\langle B\rangle$, the energy expectation value stays the same over time.}}  
\label{ncfig1}
\end{figure}

The analytical expression of the energy profile corresponding to the exponentially varying EP solution set can be derived from Eqn.(\ref{exp1}) and Eqn.(\ref{expd0}) with the constraint relation in Eqn.(\ref{expk0}). The exponentially varying energy expectation value for $n=m$ is given by,
\begin{align}
\braket{\,H(t)\,}|_{m=n}^{exp}\,=\,(2m+1)\left[\Delta\,\mu^2-\dfrac{\Gamma^2\,\mu^2}{2\,\sigma\,\left(\complement\,e^{\Gamma\,t}-1\right)^2}   \right]~;(\,\complement\,>\,1)~.\label{Eexp}
\end{align}
As the limit $\complement\rightarrow\infty$ is applied, the above expression converges to a constant value. It also represents the exponential energy expression found in \cite{SG1} when $n=m$. For this particular choice of quantum number $m=n$, the analytical form of energy value (Eqn.[\ref{Eexp}]) and the corresponding graphics (see Fig.[\ref{ncfig1}]) show that the energy grows with respect to time and ultimately saturates at a later time. The graphical representation also nicely reflects the fact that, at any given moment, due to the deduction of energy by an amount of $4\rho^2d^2$ (as detailed in Eqn.(\ref{enr})), the energy associated with the generalized Bopp-shift transformation does not exceed that associated with the standard Bopp-shift transformation. However, this comparison holds true only when the numerical values of the constants used in both cases are chosen according to the relation in Eqn.(\ref{exp1c}).

\subsubsection{Rationally varying energy profiles}
\begin{figure}[t]
\centering
\includegraphics[scale=1]{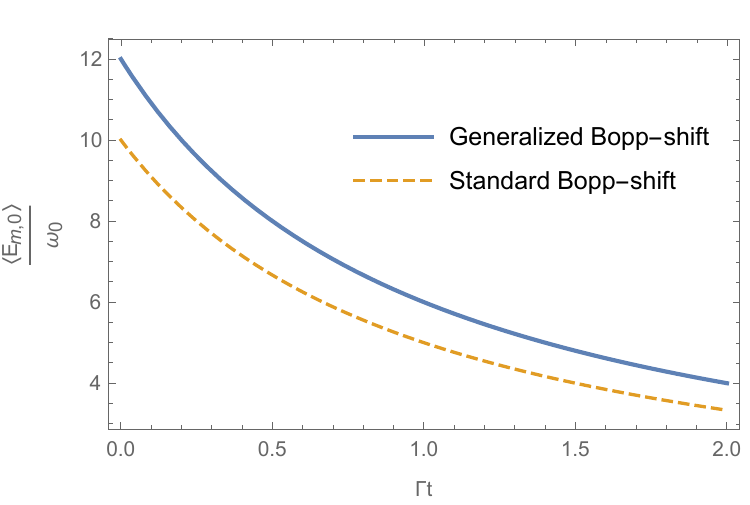}
\caption{\textit{Graph of the rational energy expectation value (divided by $\omega_0$ to eliminate the dimension), with respect to $\Gamma$t. The expectation value of 
energy $\langle E \rangle$ is computed when $\langle A\rangle$ the system belongs to the modified noncommutativity and the value of the constants  are (in accordance with Eqn.(\ref{ratc2})) chosen as $m, n, k, \mu, \Gamma, \omega_0, \delta, \chi, \sigma =1$ and $\Delta=2$; $\langle B\rangle$ the system belongs to the standard noncommutativity and the value of the constants  are (as per Eqn.(\ref{ratc1})) as $m, k, \mu, \Gamma, \sigma, \chi=1$ and $\Delta=\dfrac{10}{9}$.  Both the energy dynamics, having different numerical values of the constants, exhibits a rational decay over time and tends to zero at a large time.}}  
\label{ncfig2}
\end{figure}

The analytical expression of the energy profile corresponding to the rationally varying EP solution set can be derived from Eqn.(\ref{rat1}) and Eqn.(\ref{ratd}) with the constraint relation in Eqn.(\ref{ratc2}). The rationally varying energy expectation value for $n=m$ is given by,
\begin{align}
\braket{\,H(t)\,}|_{m=n}^{rat}\,=\,\dfrac{(2m+1)}{\left(\Gamma\,t+\chi \right)}\left[\dfrac{(k+2)^2\,\left(\sigma^2+\Delta\mu^4\sigma\right)+\Gamma^2\mu^4-4\mu^4\delta^2k^2}{2\,(k+2)\,\mu^2\,k\,\sigma\,}\right]~.\label{Erat}
\end{align}
When $\delta$ tends to zero, the rationally varying energy expression
in \cite{SG1} for $n=m$ can be reproduced from the above result. Although, both energy dynamics exhibit a rational decay, the decay rates are different since the two systems belong to different kinds of noncommutativity. In contrast to the previous example (Fig.\ref{ncfig1}), the graphical representation (Fig.\ref{ncfig2}) indicates that, at any given point, the energy associated with the generalized Bopp shift transformation exceeds that of the standard Bopp shift transformation. This discrepancy arises because the constants associated with the rational solution set adhere to the constraint relation in Eqn.(\ref{ratc2}) for the generalized system, whereas they conform to the constraint relation in Eqn.(\ref{ratc1}) for the standard system.

\subsection{Generalized uncertainty equalities for commutative and noncommutative operators}
Here, we intend to establish the uncertainty relations between the canonical coordinate operators. Although we did not compute the inequality uncertainty relations, we derived the uncertainty equality and verified our generalized findings with those in \cite{Dey} under standard noncommutativity.

\subsubsection*{Uncertainty equalities for commutative operators :}
From Eqn.(\ref{xsq}), the variances for the canonical coordinate operators (belonged to an ordinary quantum mechanical phase space) are deduced to be 
\begin{align}
\Delta\,x_1=\sqrt{\braket{x_1^2}\,-\,\braket{x_1}^2}\,=\,\dfrac{\sqrt{m+n+1}}{\sqrt{2}}\,\rho~,~
~\Delta\,x_2=\sqrt{\braket{x_2^2}\,-\,\braket{x_2}^2}\,=\,\dfrac{\sqrt{m+n+1}}{\sqrt{2}}\,\rho ~;
\end{align}
and from Eqn.(\ref{psq3}), the same for the coordinate operators $p_1$ and $p_2$ can be derived as 
\begin{align}
\Delta\,p_1=\sqrt{\braket{p_1^2}\,-\,\braket{p_1}^2}\,=\dfrac{\sqrt{m+n+1}}{a\,\rho}\sqrt{\dfrac{a^2+\rho^2\left(\dot{\rho}-2\rho\,d\right)^2}{2}}~;\nonumber\\
\Delta\,p_2=\sqrt{\braket{p_2^2}\,-\,\braket{p_2}^2}\,=\,\dfrac{\sqrt{m+n+1}}{a\,\rho}\sqrt{\dfrac{a^2+\rho^2\left(\dot{\rho}-2\rho\,d\right)^2}{2}}~. 
\end{align}
Thus, the uncertainty equalities among the coordinate operators $x_j$ and $p_j$, where $j\,=\,[1,2]$, are found as,
\begin{align}
\Delta\,x_1\,\Delta\,y_1=\,\left(m+n+1\right)\dfrac{\rho^2}{2}~&,~
\Delta\,p_1\,\Delta\,p_2=\dfrac{(n+m+1)}{2}\left[\dfrac{1}{\rho^2}+\dfrac{(\dot{\rho}-2\rho\,d)^2}{a^2} \right]\,~,\nonumber\\
\Delta\,x_1\,\Delta\,p_1=\Delta\,x_2\,\Delta\,p_2&=\dfrac{m+n+1}{2\,a}\sqrt{a^2+\rho^2\left(\dot{\rho}-2\rho\,d\right)^2}~.
\end{align}
At the limit $d(t)\,\rightarrow\,0$, the above generalized results reduce to the same obtained in \cite{Dey}.

\subsubsection*{Uncertainty equalities for noncommutative operators :}

\noindent
By using the results derived in Eqn.(\ref{xsq}, \ref{psq3}) and the coordinate transformation relations given in Eqn.(\ref{Xbop}, \ref{Pbop}), we can have that, 
\begin{align}
\braket{X_i}=0~;~\braket{P_i}=0~~~;~~~i\in [1,2]~.
\end{align}   
Similarly, by using the results in Eqn(s).(\ref{xsq}, \ref{psq3}, \ref{si}, \ref{zee1}, \ref{bi}) we also determine that
\begin{align}
\braket{X_i^2}=\dfrac{(n+m+1)}{2}\,\left[\rho^2\,\left(1-\dfrac{\theta\,\Omega}{4}\right)+\dfrac{\theta^2}{4}\,\left(\dfrac{1}{\rho^2}+\dfrac{\left(\dot{\rho}-2\rho\,d\right)^2}{a^2}\right) 
-\dfrac{\theta\sqrt{-\theta\,\Omega\,}\rho}{2\,a}\left(\dot{\rho}-2\rho\,d \right)  \right]-(m-n)\,\dfrac{\theta}{2}~,
\end{align}
\begin{align}
\braket{P_i^2}=\dfrac{(m+n+1)}{2}\,\left[\,\left(1-\dfrac{\theta\,\Omega}{4}\right)\left(\dfrac{1}{\rho^2}+\dfrac{(\dot{\rho}-2\rho\,d)^2}{a^2}\right)+\dfrac{\Omega^2\rho^2}{4}\,
+\dfrac{\Omega\sqrt{-\theta\,\Omega\,}\rho}{2\,a}
\left(\dot{\rho}-2\rho\,d \right)  \right]-(m-n)\,\dfrac{\Omega}{2}~.
\end{align}
Next, we will obtain the product of those variances for the NC canonical coordinates. The results are as follows,
\begin{align}
&\Delta\,X\,\Delta\,Y=\sqrt{\braket{X^2}\braket{Y}^2}=\braket{X^2}=\braket{Y^2}\nonumber\\
&=\dfrac{(m+n+1)}{2}\,\left[\rho^2\,\left(1-\dfrac{\theta\,\Omega}{4}\right)+\dfrac{\theta^2}{4}\,\left(\dfrac{1}{\rho^2}+\dfrac{(\dot{\rho}-2\rho\,d)^2}{a^2}\right) 
-\dfrac{\theta\sqrt{-\theta\,\Omega\,}\rho}{2\,a}\left(\dot{\rho}-2\rho\,d \right)  \right]-(m-n)\,\dfrac{\theta}{2}~,
\end{align}
\begin{align}
&\Delta\,P_X\,\Delta\,P_Y=\sqrt{\braket{P_X^2}\braket{P_Y}^2}=\braket{P_X^2}=\braket{P_Y^2}\nonumber\\
&=\dfrac{(m+n+1)}{2}\left[\,\left(1-\dfrac{\theta\,\Omega}{4}\right)\left(\dfrac{1}{\rho^2}+\dfrac{(\dot{\rho}-2\rho\,d)^2}{a^2}\right)+\dfrac{\Omega^2\rho^2}{4}
+\dfrac{\Omega\sqrt{-\theta\Omega\,}\rho}{2\,a}
\left(\dot{\rho}-2\rho\,d \right)  \right]-(m-n)\dfrac{\Omega}{2}~,
\end{align}
\begin{align}
&\Delta\,X\,\Delta\,P_X=\Delta\,Y\,\Delta\,P_Y=\sqrt{\braket{X^2}\braket{P_X^2}}=\sqrt{\braket{Y^2}\braket{P_Y^2}}\nonumber\\
&=\left\lbrace\dfrac{(m+n+1)}{2}\,\left[\rho^2\,\left(1-\dfrac{\theta\,\Omega}{4}\right)+\dfrac{\theta^2}{4}\,\left(\dfrac{1}{\rho^2}+\dfrac{\left(\dot{\rho}-2\rho\,d\right)^2}{a^2}\right) 
-\dfrac{\theta\sqrt{-\theta\,\Omega\,}\rho}{2\,a}\left(\dot{\rho}-2\rho\,d \right)  \right]-(m-n)\,\dfrac{\theta}{2} \right\rbrace^{\dfrac{1}{2}}\nonumber\\
&\times\left\lbrace\dfrac{(m+n+1)}{2}\,\left[\,\left(1-\dfrac{\theta\,\Omega}{4}\right)\left(\dfrac{1}{\rho^2}+\dfrac{(\dot{\rho}-2\rho\,d)^2}{a^2}\right)+\dfrac{\Omega^2\rho^2}{4}\,
+\dfrac{\Omega\sqrt{-\theta\,\Omega\,}\rho}{2\,a}
\left(\dot{\rho}-2\rho\,d \right)  \right]-(m-n)\,\dfrac{\Omega}{2}\right\rbrace^{\dfrac{1}{2}}~.
\end{align}
As expected, the above results align with those obtained in \cite{Dey}, as the coefficient $d$ as well as one of the NC parameters approaches to zero.

\section{Summary}
The primary objective of the work discussed in this chapter is to examine the behaviour of a two dimensional time independent simple harmonic oscillator in the framework of a dynamic noncommutativity governed by a modified version of the Bopp-shift relations newly defined in \cite{spb}. We are interested to observe the effect of this different kind of noncommutativity on the oscillator's behaviour 
when mapped via the standard Bopp-shift relations in \cite{Dey}. To achieve this, we consider a time independent model of harmonic oscillator in a two dimensional dynamic NC space. We utilize the generalized Bopp-shift relations \cite{spb} to map the system in terms of the commutative canonical coordinates. We then adopt the exact functional form of the Hamiltonian eigenfunction, as established in \cite{SG3}, since our previous work deals with a similar kind of dynamic system and obtains the time dependent hermitian invariant operator and its corresponding nonlinear differential equation, known as the Ermakov-Pinney (EP) equation. Since this EP equation plays the role of a constraint equation among the coefficients of the Hamiltonian, it is required to be solved explicitly. As this present study extends the model considered in \cite{Dey}, where the analytical form of the EP parameters was derived explicitly using a special integrability condition \cite{chill}, we extend their solution set as well. We derived the precise functional form of the additional EP variable that is associated with the EP solution set deduced in \cite{Dey}. In this regard, we noted some interesting properties in the analytical form of the additional EP parameter, which were derived for both the exponential and rational forms of EP solution set. Although, in \cite{Dey}, the first three exponential EP variables were found as three different explicit functions of time, the fourth parameter we calculated exhibits a generic analytical structure that
can adapt to form specific analytical functions based on varying values of the constant within that parameter. We illustrated this matter with an appropriate example. On the other hand, the power of the first three EP parameters were found in \cite{Dey} to depend on a real, positive constant. In contrast, the fourth EP parameter exhibits an inverse variation with respect to time. However, the consistency of both the exponential and rationally varying EP solution sets with the Chiellini integrability condition has been verified. We calculate the Hamiltonian's expectation value in a general form and demonstrate that, without the modified term from the coordinate mapping relations, it matches the result from our earlier work \cite{SG1}. In that work, we employed the standard Bopp-shift relations to transform the Hamiltonian of a damped harmonic oscillators from NC space to ordinary space. We investigate the energy expectation value in the eigenstates of the Hamiltonian for a specific quantum number, examining both the exponential and rational forms of the EP solution set. Additionally, we create graphical representations of the explicit energy profiles for both exponential and rational variations, and compared their behaviors to those related to the standard coordinate mapping relations. Although the exponential energy dynamics associated with the modified noncommutativity initially grow and then saturate after a certain time, those associated with the standard noncommutativity remain constant. On the other hand, the rationally varying energy dynamics rationally decays to zero with respect to time. Finally, we compute the variances for both commutative and NC canonical coordinate operators to establish the generalized uncertainty (equality) relations among them. As expected, our results successfully reproduce those found for standard noncommutativity \cite{Dey}.
\chapter{Conclusions}
In this thesis, our focus is on the study of various simple prototype models that are exactly solvable within the framework of noncommutativity in a time dependent background. To solve these models exactly, we consistently employ the Lewis-Riesenfeld theory which is an exact quantum theory to solve a time dependent Schr\"{o}dinger equation. The eigenfunction of a time dependent Hamiltonian, obtained through the Lewis technique, essentially combines the eigenfunction of a time dependent Hermitian invariant operator (remaining invariant with respect to time) and a time dependent phase factor known as the Lewis phase factor. While this phase factor is not to be chosen arbitrarily, it must be derived to form a Schr\"{o}dinger solution. Another notable feature of a Schr\"{o}dinger solution obtained using the Lewis technique is that it is always associated with a non linear differential equation, also known as the Ermakov-Pinney equation in the literature. Here, the benefits of choosing the method of invariants become apparent. The differential EP equation associated with this method not only enables us to generate a class of exact Schr\"{o}dinger solutions corresponding to each explicit form of EP solution set but also provides us with a very simple way to derive the geometric phase from the adiabatically approximated form of the Lewis phase. We utilized these properties in our studies concerning the exact solutions of models in time dependent NC space. In our various works, we address three key aspects related to the study of exactly solvable models. Firstly, we focus on developing effective procedures to obtain a class of explicit solutions in closed form for model systems and analyse these solutions using physical quantities such as energy expectation values, which can also be represented graphically. Secondly, the most intriguing aspect of this thesis involves separating the geometric component, under adiabatic approximation, from the Lewis phase factor which enables the construction of the eigenstate in an exact form for the periodic Hamiltonian. Lastly, we conduct a comparative study of an exactly solvable prototype model within two different types of noncommutativity. While our initial works \cite{SG1} and \cite{SG2} are related to the first aspect, the third study \cite{SG3} is under the second aspect and the final work \cite{SG4} corresponds to the last aspect.\\
\noindent In our initial work \cite{SG1}, we treated with a two dimensional quantum damped harmonic oscillator placed in time varying NC framework and solved the system using Lewis prescription. This  work, by choosing various sets of EP solutions, namely, exponentially varying solution and rationally varying solution, and various kinds of damping scenarios, mainly figures out a fruitful procedure for obtaining a class of exact solutions for a damped harmonic oscillator in a time dependent NC space. The energy dynamics corresponding to those solutions are also analysed analytically and graphically. Upon observing these behaviours, the initial remark is that the time dependence in the exponentially varying energy expression of the model is purely attributed to the time dependent noncommutativity. This is a significant NC effect since, without NC parameters, the corresponding energy expressions would reduce to a constant value. Among the three types of damping scenarios designed for the exponentially varying energy expression, the first case, where dissipation is assigned by an exponentially decaying angular frequency only, shows an initial decay up to a certain time before it starts to increase. This non trivial behaviour arises from the particular form of the solutions to the EP equation, which determines the forms of the NC parameters, resulting in the mentioned behaviour of the expectation value of the energy. In the second case, where damping is present but the frequency remains constant, the energy remains constant. This occurs because, although the NC parameters are individually time dependent, the overall time dependent effect of the NC parameters cancels out as the exponentially decaying NC parameter is balanced by the exponentially expanding NC parameter. This is another significant observation from this study. Interestingly, in the third case, where both the frequency and damping term decay exponentially over time, the energy exhibits a consistent decaying behaviour. Although the energy behaviour for the rationally decaying EP solution and the elementary EP solution also shows decay, these two cases prevent the system from being always Hermitian. Thus, by investigating these cases of damped oscillators, we conclude that the behaviour corresponding to the exponentially decaying solution, where both the frequency and damping term decay exponentially over time, is analogous to a damped oscillator in commutative space. \\
\noindent Our subsequent work, presented in \cite{SG2}, extends our initial study by incorporating an externally applied time varying magnetic field. We intend to explore how the interaction between damping and an external time dependent magnetic field affects the energy dynamics studied in \cite{SG1}. This model, like the previous one, is exactly solvable, allowing us to study the energy dynamics under various damping scenarios and applied magnetic fields. It was observed that the energy expression, which remained constant for a decaying damping factor and constant angular frequency in \cite{SG1}, begins to decay when a constant or decaying magnetic field is applied. Additionally, the exponentially varying energy expression which showed significant decaying behaviour in \cite{SG1} for both decaying damping factors and angular frequencies, ceased to decay after a certain point and began to grow over time when an exponentially expanding magnetic field was applied from an external source. Our analysis indicates that the impact of the externally applied magnetic field on the previously discussed system \cite{SG1} varies. The nature of the evolution of the exponentially varying energy expressions is found to be highly dependent on the nature of the time varying magnetic field. In contrast, the rationally varying energy expressions are found to be less affected by the magnetic field. Nevertheless, at any given moment, the expectation value of rationally decaying energy is higher in the presence of a magnetic field compared to when the field is absent, due to the absence of magnetic energy in the latter case. Another notable observation from this study is that the second type of rationally varying energy is the only one that remains physical at all times.\\
\noindent Our next work \cite{SG3} explores whether a non zero geometric phase is induced by a periodic Hamiltonian that includes a scale invariant term in a time varying NC framework. In other words, we intend to explore whether the geometric phase acquired through adiabatic evolution is influenced by the structure of the NC space in which the system is situated. For doing so, we investigate two distinct systems in different types of NC spaces. The first system features an explicit TR symmetry breaking scale invariant term, which is suggested in literature as essential for achieving a non zero Berry phase. The second system initially lacks this term in its Hamiltonian; however, it appears when the system is mapped in terms of the standard commutative variables \cite{spb}. Therefore, the presence of the Berry phase in the second system is an inherently NC phenomenon.  it will be interesting to explore whether the geometric phase acquired through adiabatic evolution is influenced by the structure of the NC space in which the system is situated. We construct the exact eigenstates for both systems using the Lewis technique and derive the general form of the geometric phase from the adiabatically approximated Lewis phase. The main conclusion is drawn by verifying the explicit existence of the Berry phase in both systems, utilizing appropriate periodic forms of the original Hamiltonian parameters and finite NC parameters. Notably, for the first system, even without noncommutativity, explicit calculations result in a non zero geometric phase. However, for certain selections of periodic Hamiltonian parameters, the geometric phase becomes zero, even in NC space. In contrast, for the second system, where a Berry phase is generally anticipated with finite NC parameters, our explicit calculations reveal that the geometric phase is zero for most configurations of periodic Hamiltonian parameters, even when considering the second system in NC space. Thus, our explicit calculation of the Berry phase across various systems and choices of periodic Hamiltonian parameters under different conditions (both with and without noncommutativity) clearly shows that having a TR symmetry breaking scale invariant term in the Hamiltonian is necessary but not sufficient for a non zero geometric phase. The periodic Hamiltonian must be carefully designed to ensure a finite (non-zero) Berry phase. This is one of the main conclusion of this work.\\
\noindent Being inspired by the fact that the newly introduced noncommutativity in \cite{spb} has led to the observation of a geometric phase shift in planar NC quantum mechanics, and that such transformations enable exact analytical solutions for quantum systems \cite{SG3}, in our final work \cite{SG4}, we intend to explore whether we can establish a quantum theory in a time varying NC framework governed by the NC algebra introduced by \cite{spb}. For doing so, we extend the work done by \cite{Dey}, where a time independent harmonic oscillator is studied in a time dependent NC space associated with the standard Bopp-shift relations. First, we present the exact solution of the model. Next, we calculate the general form of the energy expectation value and the uncertainty equality relations. To recognize the effect explicitly, for a particular choice of quantum number, we derive an explicit form of the energy expectation value corresponding to the exponential EP solution. While the energy dynamics associated with the newly deformed NC space shows initial growth over time and saturates finally, the behaviour governed by the standard Bopp-shift relations always remains constant over time. It is noteworthy to mention that all the general results deduced in our work can reproduce the same findings in \cite{Dey} when the modified version of the Bopp-shift relation reduces to the standard Bopp-shift relations. We have concluded that, for the particular choice of quantum number what we considered for graphical exploration, if one can uniformly choose the explicit form of the time dependent parameters and the numerical values of the constants for both the systems associated with the modified and standard Bopp-shift relations, the modified Bopp-shift relation would, at any given moment, prevent the system's energy from surpassing that associated with the standard Bopp-shift relation.\\
\noindent Here is the summarized version of our overall conclusion. In this thesis work, the various kinds of simple prototype models studied in the NC background are exactly solvable, and their studies are primarily based on their exact solutions. It indicates a strong possibility of developing a quantum theory in the background of time dependent NC space. Before we end, we would also like to mention that, although current technology does not permit experimental observation of Planck scale signatures, making the experimental status of noncommutativity currently out of reach, this study holds significant value. It contributes to a deeper understanding of quantum theory in time dependent NC backgrounds.

\chapter*{Appendix}
\section*{Appendix A: Explicit forms of some Lewis phases}
In chapter \ref{paper2}, we examined the eigenfunctions associated with the exponentially decaying EP solution set of a model system depicting a damped harmonic oscillator influenced by a time varying external magnetic field in a NC space. Below Eqn.(\ref{exp3incC}), it was noted that the explicit structure of the Lewis phase factor which is essential to prepare the eigenstate of a damped oscillator with an exponentially decreasing frequency, under the influence of a exponentially growing magnetic field in NC background, can be determined [Set I, Case IV] in a closed form. The phase factor is found to be,
\begin{align}
&\Theta_{\,n\,,\,l}(t)\,=
\dfrac{2(n+l)}{\Gamma} \left[\sqrt{\dfrac{\Delta}{M}-\omega_0^2}-\sqrt{\dfrac{\Delta}{M}-\omega_0^2 e^{-\Gamma\,t}} \right]+ (n+l)\left[-\dfrac{\sigma}{\mu^2}\right]t \nonumber\\
&+\dfrac{2(n+l)}{\Gamma}\left \lbrace \sqrt{\dfrac{q^2B_0^2\sigma}{4M}+\omega_0^2(M\sigma-1)}-\sqrt{\dfrac{q^2B_0^2\sigma}{4M}+\omega_0^2(M\sigma-1)e^{-\Gamma\,t}} \right \rbrace\nonumber\\
&+\dfrac{i(n+l)}{M\Gamma}\sqrt{q^2B_0^2+4M\Delta}\left [tan^{-1}\dfrac{\sqrt{q^2B_0^2+4M\Delta}}{2\sqrt{M^2\omega_0^2-M\Delta}}-tan^{-1}\dfrac{\sqrt{q^2B_0^2+4M\Delta}}{2\sqrt{M^2\omega_0^2e^{-\Gamma\,t}-M\Delta}}  \right]
\nonumber\\
&+\dfrac{qB_0(n+l)}{2M\Gamma}\,log\,\dfrac{{(4M^2\omega_0^2+q^2B_0^2) \left \lbrace 
\left(\sqrt{\omega_0^2(M\sigma-1)+\dfrac{q^2B_0^2\sigma}{4M}e^{\Gamma\,t}}+\dfrac{qB_0\sigma}{2}e^{\Gamma\,t/2}        \right)^2  +\omega_0^2(M\sigma-1)^2 \right \rbrace}}{(4M^2\omega_0^2+q^2B_0^2\,e^{\Gamma\,t})\left \lbrace 
\left(\sqrt{\omega_0^2(M\sigma-1)+\dfrac{q^2B_0^2\sigma}{4M}}+\dfrac{qB_0\sigma}{2}        \right)^2 +\omega_0^2(M\sigma-1)^2 \right \rbrace} \nonumber\\
&+\dfrac{(n+l)\sqrt{q^2B_0^2+4M\Delta}}{2M\Gamma}\times\nonumber\\
&\,log\,\dfrac{\left[(4M^2\omega_0^2+q^2B_0^2)\left \lbrace q^2B_0^2\sigma M^2\omega_0^2+4M^3\sigma\Delta\omega_0^2-2M\omega_0^2q^2B_0^2-4\omega_0^2M^2\Delta+e^{\Gamma\,t}\left(\dfrac{q^4B_0^4\sigma}{4}+q^2B_0^2\Delta\right)+  \atop e^{\Gamma\,t}q^2B_0^2\sigma M\Delta-2qB_0\sqrt{(q^2B_0^2+4M\Delta)\left(\omega_0^2[M\sigma-1]+\dfrac{q^2B_0^2\sigma}{4M}e^{\Gamma\,t} \right)(M\Delta e^{\Gamma\,t}-M^2\omega_0^2)}              
  \right\rbrace\right] }{\left[(4M^2\omega_0^2+q^2B_0^2 e^{\Gamma\,t})\left \lbrace q^2B_0^2\sigma M^2\omega_0^2+4M^3\sigma\Delta\omega_0^2-2M\omega_0^2q^2B_0^2-4\omega_0^2M^2\Delta+\dfrac{q^4B_0^4\sigma}{4}+q^2B_0^2\Delta  \atop +q^2B_0^2\sigma M\Delta-2qB_0\sqrt{(q^2B_0^2+4M\Delta)\left(\omega_0^2[M\sigma-1]+\dfrac{q^2B_0^2\sigma}{4M} \right)(M\Delta -M^2\omega_0^2)}              
  \right\rbrace\right] }\nonumber\\
&+\dfrac{(n+l)\sqrt{\sigma\Delta}}{\Gamma}\,log\,\dfrac{\left [{\omega_0^2M^2\sigma\Delta-M\Delta\omega_0^2-\dfrac{q^2B_0^2\sigma M\omega_0^2}{4}+\dfrac{q^2B_0^2\sigma\Delta}{2}e^{\Gamma\,t}\atop + \sqrt{q^2B_0^2\sigma\Delta\left(M\Delta e^{\Gamma\,t}-M^2\omega_0^2\right)\left(\omega_0^2[M\sigma-1]+\dfrac{q^2B_0^2\sigma}{4M}e^{\Gamma\,t} \right)} }\right ]}{\left[{\omega_0^2M^2\sigma\Delta-M\Delta\omega_0^2-\dfrac{q^2B_0^2\sigma M\omega_0^2}{4}+\dfrac{q^2B_0^2\sigma\Delta}{2}\atop +\sqrt{q^2B_0^2\sigma\Delta\left(M\Delta -M^2\omega_0^2\right)\left(\omega_0^2[M\sigma-1]+\dfrac{q^2B_0^2\sigma}{4M} \right)}}\right] }+\dfrac{(n+l)qB_0}{2M\Gamma}log\,\dfrac{q^2B_0^2+4M^2\omega_0^2e^{-\Gamma\,t}}
{q^2B_0^2+4M^2\omega_0^2}\nonumber\\
&+\dfrac{(n+l)iqB_0\sqrt{M\sigma-1}}{2M\Gamma} \times \nonumber\\
 &\left[log\dfrac{{M\sigma\Delta-\Delta-\dfrac{q^2B_0^2\sigma }{4}-2M\omega_0^2(M\sigma-1)e^{-\Gamma\,t}-2i\sqrt{(M\sigma-1)\left(\omega_0^2[M\sigma-1]e^{-\Gamma\,t}+\dfrac{q^2B_0^2\sigma}{4M}\right)\atop \times(M\Delta-M^2\omega_0^2e^{-\Gamma\,t})~} }}{{M\sigma\Delta-\Delta-\dfrac{q^2B_0^2\sigma }{4}-2M\omega_0^2(M\sigma-1)-2i\sqrt{(M\sigma-1)\left(\omega_0^2[M\sigma-1]+\dfrac{q^2B_0^2\sigma}{4M}\right)(M\Delta-M^2\omega_0^2)}}}\right]~.\nonumber\\
\end{align}
In the regard of the eigenstate associated with the rational EP solution set, we mentioned below Eqn.(\ref{2ratdecC}) that the the explicit structure of the Lewis phase factor which is essential to prepare the eigenstate of a damped oscillator with a rationally decreasing frequency, under the influence of a rationally decreasing magnetic field in NC background, can be determined [Set II, Case I] in a closed form. The phase factor is found to be,
\begin{align}
&\Theta_{\,n\,,\,l}(t)\,\nonumber\\
&= -\dfrac{2(n+l)}{\Gamma}\left(\dfrac{\sigma}{\mu^2}+\dfrac{qB_0M\omega_0^2}{4M^2\omega_0^2+q^2B_0^2} \right)log\,\dfrac{\Gamma\,t+\chi}{\chi}+\dfrac{4(n+l)M^2\omega_0^2}{(q^2B_0^2+4M^2\omega_0^2)\Gamma}\left[ \sqrt{\left( \dfrac{q^2B_0^2\sigma}{M}+4\omega_0^2\sigma\,M\right)\dfrac{1}{\chi^2} \atop -\omega_0^2}  \right.\nonumber\\
-&\left.\sqrt{\left( \dfrac{q^2B_0^2\sigma}{M}+4\omega_0^2\sigma\,M\right)\dfrac{1}{(\Gamma\,t+\chi)^2}-\omega_0^2}\,+\omega_0\left\lbrace tan^{-1}\dfrac{\sqrt{M}~\omega_0\chi}{\sqrt{\left(q^2B_0^2\sigma+4\omega_0^2\sigma\,M^2\right)-\omega_0^2\chi^2M}}\right.\right.\nonumber\\
&\left.\left.
-tan^{-1}\dfrac{\sqrt{M}~\omega_0(\Gamma\,t+\chi)}{\sqrt{\left(q^2B_0^2\sigma+4\omega_0^2\sigma\,M^2\right)-\omega_0^2M(\Gamma\,t+\chi)^2}}
 \right\rbrace \right]
+\dfrac{2(n+l)qB_0}{q^2B_0^2+4M^2\omega_0^2}
 \left[\dfrac{1}{\Gamma}\left \lbrace \sqrt{\left(\Delta-\dfrac{M\omega_0^2}{\chi^2}\right) \atop \times\left({4M^2\omega_0^2\sigma+\atop q^2B_0^2\sigma-M\omega_0^2\chi^2} \right)~}\right.\right.\nonumber\\
 -&\left.\left.\sqrt{\left(M\Delta-\dfrac{M^2\omega_0^2}{(\Gamma\,t+\chi)^2}\right)\left(\dfrac{q^2B_0^2\sigma}{M}+4\omega_0^2\sigma\,M-\omega_0^2(\Gamma\,t+\chi)^2 \right)} \right\rbrace -\dfrac{2iM\omega_0^2}{\Gamma}\left\lbrace EllipticE \atop \left(isinh^{-1}\left[\dfrac{i\omega_0\sqrt{M}(\Gamma\,t+\chi)}{\sqrt{q^2B_0^2\sigma \atop +4\omega_0^2\sigma M^2}}\right]\right.\right.\right.
 \nonumber\\
 ,&\left.\left.\left.\dfrac{q^2B_0^2\sigma\Delta+4\omega_0^2\sigma M^2\Delta}{M^2\omega_0^4}\right) -EllipticE\left(isinh^{-1}\left[\dfrac{i\omega_0\sqrt{M}\chi}{\sqrt{q^2B_0^2\sigma+4\omega_0^2\sigma M^2}}\right],\dfrac{q^2B_0^2\sigma\Delta+4\omega_0^2\sigma M^2\Delta}{M^2\omega_0^4}\right) \right\rbrace     \right.\nonumber\\
&-\left. \dfrac{i(q^2B_0^2\sigma\Delta+4\omega_0^2\sigma M^2\Delta-M^2\omega_0^4)}{\Gamma M\omega_0^2}\left\lbrace
EllipticF\left(isinh^{-1}\left[\dfrac{i\omega_0\sqrt{M}(\Gamma\,t+\chi)}{\sqrt{q^2B_0^2\sigma+4\omega_0^2\sigma M^2}}\right],\dfrac{q^2B_0^2\sigma\Delta+4\omega_0^2\sigma M^2\Delta}{M^2\omega_0^4}\right)\right.\right.\nonumber\\
-&\left.\left.EllipticF\left(isinh^{-1}\left[\dfrac{i\omega_0\sqrt{M}\chi}{\sqrt{q^2B_0^2\sigma+4\omega_0^2\sigma M^2}}\right],\dfrac{{q^2B_0^2\sigma\Delta+\atop 4\omega_0^2\sigma M^2\Delta}}{M^2\omega_0^4}\right)    \right\rbrace          \right] +\dfrac{(n+l)}{\Gamma}\left[{\sqrt{\dfrac{\Delta}{M}(\Gamma\,t+\chi)^2-\omega_0^2} \atop -\sqrt{\dfrac{\Delta}{M}\chi^2-\omega_0^2}}\right.\nonumber\\
+&\left.\omega_0\,tan^{-1}\sqrt{\dfrac{M\omega_0^2}{\Delta(\Gamma\,t+\chi)^2-M\omega_0^2}} -\omega_0\,tan^{-1}\sqrt{\dfrac{M\omega_0^2}{\Delta\chi^2-M\omega_0^2}} \right]-\dfrac{(n+l)q^2B_0^2}{M(4M^2\omega_0^2+q^2B_0^2)\Gamma}\left[\sqrt{M\Delta(\Gamma\,t+\chi)^2 \atop -M^2\omega_0^2}\right.
\nonumber\\ 
-&\left.\sqrt{M\Delta\chi^2-M^2\omega_0^2} 
+M\omega_0\,\left(tan^{-1}\dfrac{M\omega_0}{\sqrt{M\Delta(\Gamma\,t+\chi)^2-M^2\omega_0^2}}-\,tan^{-1}\dfrac{M\omega_0}{\sqrt{M\Delta\chi^2-M^2\omega_0^2}}   \right)\right]~.
\end{align}
In the above expression $EllipticF$ and $EllipticE$ respectively denote the incomplete elliptic integrals of the first and second kinds.
\section*{Appendix B: Associated Laguerre polynomial identity}

\noindent Here we intend to prove an identity regarding the associated Laguerre polynomial. The identity is given by
\begin{align}
(n-m+1)\,\int_0^\infty\,z^{n-m}\,e^{-z}\,L^{n-m}_m(z)\,L^{n-m+1}_{m-1}(z)\,dz\,-\,\int_0^\infty\,z^{n-m+1}\,e^{-z}\,L^{n-m}_m(z)\,L^{n-m+2}_{m-2}(z)\,dz=0~.\label{ap1}
\end{align}
First, we require to mention another identity regarding the same polynomial. That is given by
\begin{align}
\,z\,L^{n-m+1}_{m-1}(z)\,=\,n\,L^{n-m}_{m-1}(z)-\,m\,L^{n-m}_{m}(z)~.\label{ap2}
\end{align}
The left hand side of Eqn.(\ref{ap1}) can be expressed as,
\begin{align}
(n-m+1)\,\int_0^\infty\,z^{n-m}\,e^{-z}\,L^{n-m}_m(z)\,L^{n-m+1}_{m-1}(z)\,dz\,-\,\int_0^\infty\,z^{n-m}\,e^{-z}\,L^{n-m}_m(z)\,\left[z\,L^{n-m+2}_{m-2}(z)\right]\,dz=0~.\label{ap3}
\end{align}
By using the identity provided in Eqn.(\ref{ap2}), 
\begin{align}
z\,L^{n-m+2}_{m-2}(z)\,=\,n\,L^{n-m+1}_{m-2}(z)-\,(m-1)\,L^{n-m+1}_{m-1}(z)~.\label{ap4}
\end{align}
After the substitution of the above identity into Eqn.(\ref{ap3}), we derive the following relation
\begin{align}
(n-m+1)\,\int_0^\infty\,z^{n-m}\,e^{-z}\,L^{n-m}_m(z)\,L^{n-m+1}_{m-1}(z)\,dz-\,\int_0^\infty\,z^{n-m}\,e^{-z}\,L^{n-m}_m(z)\,\left[z\,L^{n-m+2}_{m-2}(z)\right]\,dz\nonumber\\
=n\,\int_0^\infty\,z^{n-m}\,e^{-z}\,L^{n-m}_{m}(z)\,\left[\,L^{n-m+1}_{m-1}(z)-L^{n-m+1}_{m-2}(z)\,\right]\,dz~.
\end{align}

\noindent Utilizing the second relation given in Eqn.(\ref{psqoi}), we simplify the above relation to be,
\begin{align}
n\,\int_0^\infty\,z^{n-m}\,e^{-z}\,L^{n-m}_{m}(z)\,\left[\,L^{n-m+1}_{m-1}(z)-L^{n-m+1}_{m-2}(z)\,\right]\,dz
=n\,\int_0^\infty\,z^{n-m}\,e^{-z}\,L^{n-m}_m(z)\,L^{n-m}_{m-1}(z)\,dz~.
\end{align}
The orthonormality relation given in Eqn.(\ref{psqoi}) leads the right hand side of the above relation to be zero. Hence,
\begin{align}
(n-m+1)\,\int_0^\infty\,z^{n-m}\,e^{-z}\,L^{n-m}_m(z)\,L^{n-m+1}_{m-1}(z)\,dz\,&-\,\int_0^\infty\,z^{n-m+1}\,e^{-z}\,L^{n-m}_m(z)\,L^{n-m+2}_{m-2}(z)\,dz\nonumber\\
&=n\,\int_0^\infty\,z^{n-m}\,e^{-z}\,L^{n-m}_m(z)\,L^{n-m}_{m-1}(z)\,dz\nonumber\\
&=0~.
\end{align}





\end{document}